%% file: paper.tex
\documentclass[12pt,a4paper,notitlepage]{article}

\usepackage{graphicx}
\usepackage{amssymb}
\usepackage{amsmath}

\usepackage{amsthm}
\usepackage{amsfonts}
\usepackage{times}

\usepackage[format=hang]{subcaption}
\usepackage{color}
\usepackage[T1]{fontenc}
\usepackage{hyperref}
\usepackage[square]{natbib}
\usepackage{upgreek}

\usepackage{pgfplots}
\usepackage{bm}

\usepackage{float}

\usepackage{todonotes}

\usepackage{subcaption} 
\usepackage{siunitx}
\usepackage{tikz}
\usepackage{tikz-3dplot}
\usetikzlibrary{calc,backgrounds,spy}

\newcommand{\re}[1]{\textcolor{black}{#1}}

\defcitealias{KlockeZeisEtAl2014}{Klocke, Zeis et~al.}
\defcitealias{KlockeKlinkEtAl2014}{Klocke, Klink et~al.}
\defcitealias{HolthusenBrepolsEtAl2022a}{Holthusen et~al.}
\defcitealias{HolthusenBrepolsEtAl2022b}{Holthusen et~al.}
\defcitealias{BarfuszBrepolsEtAl2021}{Barfusz et~al.}
\defcitealias{BarfuszvanderVeldenEtAl2021}{Barfusz et~al.}
\defcitealias{BarfuszvanderVeldenEtAl2021}{Barfusz et~al.}
\defcitealias{vanderVeldenRommesEtAl2021}{van der Velden et~al.}

\pgfrealjobname{paper} 
\pgfplotsset{compat=1.14}

\input{00_Input/inputColorsDashpatterns}
\input{00_Input/inputECM}

\input{00_Input/inputDamage}
\AtBeginDocument{}

\date{}

\graphicspath{{./02_Figures/}}

\begin{document}

\author{\large {\parbox{\linewidth}{\centering Tim van der Velden$^a$,
                                               Stefanie Reese$^{a,b}$,
                                               Hagen Holthusen$^a$,
                                               Tim Brepols$^{a,}$\footnote{Corresponding author: \\ phone: +49 (0) 241 80 25002, fax: +49 (0) 241 80 22001, email: tim.brepols@rwth-aachen.de}}}\\[0.5cm]
  \hspace*{-0.1cm}
  \normalsize{\em \parbox{\linewidth}{\centering
    \vspace{2mm}
    $^a$Institute of Applied Mechanics, RWTH Aachen University,\\Mies-van-der-Rohe-Str. 1, D-52074 Aachen, Germany\\    
    $^b$University of Siegen, Adolf-Reichwein-Str. 2a, D-57076 Siegen, Germany
    }
  }
}

\title{\LARGE An anisotropic, brittle damage model for finite strains with a generic damage tensor regularization}

\maketitle

\vspace{-3mm}

\small
{\bf Abstract.} {
\input{01_Sections/abstract.tex}
}


\normalsize

\input{01_Sections/content.tex}

\input{01_Sections/appendix.tex}

\bibliographystyle{agsm}
\bibliography{literature,literature_additionals_p4,literature_additionals_introduction}

\end{document}

%% file: 00_Input/inputColorsDashpatterns.tex
\definecolor{rwth1}{RGB}{0,84,159}      
\definecolor{rwth2}{RGB}{142,186,229}   
\definecolor{rwth3}{RGB}{0,97,101}      
\definecolor{rwth4}{RGB}{0,152,161}     
\definecolor{rwth5}{RGB}{87,171,39}     
\definecolor{rwth6}{RGB}{189,205,0}     
\definecolor{rwth7}{RGB}{255,237,0}     
\definecolor{rwth8}{RGB}{246,168,0}     
\definecolor{rwth9}{RGB}{227,0,102}     
\definecolor{rwth10}{RGB}{204,7,30}     
\definecolor{rwth11}{RGB}{161,16,53}    
\definecolor{rwth12}{RGB}{97,33,88}     
\definecolor{rwth13}{RGB}{122,111,172}  

\definecolor{modelC}{RGB}{0,84,159}      
\definecolor{modelA}{RGB}{87,171,39}     
\definecolor{modelB}{RGB}{246,168,0}     

\definecolor{sfb1}{RGB}{0,84,165}      
\definecolor{sfb2}{RGB}{201,0,35}      
\definecolor{sfb3}{RGB}{231,95,1}      
\definecolor{sfb4}{RGB}{127,127,127}   
\definecolor{sfb5}{RGB}{217,217,217}   

\tikzstyle{dashpattern0} = [dash pattern = ]
\tikzstyle{dashpattern1} = [dash pattern = on 4.25pt off 0.75pt]
\tikzstyle{dashpattern2} = [dash pattern = on 1.5pt off 0.5pt]
\tikzstyle{dashpattern3} = [dash pattern = on 0.75pt off 0.4pt]
\tikzstyle{dashpattern4} = [dash pattern = on 3pt off 1pt on 1pt off 1pt]
\tikzstyle{dashpattern5} = [dash pattern = on 3.75pt off 0.5pt on 0.75pt off 0.5pt on 0.75pt off 0.5pt]
\tikzstyle{dashpattern6} = [dash pattern = on 3.25pt off 0.5pt on 0.75pt off 0.5pt on 0.75pt off 0.5pt on 0.75pt off 0.5pt]
\tikzstyle{dashpattern7} = [dash pattern = on 3.25pt off 0.5pt on 0.75pt off 0.5pt on 0.75pt off 0.5pt on 0.75pt off 0.5pt on 0.75pt off 0.5pt]
\tikzstyle{dashpattern8} = [line cap=round, dash pattern = on 3.25pt off 2.75pt]
\tikzstyle{dashpattern9} = [line cap=round, dash pattern = on 0.01pt off 2pt]
\tikzstyle{dashpattern10}= [line cap=round, dash pattern = on 3.25pt off 2pt on 0.01pt off 2pt]
\tikzstyle{dashpattern11}= [line cap=round, dash pattern = on 3.5pt off 1.75pt on 0.01pt off 1.75pt on 0.01pt off 1.75pt]
\tikzstyle{dashpattern12}= [line cap=round, dash pattern = on 3.5pt off 1.75pt on 0.01pt off 1.75pt on 0.01pt off 1.75pt on 0.01pt off 1.75pt]
\tikzstyle{dashpattern13}= [line cap=round, dash pattern = on 3.5pt off 1.75pt on 0.01pt off 1.75pt on 0.01pt off 1.75pt on 0.01pt off 1.75pt on 0.01pt off 1.75pt]

%% file: 00_Input/inputECM.tex
\newcommand{\dV}{\mathrm{d}V}

\newcommand{\pd}[2]{\displaystyle\frac{\partial #1}{\partial #2}}

\newcommand{\coloneq}{\mathrel{\resizebox{\widthof{$\mathord{=}$}}{\height}{ $\!\!\resizebox{1.2\width}{0.8\height}{\raisebox{0.23ex}{$\mathop{:}$}}\!\!=\!\!$ }}}

\newcommand{\D}{\bm{D}}
\newcommand{\E}{\bm{E}}

\newcommand{\I}{\mathbf{I}}

\renewcommand{\d}{d}



%% file: 00_Input/inputDamage.tex
\newcommand{\Div}[1]{{\rm Div} \hspace{-0.5mm} \left( #1 \right)}
\newcommand{\dev}[1]{{\rm dev} \hspace{-0.5mm} \left( #1 \right)}
\renewcommand{\exp}[1]{{\rm exp} \hspace{-0.5mm} \left( #1 \right)}

\newcommand{\Grad}[1]{{\rm Grad} \hspace{-0.5mm} \left( #1 \right)}

\newcommand{\sym}[1]{{\rm sym} \hspace{-0.5mm} \left( #1 \right)}
\newcommand{\tr}[1]{{\rm tr} \hspace{-0.5mm} \left( #1 \right)}
\newcommand{\bigtr}[1]{{\rm tr} \hspace{-0.0mm} \big( #1 \big)}
\newcommand{\Bigtr}[1]{{\rm tr} \hspace{-0.0mm} \Big( #1 \Big)}

\newcommand{\Tzd}{^{\stackrel{23}{\mathrm{T}}}}

\newcommand{\dA}{\mathrm{d}A}
\renewcommand{\dV}{\mathrm{d}V}

\newcommand{\intGtn}{\int_{\Gamma_{t0}}}

\newcommand{\intOn}{\int_{\Omega_0}}

\renewcommand{\pd}[2]{\displaystyle\frac{\partial #1}{\partial #2}}

\renewcommand{\coloneq}{\mathrel{\resizebox{\widthof{$\mathord{=}$}}{\height}{ $\!\!\resizebox{1.2\width}{0.8\height}{\raisebox{0.23ex}{$\mathop{:}$}}\!\!=\!\!$ }}}
\newcommand{\ah}{a_h}

\newcommand{\cd}{{c_d}}
\renewcommand{\d}{\mathbf{d}}
\newcommand{\dbar}{\bar{\mathbf{d}}}
\newcommand{\dbardot}{\dot{\bar{\mathbf{d}}}}
\newcommand{\dbari}{\bar{d}_i}

\newcommand{\e}{\bm{e}}
\newcommand{\ed}{{e_d}}
\newcommand{\fiso}{f_\mathrm{iso}}
\newcommand{\fani}{f_\mathrm{ani}}
\newcommand{\fd}{{f_d}}
\newcommand{\fn}{\bm{f}_0}

\newcommand{\gddot}{\dot{\gamma}_d}
\newcommand{\gdbar}{g_{\bar{d}}}
\newcommand{\gu}{g_u}
\newcommand{\Hd}{H_d}

\newcommand{\kani}{k_\mathrm{ani}}
\newcommand{\Kh}{K_h}
\newcommand{\ndbar}{{n_{\bar{d}}}}
\newcommand{\niD}{\bm{n}_i^{\D}}

\newcommand{\nh}{n_h}
\newcommand{\niY}{\bm{n}_i^{\Y}}
\newcommand{\Phid}{\Phi_d}
\newcommand{\psidot}{\dot{\psi}}
\newcommand{\psie}{\psi_e}
\newcommand{\psid}{\psi_d}
\newcommand{\psih}{\psi_h}
\newcommand{\psidbar}{\psi_{\bar{d}}}
\newcommand{\psinh}{\psi_\mathrm{\scriptscriptstyle NH}}
\newcommand{\rd}{r_d}
\newcommand{\Rd}{R_d}

\newcommand{\sd}{s_d}
\newcommand{\tn}{\bm{t}_0}
\renewcommand{\u}{\bm{u}}
\newcommand{\vardbar}{\delta\dbar}
\newcommand{\varE}{\delta\bm{E}}
\newcommand{\varu}{\delta\bm{u}}
\newcommand{\xid}{\xi_d}
\newcommand{\xiddot}{\dot{\xi}_d}
\newcommand{\xinc}{\bm{\xi}_{0_c}}
\newcommand{\xine}{\bm{\xi}_{0_e}}
\newcommand{\xini}{\bm{\xi}_{0_i}}
\newcommand{\Xine}{\bm{\Xi}_{0_e}}
\newcommand{\Xini}{\bm{\Xi}_{0_i}}

\newcommand{\Yn}{Y_0}

\newcommand{\A}{\mathbb{A}}

\newcommand{\C}{\bm{C}}
\newcommand{\Cdot}{\dot{\bm{C}}}

\renewcommand{\D}{\bm{D}}
\renewcommand{\Ddot}{\dot{\bm{D}}}

\renewcommand{\E}{\bm{E}}
\newcommand{\F}{\bm{F}}

\renewcommand{\I}{\bm{I}}

\renewcommand{\S}{\bm{S}}

\newcommand{\Snh}{\bm{S}_\mathrm{\scriptscriptstyle NH}}
\newcommand{\Y}{\bm{Y}}
\newcommand{\Ydbar}{\bm{Y}_{\bar{d}}}
\newcommand{\Ye}{\bm{Y}_e}
\newcommand{\Yh}{\bm{Y}_h}
\newcommand{\Ypos}{\bm{Y}_+}

%% file: 01_Sections/abstract.tex
This paper establishes a universal framework for the nonlocal modeling of anisotropic damage at finite strains. By the combination of two recent works, the new framework allows for the flexible incorporation of different established hyperelastic finite strain material formulations into anisotropic damage whilst ensuring mesh-independent results by employing a generic set of micromorphic gradient-extensions. First, the anisotropic damage model, generally satisfying the damage growth criterion, is investigated for the specific choice of a Neo-Hookean material on a single element. Next, the model is applied with different gradient-extensions in structural simulations of an asymmetrically notched specimen to identify an efficient choice in the form of a volumetric-deviatoric regularization. Thereafter, the universal framework, which is without loss of generality here specified for a Neo-Hookean material with a volumetric-deviatoric gradient-extension, successfully serves for the complex simulation of a pressure loaded rotor blade.

After acceptance of the manuscript, we make the codes of the material subroutines accessible to the public at \href{https://doi.org/10.5281/zenodo.11171630}{\textit{https://doi.org/10.5281/zenodo.11171630}}.

%% file: 01_Sections/content.tex
\section{Introduction}
\label{sec:p4_intro}

\textbf{Motivation.\quad}
The prediction of structural failure requires the precise local modeling of damage and the accurate nonlocal regularization of the softening phenomenon in structural simulations. Both aspects, the modeling and regularization of softening due to damage, constitute fundamental fields of research in continuum mechanics that continuously generate enhanced local and nonlocal solutions. This motivates the proposition of a general modeling framework that combines a flexible local anisotropic damage model with different nonlocal gradient-extensions in this work. The following paragraphs only provide an overview of recent contributions to both fields. For a general overview on damage modeling, we refer to the works of e.g.~\cite{LemaitreDesmorat2005}, \cite{VoyiadjisKattan2009}, \cite{Besson2010}, and \cite{Murakami2012}.

\textbf{Modeling of softening.\quad}
\cite{MattielloDesmorat2021} investigate different evolution laws for symmetric second-order damage tensors with respect to the Lode angle dependency.
\cite{DornWulfinghoff2021} assume a multiplicative decomposition of the deformation gradient into normal and shear crack contributions that directly yields a tension-compression asymmetry.
For low cycle fatigue, \cite{FerreiraCamposEtAl2022} propose a damage evolution law depending on the stress triaxiality and test the model in material point studies for multiaxial and nonproportional loading paths.
In \cite{ReeseBrepolsEtAl2021}, a local formulation for anisotropic damage is introduced that can incorporate arbitrary hyperelastic energies and generally fulfills the damage growth criterion (\cite{WulfinghoffFassinEtAl2017}).
\cite{PetriniEstevesEtAl2023} use a dynamic phase-field model with a fourth-order degradation tensor to account for damage anisotropy at infinitesimal strains.
Based on the work by \cite{VoyiadjisKattan2017} and \cite{Basaran2023}, \cite{VoyiadjisKattan2024} present an unsymmetrical decomposition of the tensorial damage variable to separately account for crack and void induced damage.
To capture stiffness recovery after damage evolution, \cite{ShojaeiVoyiadjis2023} lay the theoretical groundwork for statistical continuum damage healing mechanics by considering the sealing and healing effects.
\cite{LoiseauOliver-LeblondEtAl2023} employ a data-driven approach to identify an accurate tensorial representation of anisotropic damage to account for micro-cracking based on a virtual set of beam-particle simulations.

\textbf{Regularization of softening.\quad}
\cite{JirasekDesmorat2019} investigate the regularization performance of a new nonlocal integral approach with a damage-dependent interaction distance for pure damage models, damage-plasticity models, and damage with inelastic strain models. 
\cite{AhmedVoyiadjisEtAl2021} present a damage model for concrete that captures tensile, compressive, and shear damage with separate scalar damage variables, where each quantity is regularized by a gradient-extension with three individual length scales.
In \cite{ZhangXuEtAl2022}, a convenient approach for the implementation of gradient-extended damage into the finite element software \textit{Abaqus} is presented and validated in complex structural simulations, where in-built \textit{Abaqus} features, like e.g.~contact or element deletion, are additionally exploited.
In \cite{HolthusenBrepolsEtAl2020,HolthusenBrepolsEtAl2022a}, a general notation for micromorphic gradient-extensions is formulated based on a micromorphic tuple, which is investigated for anisotropic damage with different gradient-extensions in \cite{vanderVeldenBrepolsEtAl2024}.
\cite{SpraveMenzel2023} formulate a ductile anisotropic damage model at finite strains and analyze the isolated and combined regularization of damage and plasticity.
The regularization of ductile damage in the logarithmic strain space is studied in \cite{FriedleinMergheimEtAl2023} with a gradient-extension of the plastic hardening variable or the local damage variable.
For damage in semicrystalline polymers, \cite{SatouriChatzigeorgiouEtAl2022} examine the regularization effects of the damage variable and the hardening state variable.

\textbf{Current and future works.\quad}
In this work, we combine the elastic energy of the anisotropic damage model of \cite{ReeseBrepolsEtAl2021} with the micromorphic gradient-extensions investigated in \cite{vanderVeldenBrepolsEtAl2024} to obtain a universal framework for regularized anisotropic damage at finite strains. The formulation enables the incorporation of arbitrary hyperelastic finite strain energies and, thereby, yields a flexible local damage model. The generic gradient-extensions provide additionally a nonlocal flexibility in terms of the regularized quantities and the number of nonlocal degrees of freedom. In future works, the \textit{iCANN} framework of \cite{HolthusenLammEtAl2024} can provide a functional basis for the elastic energy as well as the gradient-extension and can identify the optimal set of parameters.

\textbf{Outline of the work.\quad}
The constitutive framework is presented in Section~\ref{sec:p4_modeling}. The numerical examples are provided in Section~\ref{sec:p4_examples} with single element studies in Section~\ref{sec:p4_ses}, the study of an asymmetrically notched specimen in Section~\ref{sec:p4_ANotched}, and the study of a rotor blade specimen in Section~\ref{sec:p4_RB}. The conclusions are drawn in Section~\ref{sec:p4_conclusion}.

\textbf{Notational conventions.\quad}
In this work, italic characters $a$, $A$ denote scalars (i.e.~zeroth-order tensors) and bold-face italic characters $\bm{b}$, $\bm{B}$ refer to first- and second-order tensors. The operators $\Div{\bullet}$ and $\Grad{\bullet}$ denote the divergence and gradient operation of a quantity with respect to the reference configuration. A $\cdot$ defines the single contraction and a $:$ the double contraction of two tensors. The time derivative of a quantity is given by $\dot{(\bullet)}$ and a fixed variable value by $(\bullet)'$.

\section{Constitutive modeling}
\label{sec:p4_modeling}

\subsection{Balance equations}

The constitutive framework is based on the micromorphic approach of \cite{Forest2009,Forest2016} and is formulated in the reference configuration. It comprises the balance of linear momentum
\begin{align}
  \Div{\F\S} + \bm{f}_0 &= \bm{0}    \hphantom{ \bm{t}_0 \bm{u}'  } \text{in}~\Omega_0    \label{eq:p4_blm}\\
  \F\S \cdot \bm{n}_0   &= \bm{t}_0  \hphantom{ \bm{0}   \bm{u}'  } \text{on}~\Gamma_{t0} \label{eq:p4_blmNBC}\\
  \u                    &= \bm{u}'   \hphantom{ \bm{0}   \bm{t}_0 } \text{on}~\Gamma_{u0} \label{eq:p4_blmDBC}
\end{align}
and the micromorphic balance equation
\begin{align}
  \Div{\Xini - \Xine} - \xini + \xine           &= \bm{0}  \hphantom{ \xinc \dbar'  } \text{in}~\Omega_0          \label{eq:p4_bmm}\\
  \left( \Xini - \Xine \right) \cdot \bm{n}_0   &= \xinc   \hphantom{ \bm{0} \dbar' } \text{on}~\Gamma_{c0}       \label{eq:p4_bmmNBC}\\
  \dbar                                         &= \dbar'  \hphantom{ \bm{0} \xinc  } \text{on}~\Gamma_{\bar{d}0} \label{eq:p4_bmmDBC}
\end{align}
with their corresponding Neumann and Dirichlet boundary conditions. In Eqs.~\ref{eq:p4_blm}-\ref{eq:p4_bmmDBC}, $\F$ denotes the deformation gradient, $\S$ the second Piola-Kirchhoff stress tensor, $\bm{f}_0$ the volume forces, $\bm{n}_0$ the outward normal vector, $\bm{t}_0$ the prescribed tractions on the mechanical Neumann boundary $\Gamma_{t0}$, $\u$ the mechanical displacements, and $\bm{u}'$ their fixed values on the mechanical Dirichlet boundary $\Gamma_{u0}$. Furthermore, $\Xini$ and $\xini$ denote the micromorphic internal forces, $\Xine$ and $\xine$ the micromorphic external forces, $\xinc$ the micromorphic contact forces on the micromorphic Neumann boundary $\Gamma_{c0}$, $\dbar$ the tuple of the nonlocal field variables, and $\dbar'$ its fixed value on the micromorphic Dirichlet boundary $\Gamma_{\bar{d}0}$. Analogously to \cite{BrepolsWulfinghoffEtAl2020}, \cite{HolthusenBrepolsEtAl2020,HolthusenBrepolsEtAl2022a}, the micromorphic external and contact forces are neglected in this work, i.e.~$\Xine=\bm{0}$, $\xine=\bm{0}$ and $\xinc=\bm{0}$, and micromorphic Dirichlet boundary conditions are not considered, i.e.~$\Gamma_{\bar{d}0}=\emptyset$. With these simplifications and using the test functions $\varu$ and $\vardbar$, the following weak forms with the virtual Green-Lagrange strain $\varE$ are obtained (cf.~\cite{HolthusenBrepolsEtAl2022a})
\begin{align}
  \gu \left( \u,\dbar,\varu \right)       &:= \intOn \S : \varE \, \dV - \intOn \fn \cdot \varu \, \dV - \intGtn \tn \cdot \varu \, \dA = 0, \\
  \gdbar \left( \u,\dbar,\vardbar \right) &:=  \intOn \xini \cdot \vardbar \, \dV + \intOn \Xini : \Grad{\vardbar} \, \dV = 0.
\end{align}

\subsection{Helmholtz free energy}

The Helmholtz free energy consists of four parts that additively compose the total energy
\begin{equation}
  \psi \left( \C, \D, \xid, \d, \dbar, \Grad{\dbar} \right)
  =
  \psie \left( \C, \D \right) + \psid \left( \xid \right) + \psih \left( \D \right) + \psidbar \left( \d, \dbar, \Grad{\dbar} \right)
\end{equation}
where $\psie$ denotes the elastic energy part depending on the right Cauchy-Green stretch tensor $\C$ and the second-order damage tensor $\D$, $\psid$ denotes the isotropic damage hardening energy part depending on the accumulated damage hardening variable $\xid$, $\psih$ denotes the additional kinematic damage hardening energy part depending on $\D$, and $\psidbar$ denotes the micromorphic energy contribution part depending on the local micromorphic tuple $\d$, the nonlocal counterpart $\dbar$ and its gradient $\Grad{\dbar}$.

\textbf{Remark -- Novelty.\quad}
This work combines two recently developed research results to this new model, where both findings are reflected in the Helmholtz free energy. The first core element is the elastic energy $\psie \left( \C, \D \right)$ formulated in line with \cite{ReeseBrepolsEtAl2021}, which fulfills the damage growth criterion of \cite{WulfinghoffFassinEtAl2017} at finite strains for arbitrary hyperelastic material models. In \cite{ReeseBrepolsEtAl2021}, a local anisotropic damage model is utilized. The second core element is the generic formulation of the micromorphic gradient-extension reflected by $\psidbar \left( \d, \dbar, \Grad{\dbar} \right)$ that is studied in \cite{vanderVeldenBrepolsEtAl2024}. However, in the latter work, the elastic energy formulation for anisotropic damage is based on a logarithmic strain formulation. Now, the new model combines the versatile elastic energy for anisotropic damage depending on the right Cauchy-Green stretch tensor $\C$ with a flexible gradient-extension at finite strains.

\subsection{Isothermal Clausius-Duhem inequality}
\label{sec:p4_CD}

The isothermal Clausius-Duhem inequality including the micromorphic parts (cf.~\cite{Forest2009,Forest2016}) reads
\begin{equation}
  - \psidot + \frac{1}{2} \S : \Cdot + \xini \cdot \dbardot + \Xini : \Grad{ \dbardot } \geq 0
  \label{eq:p4_CD}
\end{equation}
with the rate of the Helmholtz free energy being
\begin{equation}
  \psidot
  =
  \pd{\psie}{\C} : \Cdot
  +
  \bigg(
    \underbrace{\pd{\psie}{\D}}_{=: - \Ye}
    +
    \underbrace{\pd{\psih}{\D}}_{=: \Yh}
    +
    \underbrace{\pd{\psidbar}{\D}}_{=: \Ydbar}
  \bigg) : \Ddot
  +
  \underbrace{\pd{\psid}{\xid}}_{=: - \Rd} \xiddot
  +
  \pd{\psidbar}{\dbar} \cdot \dbardot
  +
  \pd{\psidbar}{\Grad{\dbar}} : \Grad{ \dbardot }.
  \label{eq:p4_psidot}
\end{equation}
The insertion of Eq.~\eqref{eq:p4_psidot} in Eq.~\eqref{eq:p4_CD} yields
\begin{align}
  \left( \frac{1}{2} \S - \pd{\psie}{\C} \right) : \Cdot
  +
  (\underbrace{\Ye - \Yh - \Ydbar}_{=: \Y}) : \Ddot
  +
  \Rd \, \xiddot \hspace{18mm} \notag \\
  +
  \left( \xini - \pd{\psidbar}{\dbar} \right) \cdot \dbardot
  +
  \left( \Xini - \pd{\psidbar}{\Grad{\dbar}} \right) : \Grad{ \dbardot }
  \geq 0
  \label{eq:p4_CDreformulated}
\end{align}
from which the state laws (cf.~\cite{BrepolsWulfinghoffEtAl2020},\cite{HolthusenBrepolsEtAl2022a}) for the mechanical and generalized micromorphic stresses are obtained as 
\begin{equation}
  \S = 2 \, \pd{\psie}{\C},
  \label{eq:p4_SPK}
\end{equation}
\begin{equation}
  \xini = \pd{\psidbar}{\dbar},
  \label{eq:p4_xini}
\end{equation}
\begin{equation}
  \Xini = \pd{\psidbar}{\Grad{\dbar}}.
  \label{eq:p4_Xini}
\end{equation}
After the definition of the state laws, the reduced dissipation inequality follows from Eq.~\eqref{eq:p4_CDreformulated} with the thermodynamic driving forces  $\Y := -\partial\psi/\partial\D$ and $\Rd$ as
\begin{equation}
  \Y : \Ddot + \Rd \, \xiddot \geq 0.
\end{equation}

\subsection{Damage onset criterion and evolution equations}

The damage onset criterion incorporates distortional damage hardening and is formulated analogously to \cite{HolthusenBrepolsEtAl2022a,HolthusenBrepolsEtAl2022b} reading
\begin{equation}
  \Phid \coloneq \sqrt{3} \sqrt{\Ypos : \A : \Ypos} - ( \Yn - \Rd ) \leq 0
  \label{eq:Phid}
\end{equation}
where $\A$ is a fourth-order interaction tensor
\begin{equation}
  \A = \left( \left( \I - \D \right)^\cd \otimes \left( \I - \D \right)^\cd \right)\Tzd
\end{equation}
and $\Ypos$ the positive semi-definite part of the damage driving force with $\left< \bullet \right> = \mathrm{max}( \bullet, 0 )$ 
\begin{equation}
  \Ypos = \sum_{i=1}^3 \left< Y_i \right> \niY \otimes \niY
\end{equation}
where $Y_i$ and $\niY$ denote the eigenvalues and eigenvectors of $\Y$. The associative evolution equations for the internal variables read
\begin{equation}
  \Ddot = \gddot \, \pd{\Phid}{\Y},
\end{equation}
\begin{equation}
  \xiddot = \gddot \, \pd{\Phid}{\Rd}
\end{equation}
with the Karush-Kuhn-Tucker conditions
\begin{equation}
  \gddot \geq 0, \quad \Phid \leq 0, \quad \gddot \, \Phid = 0.
\end{equation}

\subsection{Specific Helmholtz free energies and micromorphic tuples}
\label{sec:p4_specificforms}

The elastic Helmholtz free energy is formulated in line with \cite{ReeseBrepolsEtAl2021}, where a hyperelastic energy formulation is multiplied with a linear combination of two degradation functions, and reads
\begin{equation}
  \psie = \left( \left( 1 - \kani \right) \fiso + \kani \, \fani \right) \psi_\star
  \label{eq:p4_psie}
\end{equation}
where the material parameter $\kani \in [0,1]$ defines the degree of damage anisotropy and where the degradation functions $\fiso(\D)$ and $\fani(\C,\D)$ account for isotropic and anisotropic damage, respectively. This versatile formulation allows for the straightforward incorporation of different established hyperelastic energies $\psi_\star$ into the framework of anisotropic damage. Here, a Neo-Hookean energy is considered with
\begin{equation}
  \psi_\star = \psinh = \frac{\mu}{2} \left( \tr{\C} - 3 - 2 \, \mathrm{ln}\left( \sqrt{\det{\C}} \, \right) \right) + \frac{\Lambda}{4} \left( \det{\C} - 1 - 2 \, \mathrm{ln}\left( \sqrt{\det{\C}} \, \right) \right)
\end{equation}
where $\Lambda$ and $\mu$ denote the first and second Lamé constant, respectively. The degradation functions read
\begin{equation}
  \fiso \coloneq \left( 1 - \frac{\tr{\D}}{3} \right)^\ed
  \label{eq:p4_fiso}
\end{equation}
and
\begin{equation}
  \fani \coloneq \left( 1 - \frac{ \bigtr{ \C^2 \D } }{ \bigtr{ \C^2 } } \right)^\fd
  \label{eq:p4_fani}
\end{equation}
with the exponents $\ed$ and $\fd$ being additional material parameters introduced for further flexibility. In contrast to \cite{ReeseBrepolsEtAl2021}, the anisotropic degradation function $\fani(\C,\D)$ is in this work formulated with respect to the right Cauchy-Green stretch tensor $\C$ instead of the Green-Lagrange strain tensor $\E$ to avoid a division by zero after an unloading of the material.

\textbf{Remark -- Fulfillment of damage growth criterion.\quad}
This ansatz for formulating the elastic energy yields a general fulfillment of the damage growth criterion as derived in Appendix~\ref{sec:p4_app1}, which may not be achieved by an isochoric-volumetric energy split (see Appendix~\ref{sec:p4_app2}) without considering a logarithmic strain energy formulation (cf.~\cite{HolthusenBrepolsEtAl2022a}). 

The isotropic damage hardening energy contains a nonlinear (cf.~\cite{ReeseBrepolsEtAl2021}) and a linear contribution
\begin{equation}
  \psid = \rd \left( \xid + \frac{\exp{-\sd \, \xid}-1}{\sd} \right) + \frac{1}{2} \Hd \, \xid^2
  \label{eq:p4_psid}
\end{equation}
with the damage hardening material parameters $\rd$, $\sd$, and $\Hd$. The additional damage hardening energy $\psih$ results in kinematic damage hardening. Analogously to \cite{FassinEggersmannEtAl2019a,FassinEggersmannEtAl2019b} and \cite{HolthusenBrepolsEtAl2022a}, the energy is formulated in terms of the eigenvalues $D_i$ of the damage tensor and ensures that 
these do not exceed a value of one. It reads
\begin{equation}
  \psih = \Kh \sum_{i=1}^3
  \left( - \frac{ \left( 1 - D_i \right)^{1-\frac{1}{\nh}} }{ 1-\frac{1}{\nh} } - D_i + \frac{1}{1-\frac{1}{\nh}} \right)
  \label{eq:p4_psih}
\end{equation}
with the material parameters $\Kh$ and $\nh$. The micromorphic energy contribution contains a summation over the number of nonlocal degrees of freedom $\ndbar$ according to the size of the micromorphic tuple
\begin{equation}
  \psidbar
  =
  \frac{1}{2} \sum_{i=1}^{\ndbar} H_i \left( d_i - \dbari \right)^2
  +
  \frac{1}{2} \sum_{i=1}^{\ndbar} A_i \, \Grad{ \dbari } \cdot \Grad{ \dbari }
  \label{eq:p4_psidbar}
\end{equation}
and includes the micromorphic penalty parameters $H_i$ and the micromorphic gradient parameters $A_i$.

The micromorphic tuples used in this work are investigated in \cite{vanderVeldenBrepolsEtAl2024} and describe a full and two reduced regularizations of the damage tensor. The micromorphic tuple of model~A reads
\begin{equation}
  \d^\text{A} = \left(
  \bigtr{\D\bm{M}_1},
  \bigtr{\D\bm{M}_2},
  \bigtr{\D\bm{M}_3},
  \bigtr{\D\bm{M}_4},
  \bigtr{\D\bm{M}_5},
  \bigtr{\D\bm{M}_6}
  \right)
  \label{eq:p4_IlocA}
\end{equation}
where the structural tensors are defined using the Cartesian basis vectors $\e_1$, $\e_2$, and $\e_3$ as
\begin{align}
 \bm{M}_1 &= \e_1 \otimes \e_1, \quad \bm{M}_2 = \e_2 \otimes \e_2, \quad \bm{M}_3 = \e_3 \otimes \e_3, \notag \\
 \bm{M}_4 &= \e_1 \otimes \e_2, \quad \bm{M}_5 = \e_1 \otimes \e_3, \quad \bm{M}_6 = \e_2 \otimes \e_3.
\end{align}
Model~A controls all six independent components of the second-order damage tensor by a nonlocal regularization field and, thus, serves with the full regularization as a reference solution for the reduced micromorphic tuples. Model~B employs a reduced principal traces regularization with three nonlocal degrees of freedom and stems from \cite{HolthusenBrepolsEtAl2022a} with
\begin{equation}
  \d^\text{B} = \left(
  \bigtr{\D},
  \bigtr{\D^2},
  \bigtr{\D^3} 
  \right).
  \label{eq:p4_IlocB}
\end{equation}
Model~C utilizes a reduced volumetric-deviatoric regularization with two nonlocal degrees of freedom as proposed by \cite{HolthusenBrepolsEtAl2022b} and investigated in \cite{vanderVeldenBrepolsEtAl2024} with
\begin{equation}
  \d^\text{C} = \left(
  \frac{\bigtr{\D}}{3},
  \Bigtr{\dev{\D}^2}
  \right).
  \label{eq:p4_IlocC}
\end{equation}

\subsection{Explicit thermodynamic conjugate driving forces}

Considering the specific forms of the Helmholtz free energy and the micromorphic tuples given in Section~\ref{sec:p4_specificforms}, the explicit forms of the thermodynamic conjugate driving forces are presented here according to the state laws and definitions of Section~\ref{sec:p4_CD}. The second Piola-Kirchhoff stress follows from Eqs.~\eqref{eq:p4_SPK} and \eqref{eq:p4_psie}
\begin{equation}
  \S  = \left( \left( 1 - \kani \right) \fiso + \kani \, \fani \right) \Snh + 2 \, \kani \, \pd{\fani}{\C} \, \psinh
  \label{eq:p4_SPKexplicit}
\end{equation}
with $\Snh := 2 \, \partial \psinh / \partial \C$. As stated in \cite{ReeseBrepolsEtAl2021}, the partial derivative $\partial \fani / \partial \C$ should vanish in the undamaged state, i.e.~$\D=\bm{0}$, and completely damaged state, i.e.~$\D=\I$. Due to the choice of a modified anisotropic degradation function in Eq.~\eqref{eq:p4_fani}, the analytical solution of $\partial \fani / \partial \C|_{\D=\bm{0}}$ and $\partial \fani / \partial \C|_{\D=\bm{I}}$ is presented in Appendix~\ref{sec:p4_app3}.

The elastic damage driving force follows from Eqs.~\eqref{eq:p4_psidot} and \eqref{eq:p4_psie}
\begin{equation}
  \Ye = - \left( \left( 1 - \kani \right) \pd{\fiso}{\D} + \kani \, \pd{\fani}{\D} \right) \psinh
\end{equation}
where the partial derivatives of the degradation functions with respect to the damage tensor are given in Eqs.~\eqref{eq:p4_dfisodD} and \eqref{eq:p4_dfanidD}. The kinematic damage hardening driving force is formulated in the eigensystem of the damage tensor, implemented analogously to \cite{HolthusenBrepolsEtAl2022a}, and follows from Eqs.~\eqref{eq:p4_psidot} and \eqref{eq:p4_psih}
\begin{equation}
  \Yh = \Kh \sum_{i=1}^3 \left( \frac{1}{\left( 1 - D_i \right)^{1/\nh}} - 1 \right) \niD \otimes \niD
\end{equation}
where $D_i$ and $\niD$ denote the eigenvalues and eigenvectors of $\D$. The nonlocal damage driving force reads generally (see \cite{vanderVeldenBrepolsEtAl2024}) with the definition in Eq.~\eqref{eq:p4_psidot}
\begin{equation}
  \Ydbar = \sum_{i=1}^\ndbar H_i \left( d_i - \dbari \right) \pd{d_i}{\D}.
\end{equation}
The explicit forms follow with the definitions for the micromorphic tuples in Eqs.~\eqref{eq:p4_IlocA}, \eqref{eq:p4_IlocB}, and \eqref{eq:p4_IlocC} and read for model~A
\begin{align}
  \Ydbar^\text{A} = & H_1 \left( \bigtr{\D\bm{M}_1} - \bar{d}_1 \right) \sym{ \bm{M}_1 }
                  +   H_2 \left( \bigtr{\D\bm{M}_2} - \bar{d}_2 \right) \sym{ \bm{M}_2 } \notag \\
                  + & H_3 \left( \bigtr{\D\bm{M}_3} - \bar{d}_3 \right) \sym{ \bm{M}_3 }
                  +   H_4 \left( \bigtr{\D\bm{M}_4} - \bar{d}_4 \right) \sym{ \bm{M}_4 } \notag \\
                  + & H_5 \left( \bigtr{\D\bm{M}_5} - \bar{d}_5 \right) \sym{ \bm{M}_5 }
                  +   H_6 \left( \bigtr{\D\bm{M}_6} - \bar{d}_6 \right) \sym{ \bm{M}_6 },
  \label{eq:p4_YdbarA}                                    
\end{align}
for model~B
\begin{equation}
  \Ydbar^\text{B} = H_1 \left( \bigtr{\D}   - \bar{d}_1 \right) \I   
                  + H_2 \left( \bigtr{\D^2} - \bar{d}_2 \right) 2 \D 
                  + H_3 \left( \bigtr{\D^3} - \bar{d}_3 \right) 3 \D^2,
  \label{eq:p4_YdbarB}                                    
\end{equation}
and for model~C
\begin{equation}
  \Ydbar^\text{C} = \frac{H_1}{3} \left( \frac{\bigtr{\D}}{3} - \bar{d}_1 \right) \I
                  + H_2 \left( \Bigtr{\dev{\D}^2}   - \bar{d}_2 \right) \left( 2 \D -  \frac{2}{3} \tr{\D} \I \right).
  \label{eq:p4_YdbarC}                  
\end{equation}
The isotropic damage hardening driving force follows from Eqs.~\eqref{eq:p4_psidot} and \eqref{eq:p4_psid}
\begin{equation}
  \Rd = - \left[ \rd \left( 1 - \exp{ - \sd \, \xid} \right) + \Hd \, \xid \right]
\end{equation}
and the generalized micromorphic stresses from Eqs.~\eqref{eq:p4_xini}, \eqref{eq:p4_Xini} and \eqref{eq:p4_psidbar}
\begin{equation}
  \left( \xini \right)_k = - H_k \left( d_k - \bar{d}_k \right), \qquad k \in \{ 1, ..., \ndbar \},
\end{equation}
\begin{equation}
  \left( \Xini \right)_k = A_k \, \Grad{ \bar{d}_k }, \qquad k \in \{ 1, ..., \ndbar \}.
\end{equation}

\section{Numerical examples}
\label{sec:p4_examples}

First, the numerical examples show the local study without using any gradient-extension of a single finite element that is loaded by uniaxial tension, uniaxial strain, simple shear, and torsion in Section \ref{sec:p4_ses} to analyze the behavior at the material point level. Thereafter in Section \ref{sec:p4_ANotched}, the gradient-extended finite element formulation is applied for the structural simulation of an asymmetrically notched specimen using models~A, B, and C to confirm the accuracy and efficiency of the reduced volumetric-deviatoric regularization of model~C. Additionally, this example is investigated using a local formulation without gradient-extension and also using further reduced regularizations based on a single component of the damage tensor. Finally in Section \ref{sec:p4_RB}, model~C is employed for the complex three-dimensional structural simulation of a pressure loaded rotor blade specimen.

\begin{table}[htbp]
  \centering
  \caption{Material and numerical parameters}
  \label{tab:p4_matpar}
  \begin{tabular}{l l r r r}
    \hline
    Symbol      & Material parameter                            & Set~1     & Set~2          & Unit        \\ 
    \hline \hline                                               
    $\Lambda$   & First Lamé constant                           & 5000      & 25000          & MPa         \\
    $\mu$       & Second Lamé constant                          & 7500      & 55000          & MPa         \\
    $\kani$     & Degree damage anisotropy                      & 0 - 1     & 1              & -           \\
    $\ed$       & Exponent $\fiso$                              & 2         & 2              & -           \\
    $\fd$       & Exponent $\fani$                              & 1         & 1              & -           \\
    $\Yn$       & Initial damage threshold                      & 10        & 2.5            & MPa         \\
    $\cd$       & Distortional hardening exponent               & 1         & 1              & -           \\
    $\Hd$       & Linear isotropic hardening prefactor          & 1         & 1              & MPa         \\  
    $\rd$       & Nonlinear isotropic hardening prefactor       & 10        & 5              & MPa         \\
    $\sd$       & Nonlinear isotropic hardening scaling factor  & 100       & 100            & -           \\
    $\Kh$       & Kinematic hardening prefactor                 & 0.1       & 0.1            & MPa         \\
    $\nh$       & Kinematic hardening exponent                  & 2         & 2              & -           \\  
    $A_i$       & Internal length scales                        & 0         & 300 - 3000     & MPa mm$^2$  \\
    \hline                                                      
    \vspace{0mm}\\                                              
    \hline                                                      
    Symbol      & Numerical parameter                           & Value     & Value          & Unit        \\ 
    \hline \hline                                               
    $\ah$       & Taylor series sampling point                  & 0.999999  & 0.999999       & -           \\  
    $H_i$       & Micromorphic penalty parameters               & 0         & 10$^\text{4}$  & MPa         \\  
    $\eta_v$    & Artificial viscosity                          & 1         & 1              & MPa s       \\
    \hline
  \end{tabular}
\end{table}

The material parameters are provided in Table \ref{tab:p4_matpar}, where the parameters of Set~1 are utilized for the single element studies in Section \ref{sec:p4_ses} and the parameters of Set~2 for the structural simulations in Sections \ref{sec:p4_ANotched} and \ref{sec:p4_RB}. The Taylor series sampling point $\ah$ is not introduced during the presentation of the constitutive modeling in Section \ref{sec:p4_modeling}, but is required for the implementation of the kinematic damage driving force $\Yh$ as elaborated in \cite{HolthusenBrepolsEtAl2022a}.

The finite elements are eight-node hexahedrons with full integration that could be substituted in future investigations with the reduced integration-based elements of \citetalias{BarfuszBrepolsEtAl2021}[\citeyear{BarfuszBrepolsEtAl2021}, \citeyear{BarfuszvanderVeldenEtAl2021}, \citeyear{BarfuszvanderVeldenEtAl2022}]. For the finite element simulations, we utilize the software \textit{FEAP} (\cite{TaylorGovindjee2020}), for the finite element meshes of the rotor blade in Section~\ref{sec:p4_RB} the software \textit{HyperMesh} (\cite{HyperWorks2022}), and for processing the contour plots of the simulations the software \textit{ParaView} (\cite{AhrensGeveciEtAl2005}).

\subsection{Single element studies}
\label{sec:p4_ses}

The studies of a single finite element serve for the investigation of the material behavior under a homogeneous loading state and, thus, employ its local formulation by neglecting the micromorphic regularization by setting $A_i = 0~[\si{\MPa\mm\squared}]$ and $H_i = 0~[\si{\MPa}]$. 

\begin{figure}[htbp]
  \centering
  
  \begin{subfigure}{\textwidth}
    \centering
    \includegraphics{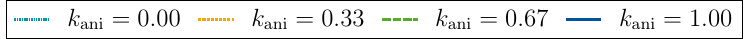}
  \end{subfigure}  
  \vspace{3mm}
  
  \begin{subfigure}{.48\textwidth}
    \centering
    \includegraphics{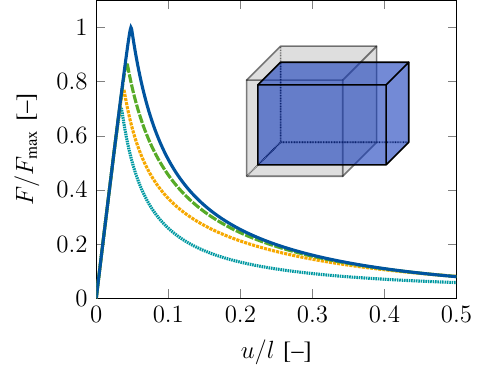}
    \caption{Uniaxial tension}
    \label{fig:p4_SEFuUT}
  \end{subfigure}
  \quad
  \begin{subfigure}{.48\textwidth}
    \centering
    \includegraphics{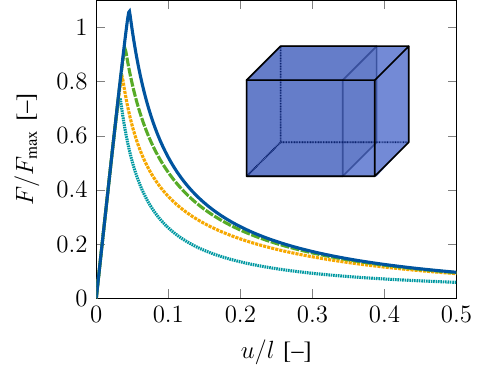}
    \caption{Uniaxial strain}
    \label{fig:p4_SEFuUS}
  \end{subfigure}%
  \vspace{5mm} 
  \begin{subfigure}{.48\textwidth}
    \centering
    \includegraphics{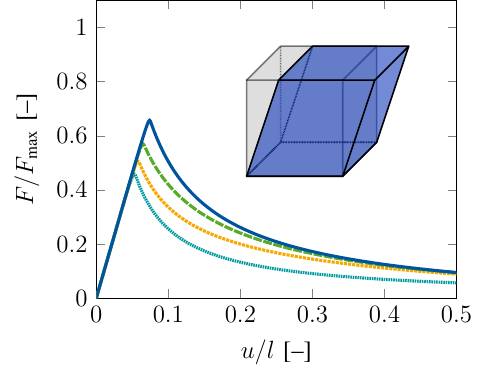}
    \caption{Simple shear}
    \label{fig:p4_SEFuSS}
  \end{subfigure}
  \quad
  \begin{subfigure}{.48\textwidth}
    \centering
    \includegraphics{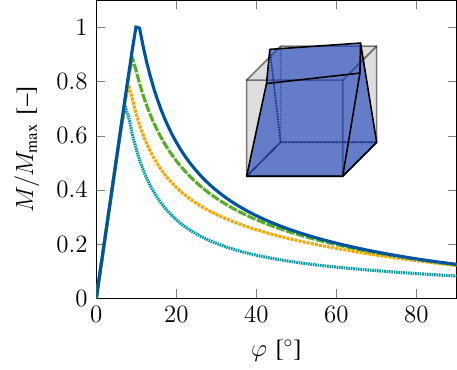}
    \caption{Torsion}
    \label{fig:p4_SEFuTfi}
  \end{subfigure}
  \caption{Single element studies for uniaxial tension, uniaxial strain, simple shear and torsion. The force-displacement curves are normalized with respect to the maximum force for uniaxial tension for $\kani = 1.00$ with  $F_\text{max} = 8.1251 \times 10^2~[\si{\newton}]$. The moment-twist curves are normalized with respect to the maximum moment for $\kani = 1.00$ with $M_\text{max} =-2.1427 \times 10^2~[\si{\newton\mm}]$.}
  \label{fig:p4_SEFu}
\end{figure}

Uniaxial tension is applied in Fig.~\ref{fig:p4_SEFuUT} for different values of the damage anisotropy parameter with $\kani \in (0.00~[\si{-}], 0.33~[\si{-}], 0.67~[\si{-}], 1.00~[\si{-}])$, where $\kani = 0.00~[\si{-}]$ models purely isotropic and $\kani = 1.00~[\si{-}]$ purely anisotropic material degradation. The normalized force-displacement curves show an increasing maximum retention force followed by a steeper force reduction for an increase of the damage anisotropy parameter $\kani$.
A uniaxial strain state is applied in Fig.~\ref{fig:p4_SEFuUS}, where the curves show a qualitatively similar behavior to Fig.~\ref{fig:p4_SEFuUT}. However quantitatively, the maximum retention force for $\kani = 1.00~[\si{-}]$ is $5.93~[\si{\percent}]$ higher compared uniaxial tension due to the constrained lateral contraction.
Simple shear is applied in Fig.~\ref{fig:p4_SEFuSS} and yields a response qualitatively similar to uniaxial tension and uniaxial strain, but the maximum retention force for $\kani = 1.00~[\si{-}]$ is $34.27~[\si{\percent}]$ smaller compared to uniaxial tension.
Finally, a torsional load is applied in Fig.~\ref{fig:p4_SEFuTfi} also for different values of the damage anisotropy parameter and confirms the previous observations of Figs.~\ref{fig:p4_SEFuUT}-\ref{fig:p4_SEFuSS} that for this model a higher material strength corresponds to higher values of damage anisotropy.

\subsection{Asymmetrically notched specimen}
\label{sec:p4_ANotched}

\begin{figure}
    \centering 
    \includegraphics{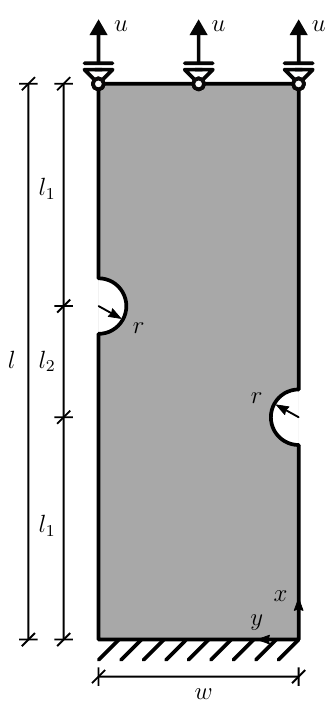}    
    \caption{Geometry and boundary value problem for the asymmetrically notched specimen.}
    \label{fig:p4_ANgeombvp}
\end{figure}

The first structural example considers an asymmetrically notched specimen in Fig.~\ref{fig:p4_ANgeombvp} under plane strain conditions with the dimensions $l = 100~[\si{\mm}]$, $l_1 = 40~[\si{\mm}]$, $l_2 = 20~[\si{\mm}]$, $w = 36~[\si{\mm}]$, $r = 5~[\si{\mm}]$ and a thickness of $1~[\si{\mm}]$ that was previously studied in e.g.~\cite{BrepolsWulfinghoffEtAl2017}. Analogously to \cite{vanderVeldenBrepolsEtAl2024}, this example is investigated using the gradient-extensions of model~A, B, and C. The gradient parameter of model~A is arbitrarily chosen as $A_i^\text{A} = 1000~[\si{\MPa\mm\squared}]$ and the parameters of model~B and C are identified as $A_i^\text{B} = 300~[\si{\MPa\mm\squared}]$ and $A_i^\text{C} = 3000~[\si{\MPa\mm\squared}]$ to obtain the same structural load bearing capacity.

\begin{figure}[htbp]
  \centering
  \begin{subfigure}{\textwidth}
  \centering
    \includegraphics{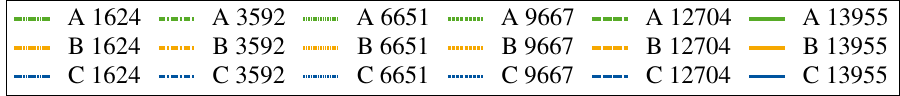}
  \end{subfigure}

  \vspace{5mm}
  
  \begin{subfigure}{.48\textwidth}
    \centering
    \includegraphics{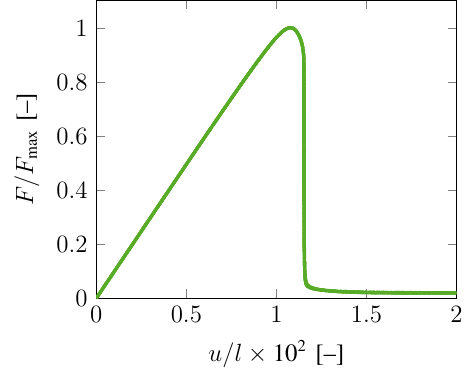}
    \vspace{-7mm}
    \caption{Model~A}
    \label{fig:p4_ANotchedFuA}
  \end{subfigure}
  \quad
  \begin{subfigure}{.48\textwidth}
    \centering
    \includegraphics{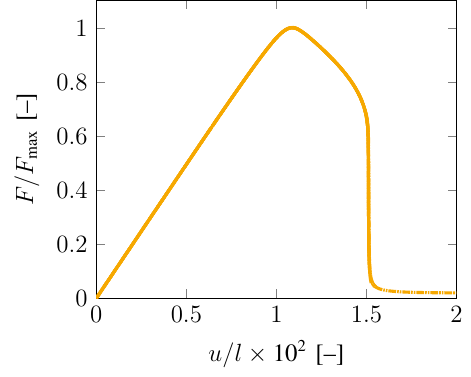}
    \vspace{-7mm}
    \caption{Model~B}
    \label{fig:p4_ANotchedFuB}
  \end{subfigure}%
  \vspace{5mm} 
  \begin{subfigure}{.48\textwidth}
    \centering
    \includegraphics{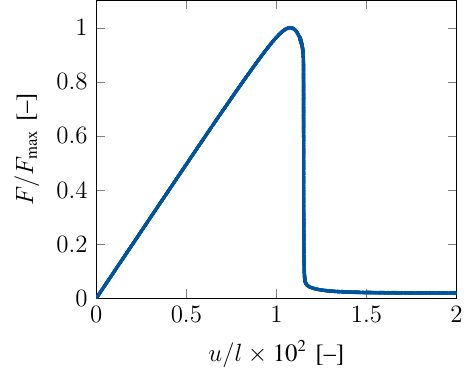}
    \vspace{-7mm}
    \caption{Model~C}
    \label{fig:p4_ANotchedFuC}
  \end{subfigure}
  \quad
  \begin{subfigure}{.48\textwidth}
    \centering
    \includegraphics{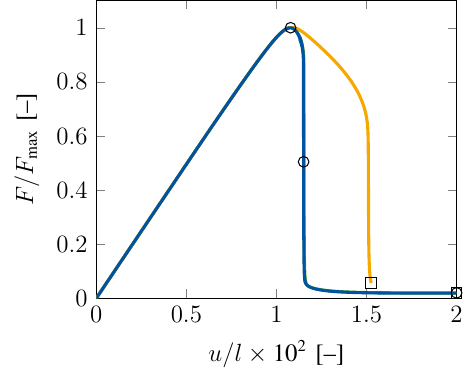}
    \vspace{-7mm}
    \caption{Model comparison (13955 elements)}
    \label{fig:p4_ANotchedFuComp}
  \end{subfigure}
  \caption{Force-displacement curves for the mesh convergence study and the model comparison of the asymmetrically notched specimen.
  The forces are normalized with respect to the maximum force of model~C (13955 elements) with $F_\text{max} = 4.4031 \times 10^4~[\si{\newton}]$.
  In Figs.~\ref{fig:p4_ANotchedFuA}, \ref{fig:p4_ANotchedFuB}, and \ref{fig:p4_ANotchedFuC}, the curves obtained with the coarse and fine meshes are essentially congruent.
  In Fig.~\ref{fig:p4_ANotchedFuComp}, the squares indicate the snapshots of Fig.~\ref{fig:p4_AnotchedDfinal} and the circles indicate the snapshots of Fig.~\ref{fig:p4_AnotchedDevolution}.}
  \label{fig:p4_ANotchedFu}
\end{figure}

The force-displacement curves in Figs.~\ref{fig:p4_ANotchedFuA}, \ref{fig:p4_ANotchedFuB}, and \ref{fig:p4_ANotchedFuC} show the mesh convergence studies for model~A, B, and C with the meshes steming from \cite{HolthusenBrepolsEtAl2022a}. All models yield excellent convergence behavior with negligible differences between the results of the coarsest and finest mesh. A model comparison of the structural response with respect to the force-displacement curves is provided in Fig.~\ref{fig:p4_ANotchedFuComp} and shows a fine agreement between model~A with the full regularization and model~C with the reduced volumetric-deviatoric regularization. For model~B, the vertical drop in the force-displacement curve is shifted significantly to the right compared to models~A and C and, thus, entails a larger amount of dissipated energy in the failure process, which is in line with the results of \cite{vanderVeldenBrepolsEtAl2024}.

\begin{figure}
  \centering 
  \begin{subfigure}{.18\textwidth} 
    \centering 
    \includegraphics[height=23mm,angle=90]{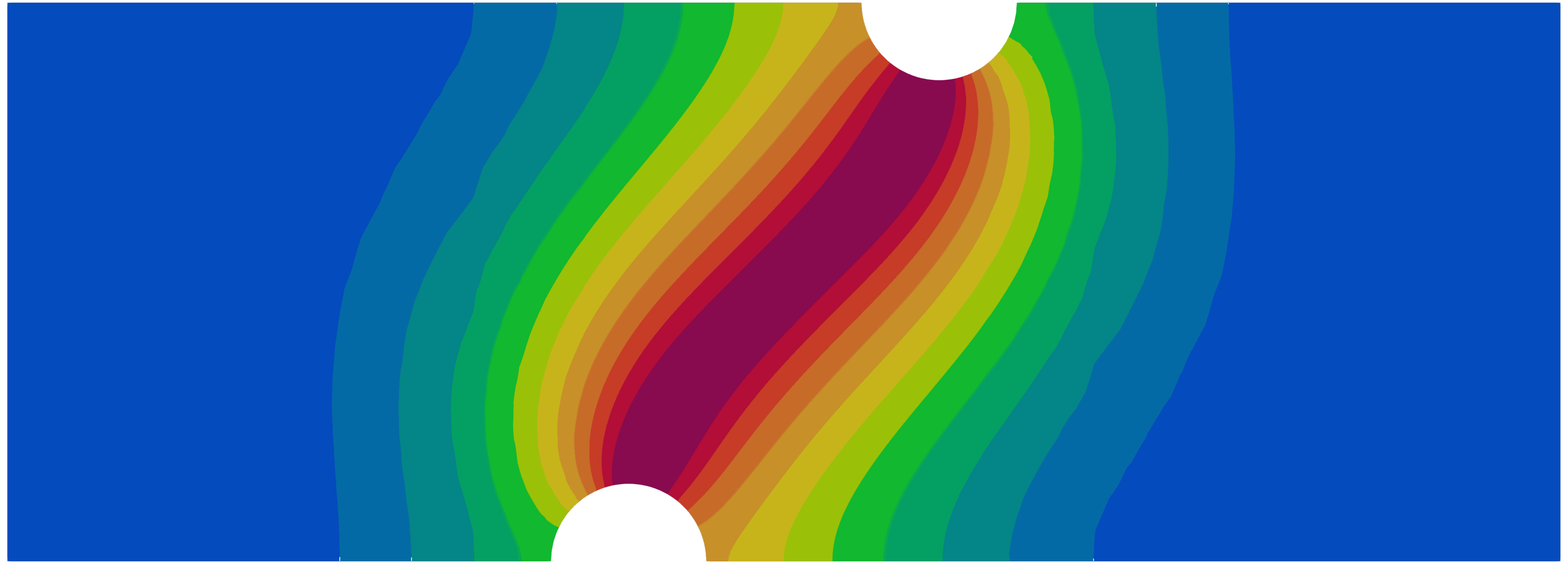}
  \end{subfigure}
  \begin{subfigure}{.18\textwidth} 
    \centering 
    \includegraphics[height=23mm,angle=90]{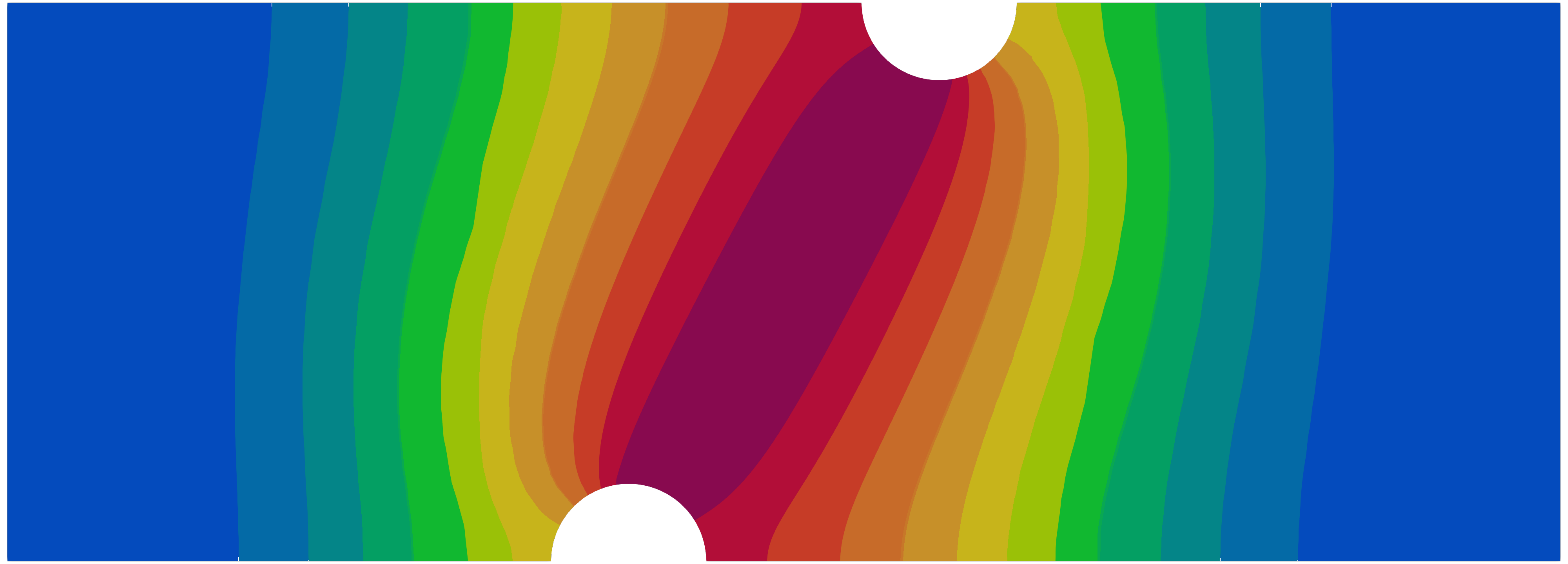}
  \end{subfigure}
  \begin{subfigure}{.18\textwidth} 
    \centering 
    \includegraphics[height=23mm,angle=90]{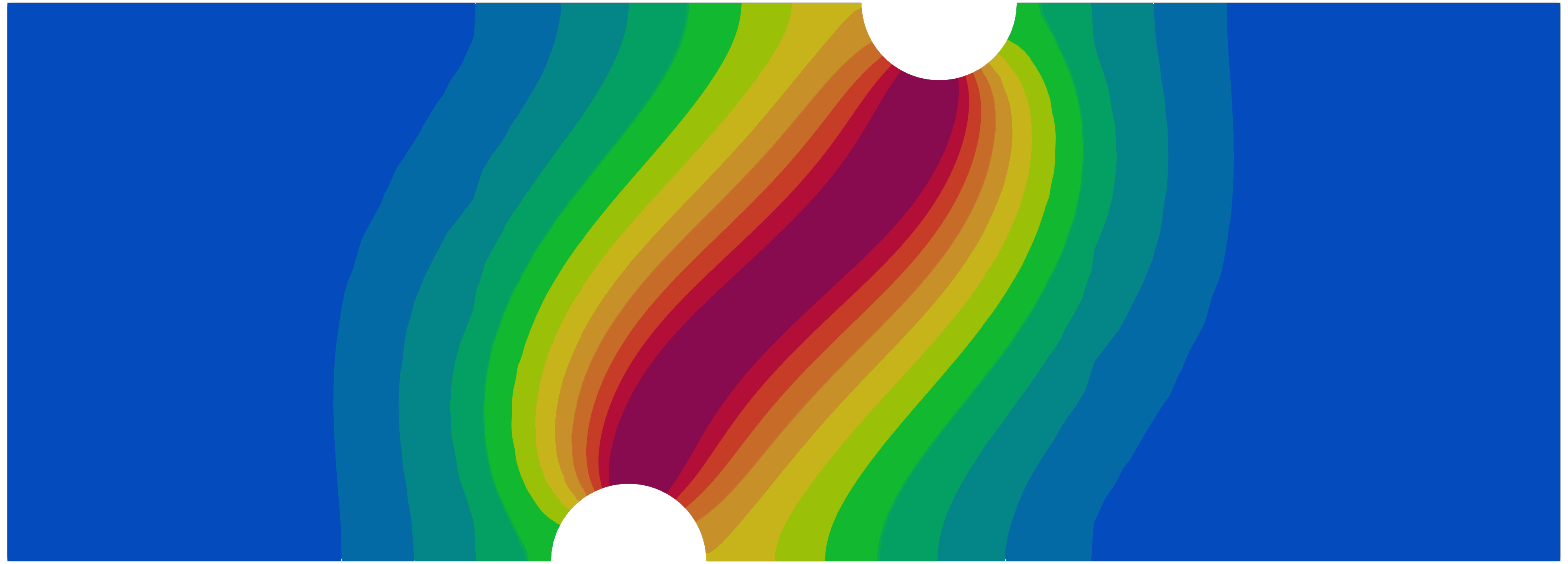}
  \end{subfigure}
  \begin{subfigure}{.08\textwidth} 
    \centering 
    \begin{tikzpicture}
      \node[inner sep=0pt] (pic) at (0,0) {\includegraphics[height=40mm, width=5mm]
      {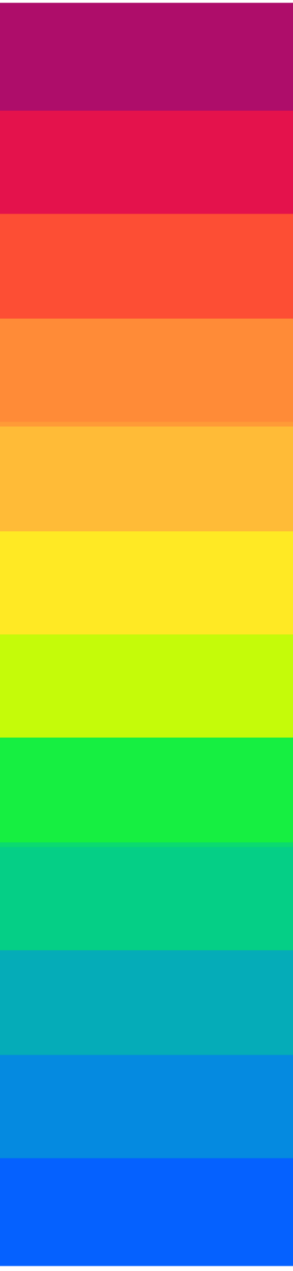}};
      \node[inner sep=0pt] (0)   at ($(pic.south)+( 0.50, 0.15)$)  {$0$};
      \node[inner sep=0pt] (1)   at ($(pic.south)+( 0.50, 3.80)$)  {$1$};
      \node[inner sep=0pt] (d)   at ($(pic.south)+( 0.00, 4.35)$)  {$D_{xx}~\si{[-]}$};
    \end{tikzpicture} 
  \end{subfigure}

  \vspace{1mm}

  \begin{subfigure}{.18\textwidth} 
    \centering 
    \includegraphics[height=23mm,angle=90]{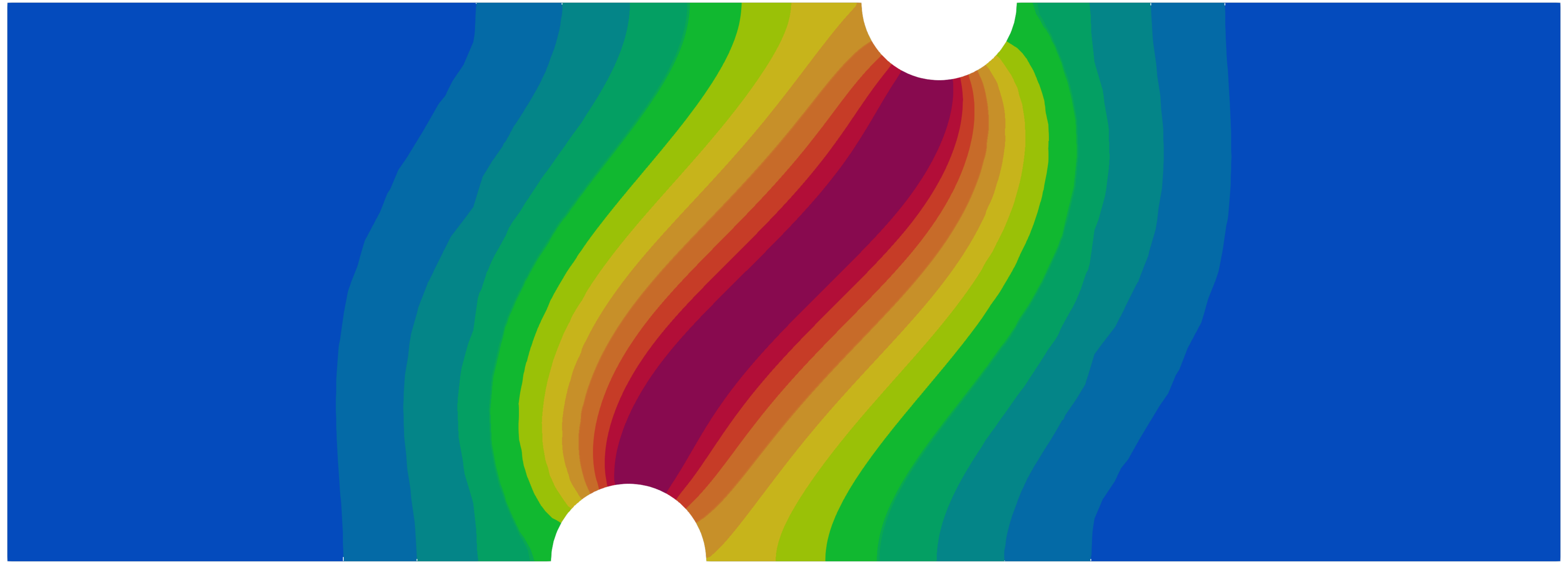}
  \end{subfigure}
  \begin{subfigure}{.18\textwidth} 
    \centering 
    \includegraphics[height=23mm,angle=90]{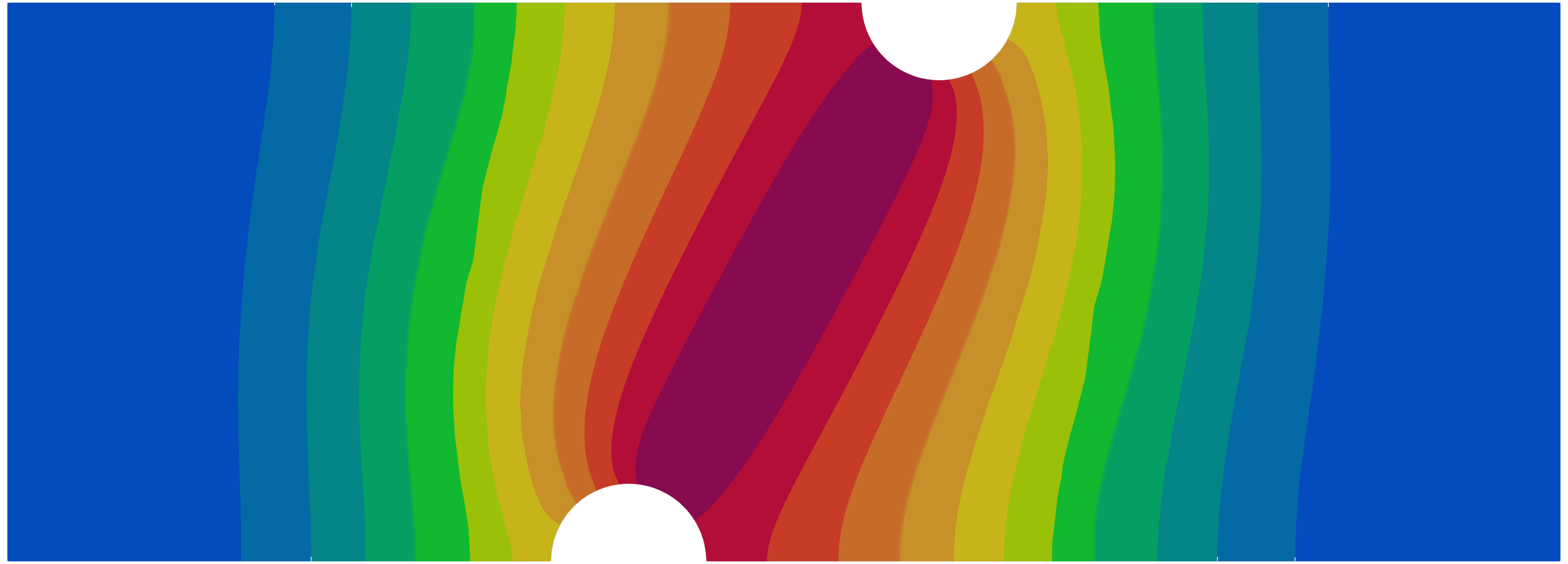}
  \end{subfigure}
  \begin{subfigure}{.18\textwidth} 
    \centering 
    \includegraphics[height=23mm,angle=90]{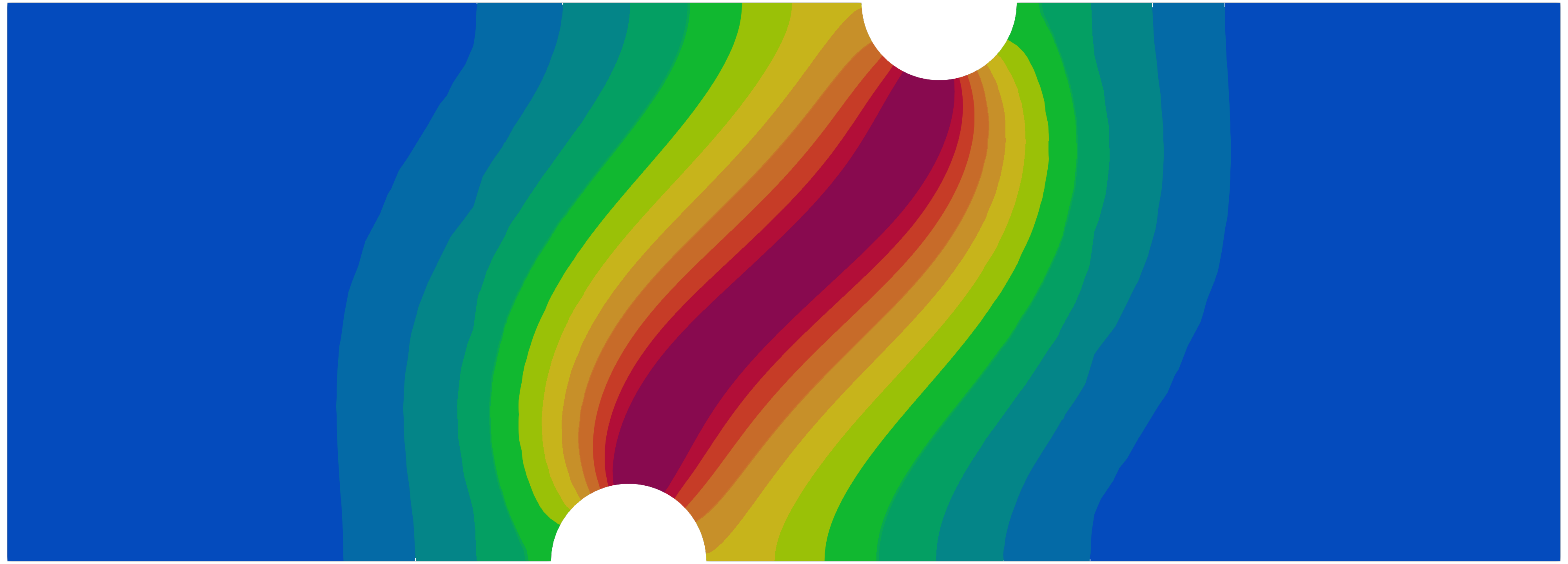}
  \end{subfigure}
  \begin{subfigure}{.08\textwidth} 
    \centering 
    \begin{tikzpicture}
      \node[inner sep=0pt] (pic) at (0,0) {\includegraphics[height=40mm, width=5mm]
      {03_Contour/00_Color_Maps/Damage_Step_Vertical.pdf}};
      \node[inner sep=0pt] (0)   at ($(pic.south)+( 0.50, 0.15)$)  {$0$};
      \node[inner sep=0pt] (1)   at ($(pic.south)+( 0.50, 3.80)$)  {$1$};
      \node[inner sep=0pt] (d)   at ($(pic.south)+( 0.00, 4.35)$)  {$D_{yy}~\si{[-]}$};
    \end{tikzpicture} 
  \end{subfigure}

  \vspace{1mm}

  \begin{subfigure}{.18\textwidth} 
    \centering 
    \includegraphics[height=23mm,angle=90]{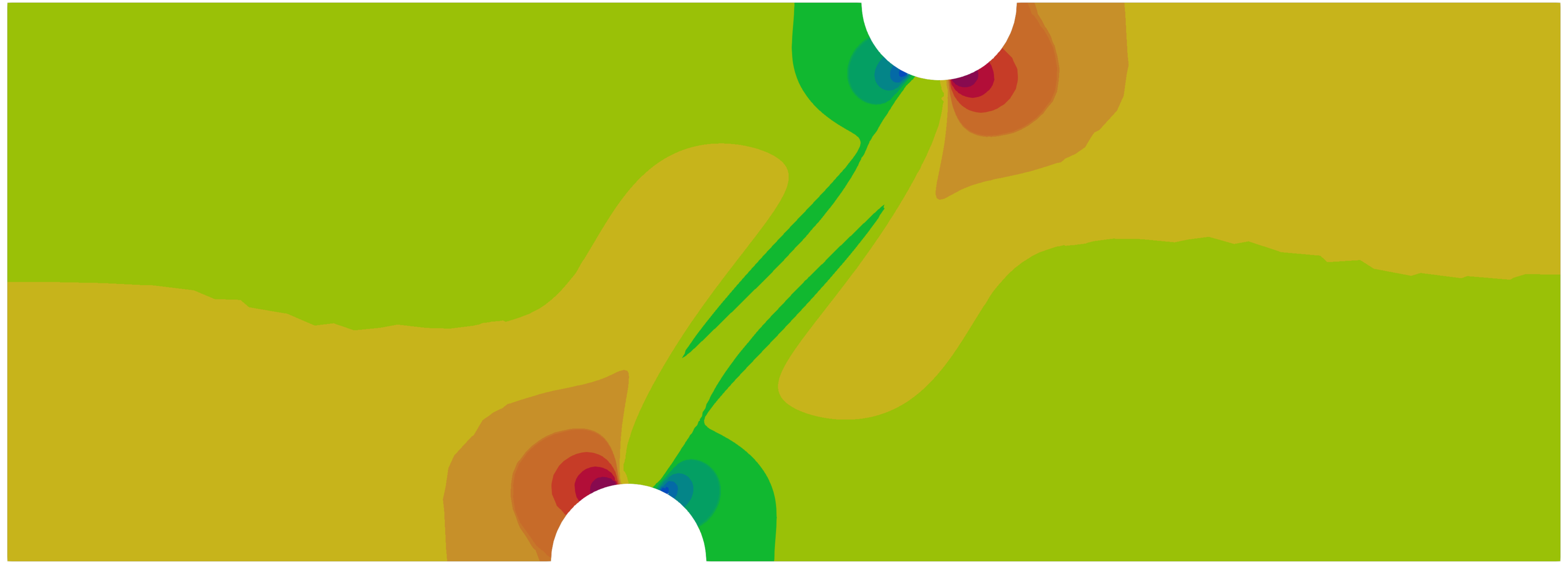}
    \caption{Model~A}
    \label{fig:p4_AnotchedDfinalA}    
  \end{subfigure}
  \begin{subfigure}{.18\textwidth} 
    \centering 
    \includegraphics[height=23mm,angle=90]{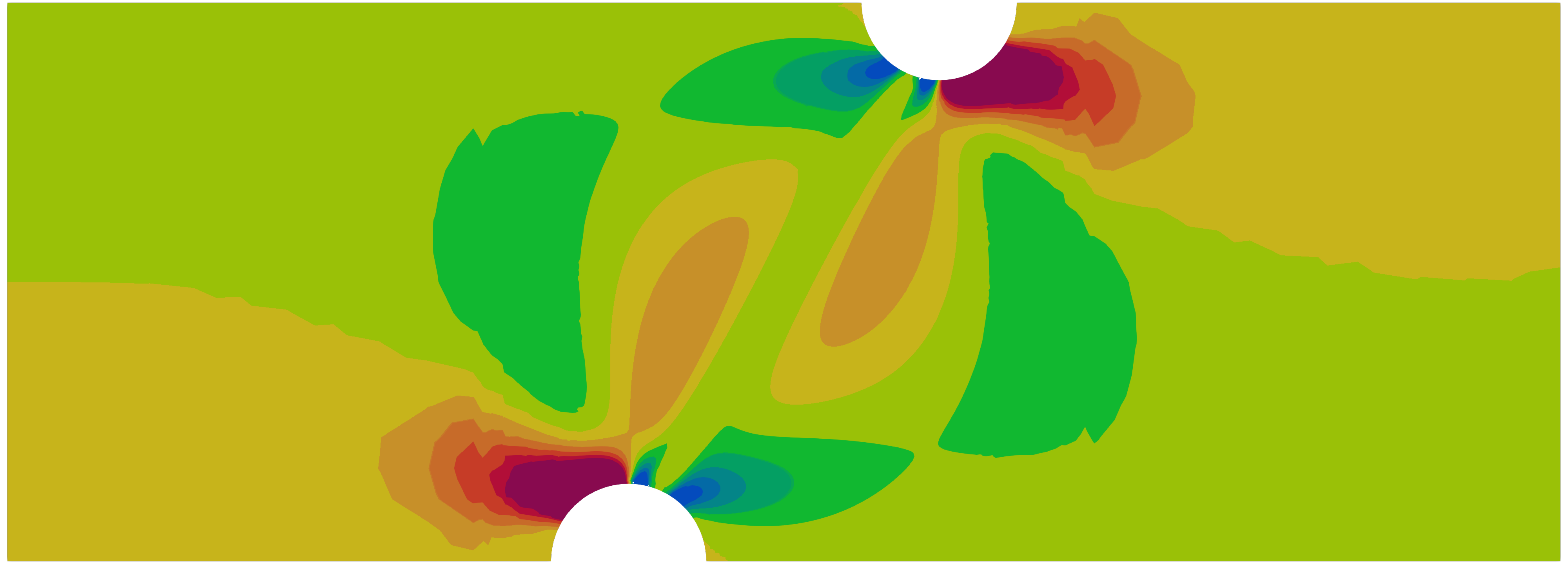}
    \caption{Model~B}
    \label{fig:p4_AnotchedDfinalB}    
  \end{subfigure}
  \begin{subfigure}{.18\textwidth} 
    \centering 
    \includegraphics[height=23mm,angle=90]{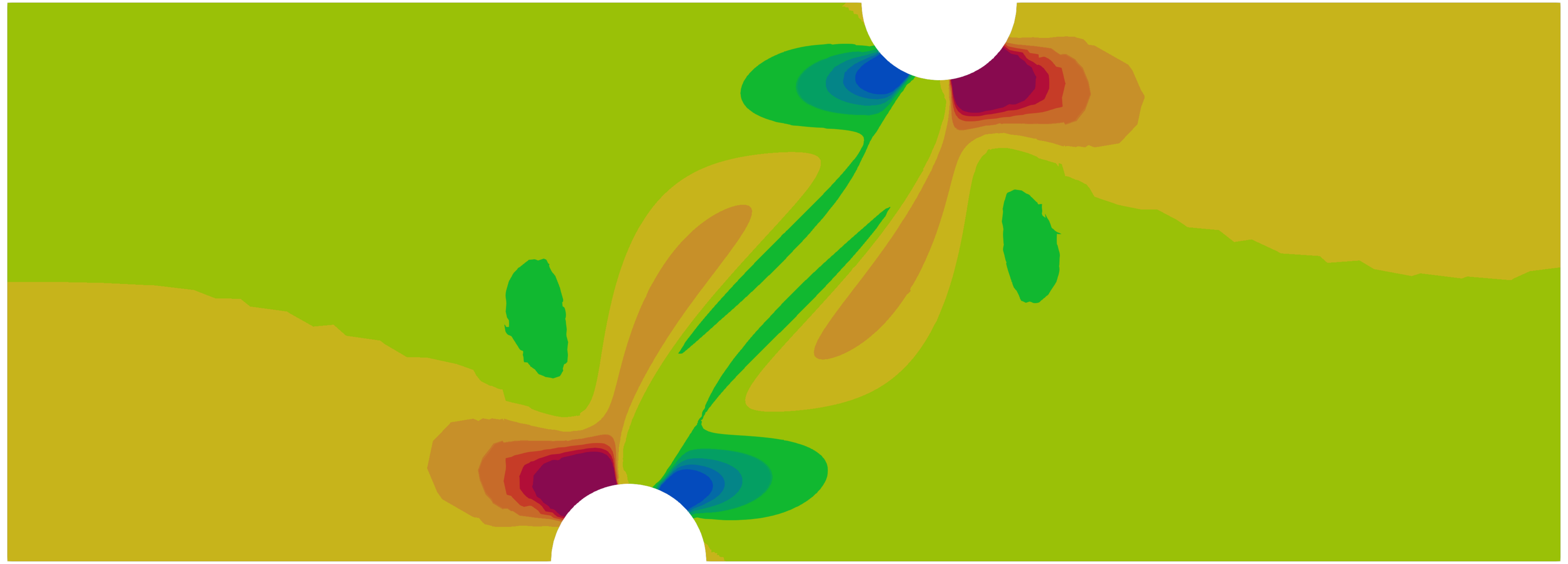}
    \caption{Model~C}
    \label{fig:p4_AnotchedDfinalC}    
  \end{subfigure}
  \begin{subfigure}{.08\textwidth} 
    \centering 
    \begin{tikzpicture}
      \node[inner sep=0pt] (pic) at (0,0) {\includegraphics[height=40mm, width=5mm]
      {03_Contour/00_Color_Maps/Damage_Step_Vertical.pdf}};
      \node[inner sep=0pt] (0)   at ($(pic.south)+( 1.40, 0.15)$)  {$\leq -0.0045$};
      \node[inner sep=0pt] (1)   at ($(pic.south)+( 1.40, 3.80)$)  {$\geq +0.0045$};
      \node[inner sep=0pt] (d)   at ($(pic.south)+( 0.00, 4.35)$)  {$D_{xy}~\si{[-]}$};
    \end{tikzpicture} 
    \hphantom{Model~C}
  \end{subfigure}
  
  \caption{Damage contour plots for the asymmetrically notched specimen with models~A, B, and C (13955 elements) at the end of the simulation.}
  \label{fig:p4_AnotchedDfinal}     
\end{figure}

A comparison of the damage contour plots of models~A, B, and C for the asymmetrically notched specimen is given in Fig.~\ref{fig:p4_AnotchedDfinal}. The crack width and damage affected zone of model~B (Fig.~\ref{fig:p4_AnotchedDfinalB}) are thicker than those of models~A and C (Figs.~\ref{fig:p4_AnotchedDfinalA} and \ref{fig:p4_AnotchedDfinalC}) and, thereby, yield a higher energy dissipation. Nevertheless, close agreement between the damage patterns of models~A and C is observed.

\begin{figure}
  \centering 
  \begin{subfigure}{.18\textwidth} 
    \centering 
    \includegraphics[height=23mm,angle=90]{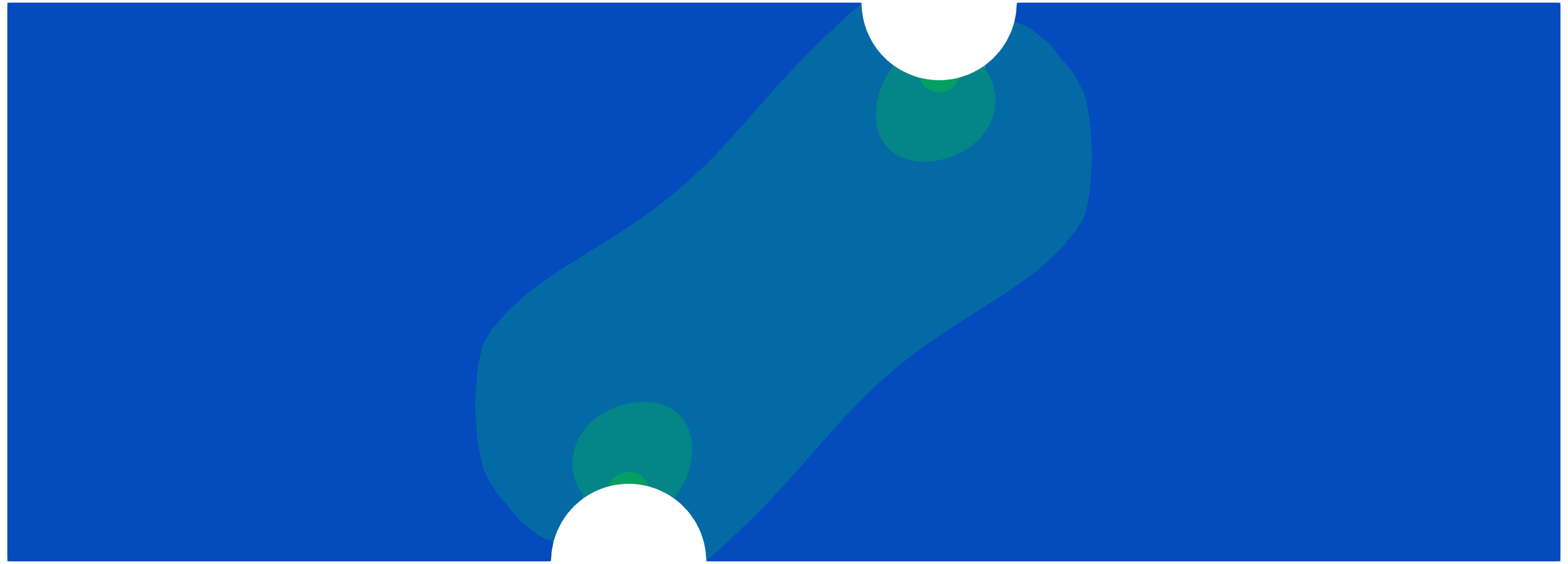}
  \end{subfigure}
  \begin{subfigure}{.18\textwidth} 
    \centering 
    \includegraphics[height=23mm,angle=90]{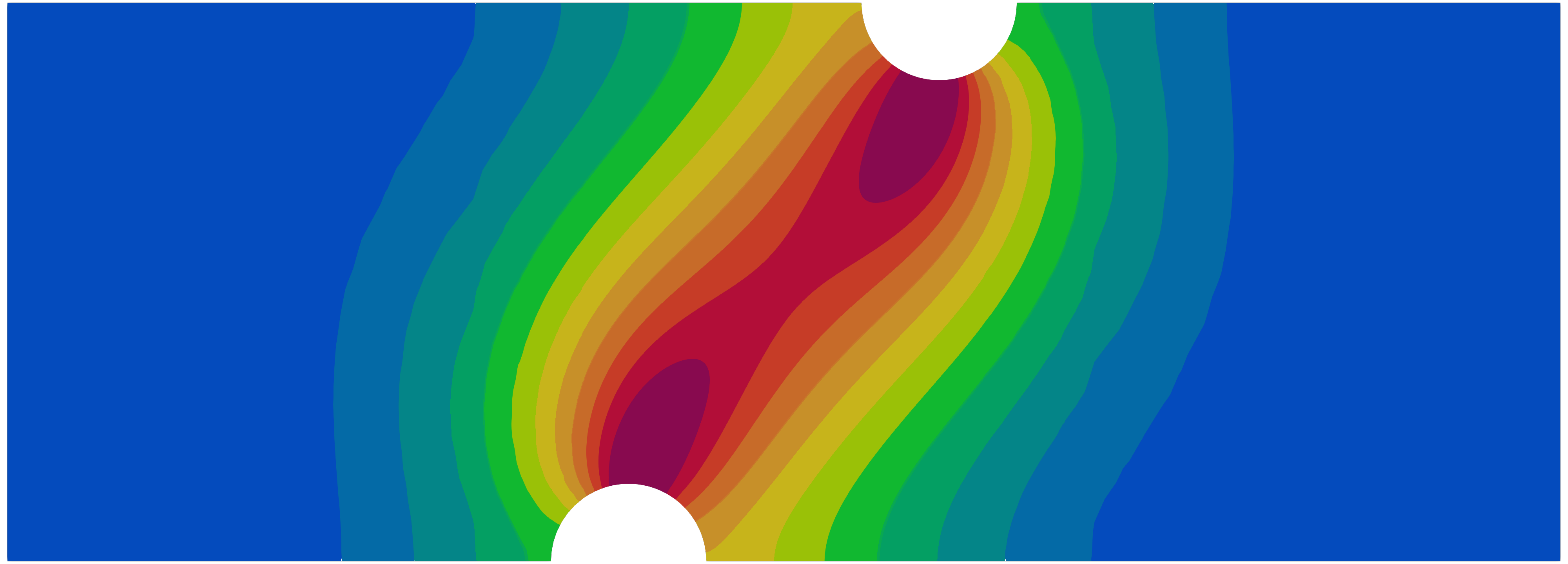}
  \end{subfigure}
  \begin{subfigure}{.18\textwidth} 
    \centering 
    \includegraphics[height=23mm,angle=90]{03_Contour/ANotched/mesh_convergence/C-13955_Dxx.pdf}
  \end{subfigure}
  \begin{subfigure}{.08\textwidth} 
    \centering 
    \begin{tikzpicture}
      \node[inner sep=0pt] (pic) at (0,0) {\includegraphics[height=40mm, width=5mm]
      {03_Contour/00_Color_Maps/Damage_Step_Vertical.pdf}};
      \node[inner sep=0pt] (0)   at ($(pic.south)+( 0.50, 0.15)$)  {$0$};
      \node[inner sep=0pt] (1)   at ($(pic.south)+( 0.50, 3.80)$)  {$1$};
      \node[inner sep=0pt] (d)   at ($(pic.south)+( 0.00, 4.35)$)  {$D_{xx}~\si{[-]}$};
    \end{tikzpicture} 
  \end{subfigure}

  \vspace{1mm}

  \begin{subfigure}{.18\textwidth} 
    \centering 
    \includegraphics[height=23mm,angle=90]{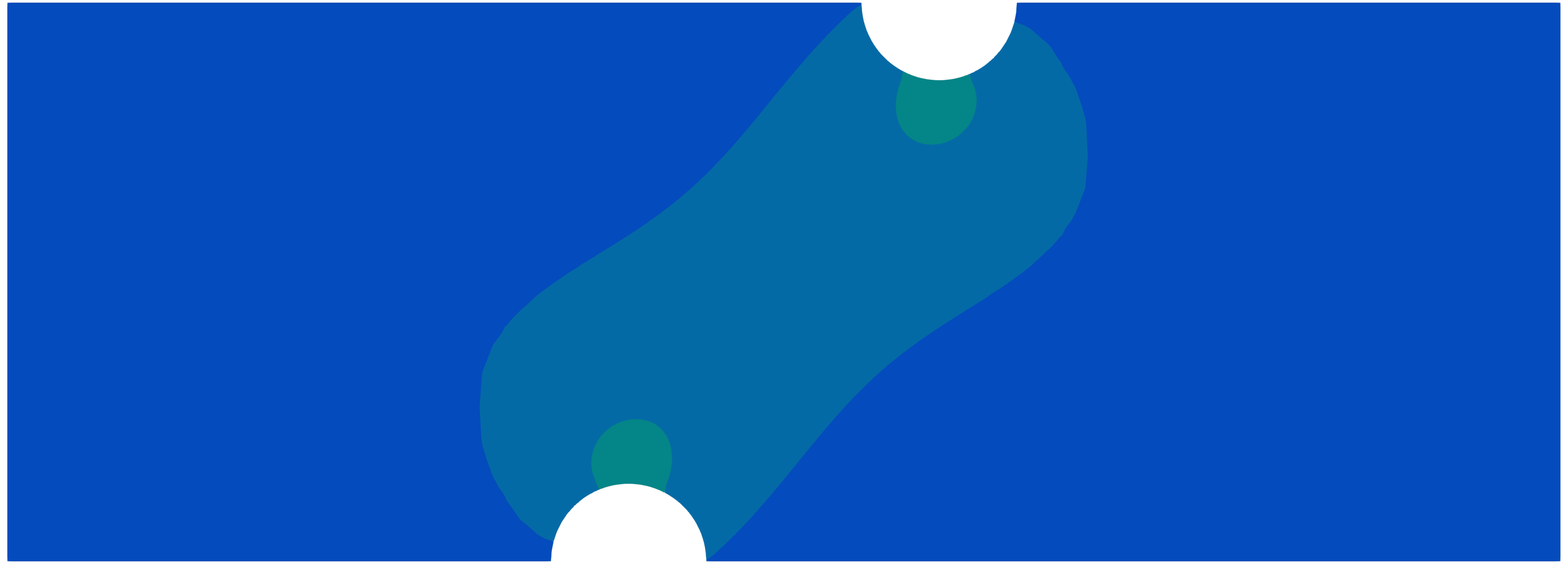}
  \end{subfigure}
  \begin{subfigure}{.18\textwidth} 
    \centering 
    \includegraphics[height=23mm,angle=90]{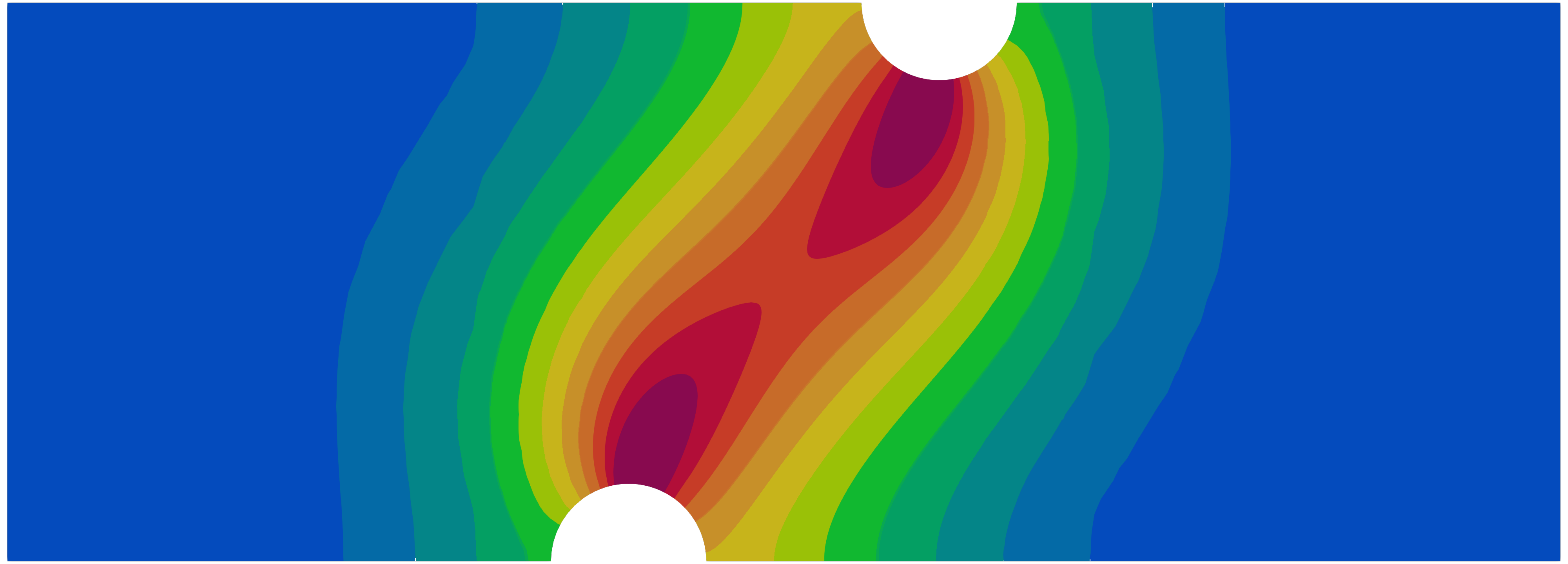}
  \end{subfigure}
  \begin{subfigure}{.18\textwidth} 
    \centering 
    \includegraphics[height=23mm,angle=90]{03_Contour/ANotched/mesh_convergence/C-13955_Dyy.pdf}
  \end{subfigure}
  \begin{subfigure}{.08\textwidth} 
    \centering 
    \begin{tikzpicture}
      \node[inner sep=0pt] (pic) at (0,0) {\includegraphics[height=40mm, width=5mm]
      {03_Contour/00_Color_Maps/Damage_Step_Vertical.pdf}};
      \node[inner sep=0pt] (0)   at ($(pic.south)+( 0.50, 0.15)$)  {$0$};
      \node[inner sep=0pt] (1)   at ($(pic.south)+( 0.50, 3.80)$)  {$1$};
      \node[inner sep=0pt] (d)   at ($(pic.south)+( 0.00, 4.35)$)  {$D_{yy}~\si{[-]}$};
    \end{tikzpicture} 
  \end{subfigure}

  \vspace{1mm}

  \begin{subfigure}{.18\textwidth} 
    \centering 
    \includegraphics[height=23mm,angle=90]{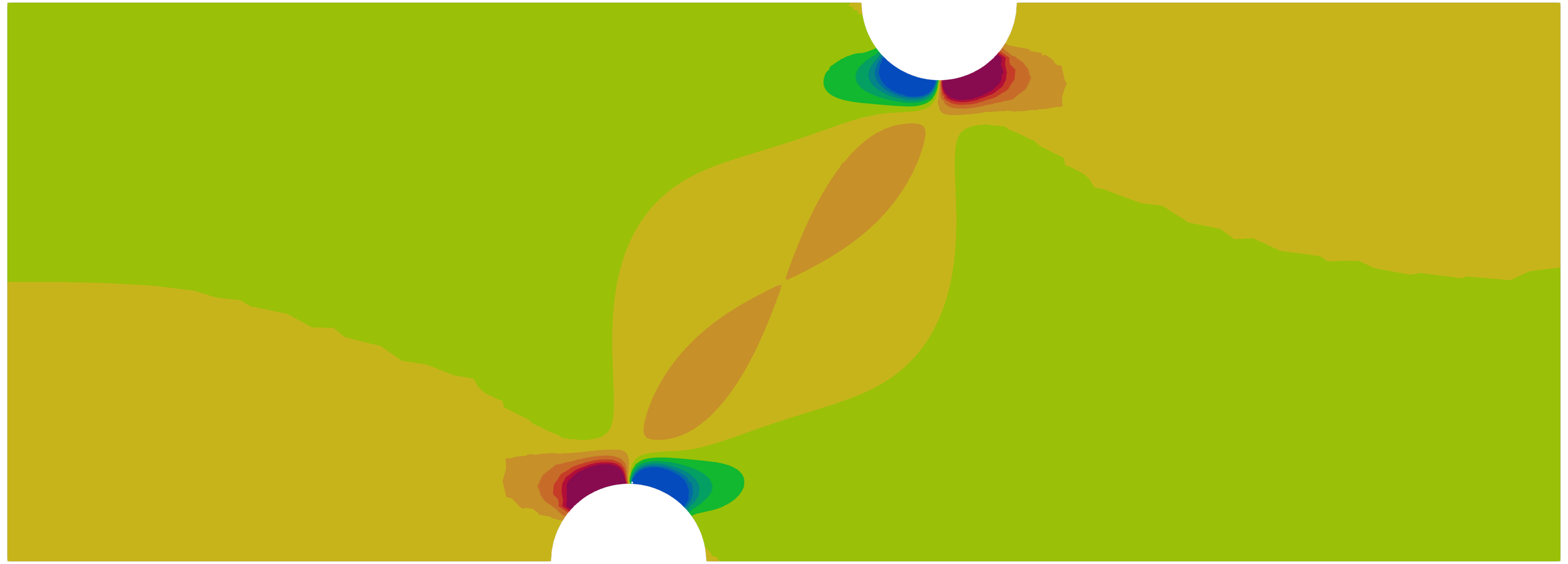}
    \caption{initial}
    \label{fig:p4_AnotchedDevolutionInitial}    
  \end{subfigure}
  \begin{subfigure}{.18\textwidth} 
    \centering 
    \includegraphics[height=23mm,angle=90]{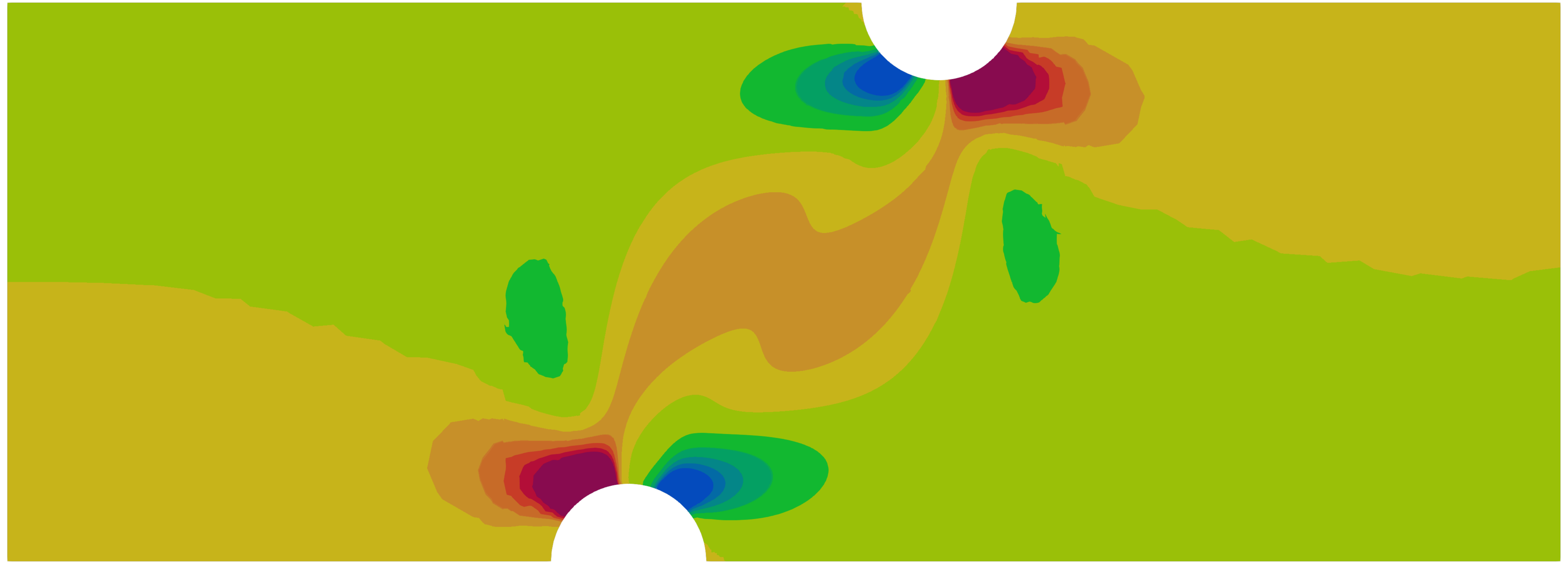}
    \caption{intermediate}
    \label{fig:p4_AnotchedDevolutionIntermediate}    
  \end{subfigure}
  \begin{subfigure}{.18\textwidth} 
    \centering 
    \includegraphics[height=23mm,angle=90]{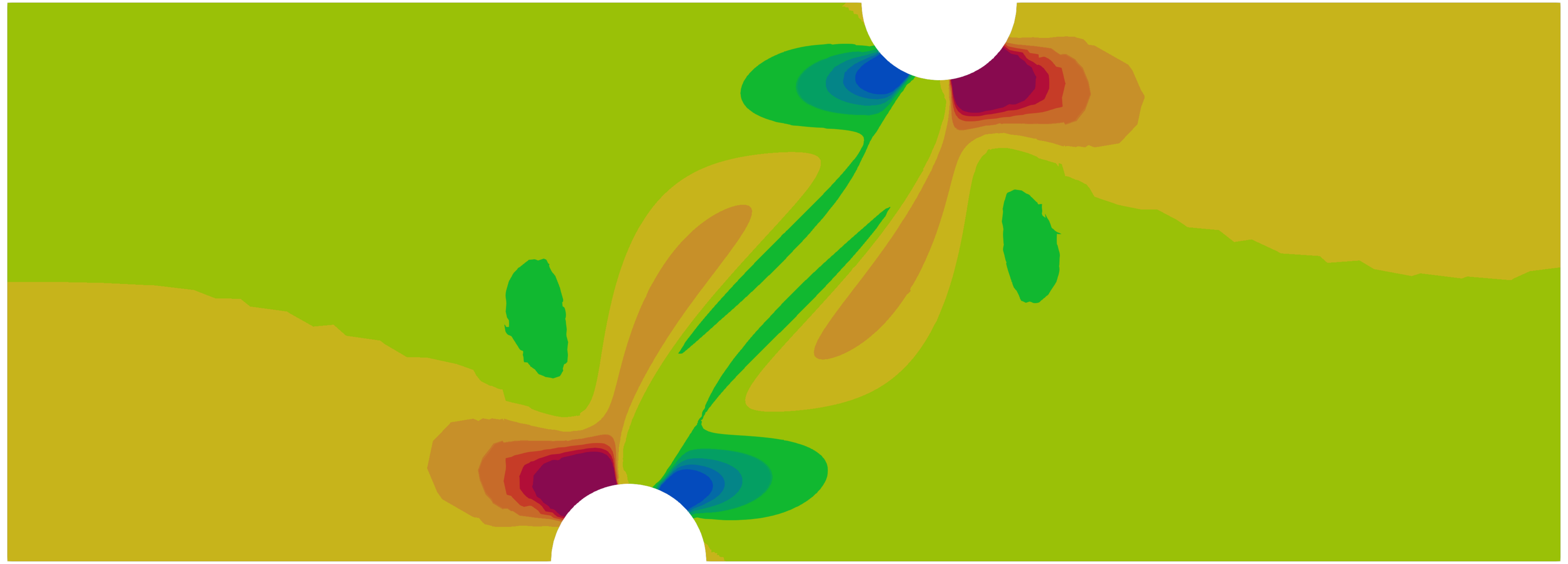}
    \caption{final}
    \label{fig:p4_AnotchedDevolutionFinal}    
  \end{subfigure}
  \begin{subfigure}{.08\textwidth} 
    \centering 
    \begin{tikzpicture}
      \node[inner sep=0pt] (pic) at (0,0) {\includegraphics[height=40mm, width=5mm]
      {03_Contour/00_Color_Maps/Damage_Step_Vertical.pdf}};
      \node[inner sep=0pt] (0)   at ($(pic.south)+( 1.40, 0.15)$)  {$\leq -0.0045$};
      \node[inner sep=0pt] (1)   at ($(pic.south)+( 1.40, 3.80)$)  {$\geq +0.0045$};
      \node[inner sep=0pt] (d)   at ($(pic.south)+( 0.00, 4.35)$)  {$D_{xy}~\si{[-]}$};
    \end{tikzpicture} 
    \hphantom{Model~C}
  \end{subfigure}
  
  \caption{Damage evolution contour plots for the asymmetrically notched specimen with model~C (13955 elements).}
  \label{fig:p4_AnotchedDevolution}     
\end{figure}

The evolution of the normal ($D_{xx}$, $D_{yy}$) and shear ($D_{xy}$) components of the damage tensor for model~C is presented in Fig.~\ref{fig:p4_AnotchedDevolution}, where the position of the snapshots is indicated by the circles in Fig.~\ref{fig:p4_ANotchedFuComp}. Damage initiates at both notches (Fig.~\ref{fig:p4_AnotchedDevolutionInitial}), then it forms a damage shear band zone (Fig.~\ref{fig:p4_AnotchedDevolutionIntermediate}), in which finally the crack forms (Fig.~\ref{fig:p4_AnotchedDevolutionFinal}).

\begin{figure}[htbp]
  \centering
  \includegraphics{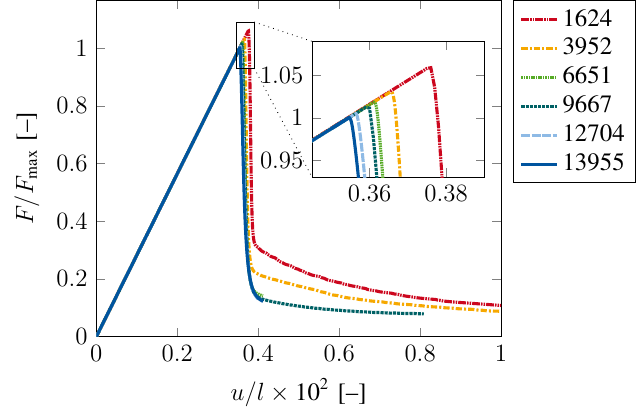}
  \caption{Force-displacement curves for the local damage model for the asymmetrically notched specimen. The forces are normalized with respect to the maximum force of the finest mesh (13955 elements) with $F_\text{max} = 1.5477 \times 10^4~[\si{\newton}]$.}
  \label{fig:p4_ANotchedFuLocal}
\end{figure}

Thereafter, the behavior of the local anisotropic damage without regularization is investigated for the asymmetrically notched specimen. Analogously to \cite{FassinEggersmannEtAl2019a}, the micromorphic parameters are set to $A_i = 0~[\si{\MPa\mm\squared}]$ and $H_i = 0~[\si{\MPa}]$ and, moreover, zero Dirichlet conditions are applied to all nonlocal degrees of freedom, i.e.~$\dbar=\bm{0}$. Fig.~\ref{fig:p4_ANotchedFuLocal} shows the corresponding force-displacement curves for different finite element mesh discretizations. Using the local formulation, a converged solution is neither achieved with respect to the maximum load bearing capacity of the specimen nor with respect to the amount of dissipated energy. 

\begin{figure}
  \centering 
  
  \begin{subfigure}{.25\textwidth} 
    \centering 
    \includegraphics[height=23mm,angle=90]{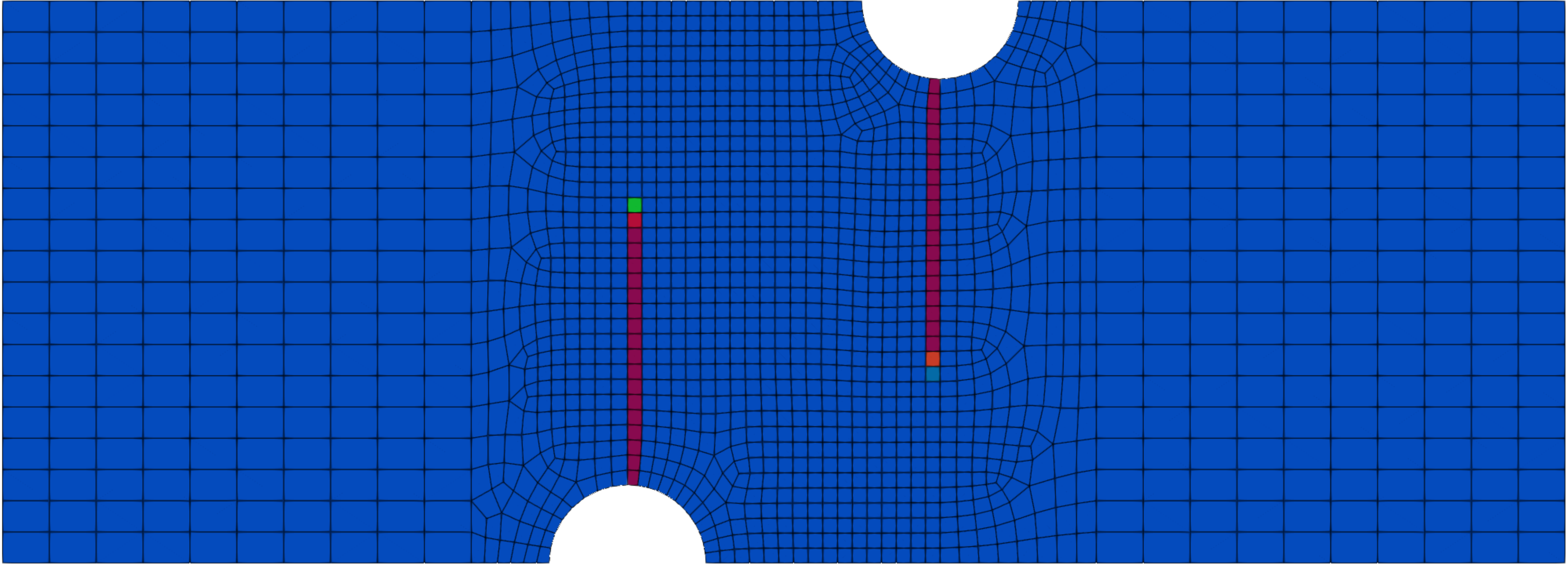}
    \caption{1624 elements}
    \label{fig:p4_AnotchedDlocal_full1624}     
  \end{subfigure}
  \begin{subfigure}{.25\textwidth} 
    \centering 
    \includegraphics[height=23mm,angle=90]{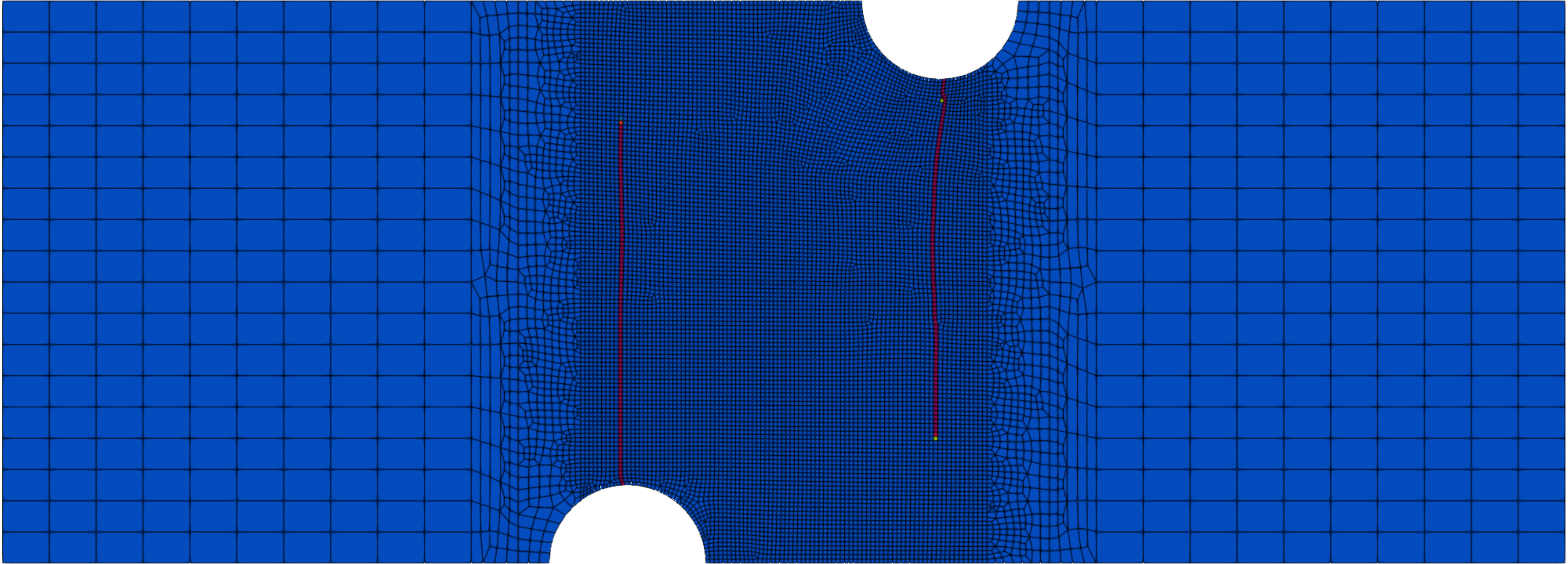}
    \caption{13955 elements}
    \label{fig:p4_AnotchedDlocal_full13955}     
  \end{subfigure}
  \begin{subfigure}{.08\textwidth} 
    \centering 
    \begin{tikzpicture}
      \node[inner sep=0pt] (pic) at (0,0) {\includegraphics[height=40mm, width=5mm]
      {03_Contour/00_Color_Maps/Damage_Step_Vertical.pdf}};
      \node[inner sep=0pt] (0)   at ($(pic.south)+( 0.50, 0.15)$)  {$0$};
      \node[inner sep=0pt] (1)   at ($(pic.south)+( 0.50, 3.80)$)  {$1$};
      \node[inner sep=0pt] (d)   at ($(pic.south)+( 0.00, 4.35)$)  {$D_{xx}~\si{[-]}$};
    \end{tikzpicture} 
    \vspace{5mm}
  \end{subfigure}

  \vspace{8mm}

  \begin{subfigure}{.33\textwidth} 
    \centering 
    \includegraphics[height=50mm,angle=90]{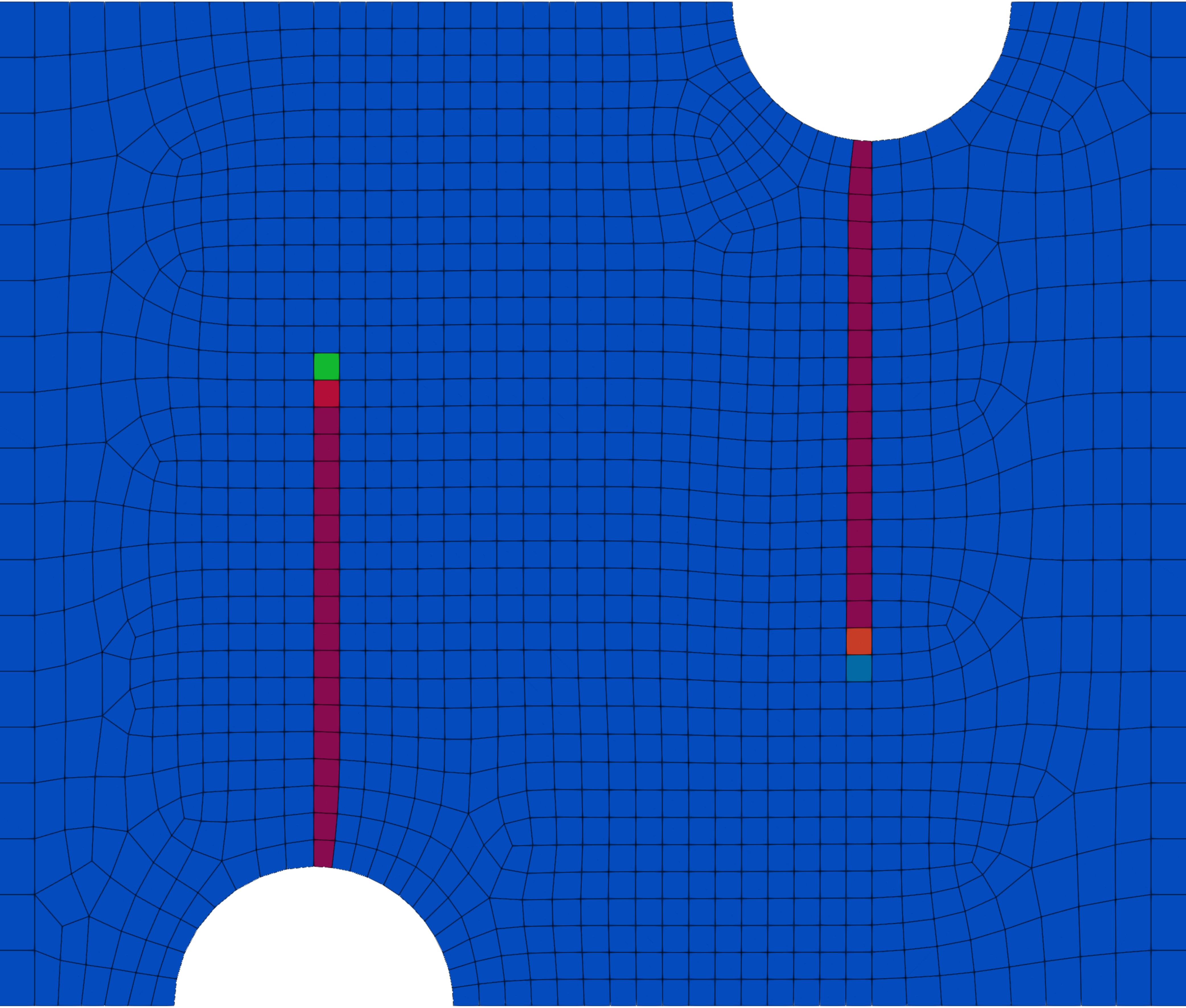}
    \caption{1624 elements}
    \label{fig:p4_AnotchedDlocal_zoom1624}     
  \end{subfigure}%
  \begin{subfigure}{.33\textwidth} 
    \centering 
    \includegraphics[height=50mm,angle=90]{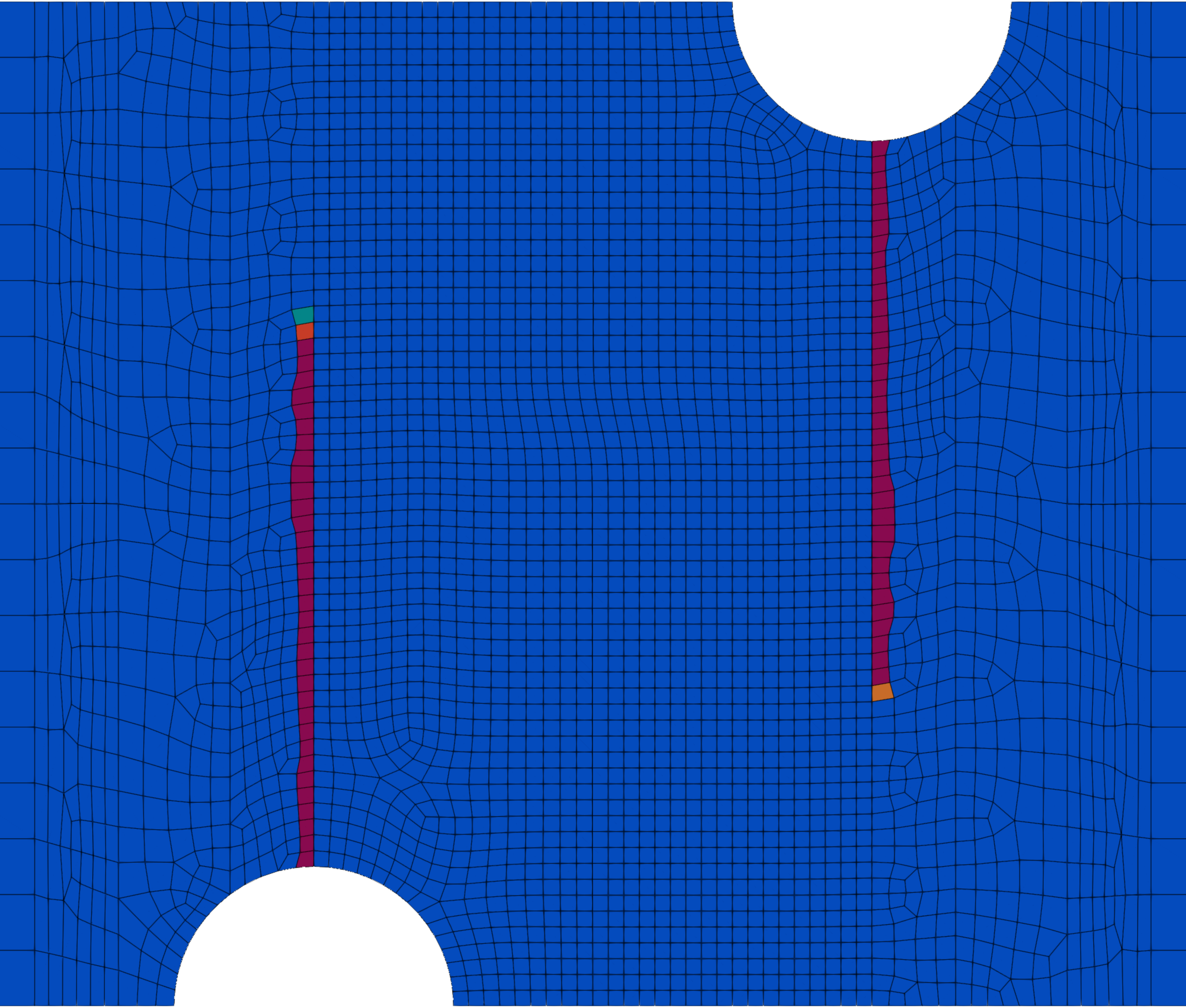}
    \caption{3592 elements}
  \end{subfigure}%
  \begin{subfigure}{.33\textwidth} 
    \centering 
    \includegraphics[height=50mm,angle=90]{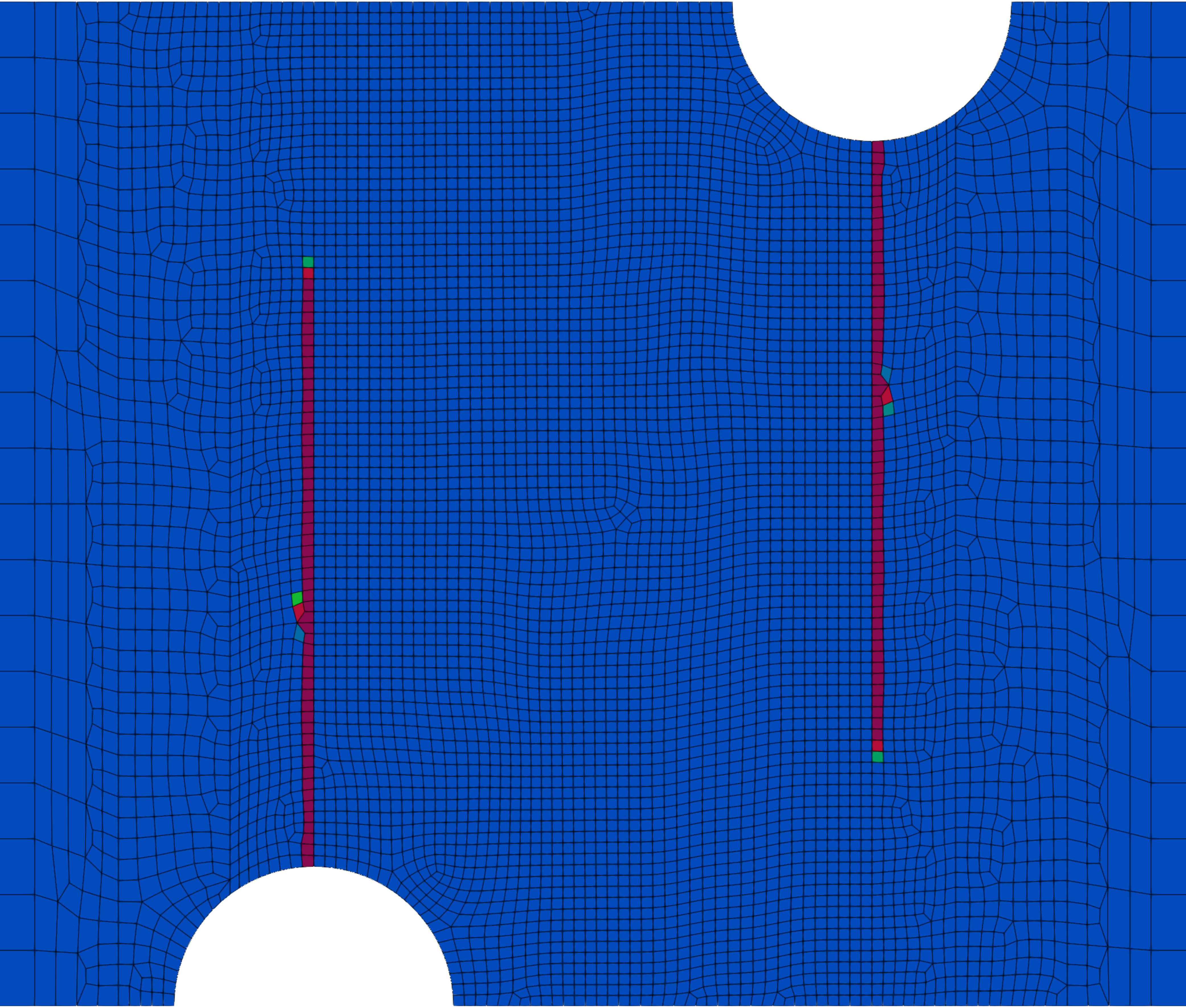}
    \caption{6651 elements}
  \end{subfigure}

  \vspace{1mm}

  \begin{subfigure}{.33\textwidth} 
    \centering 
    \includegraphics[height=50mm,angle=90]{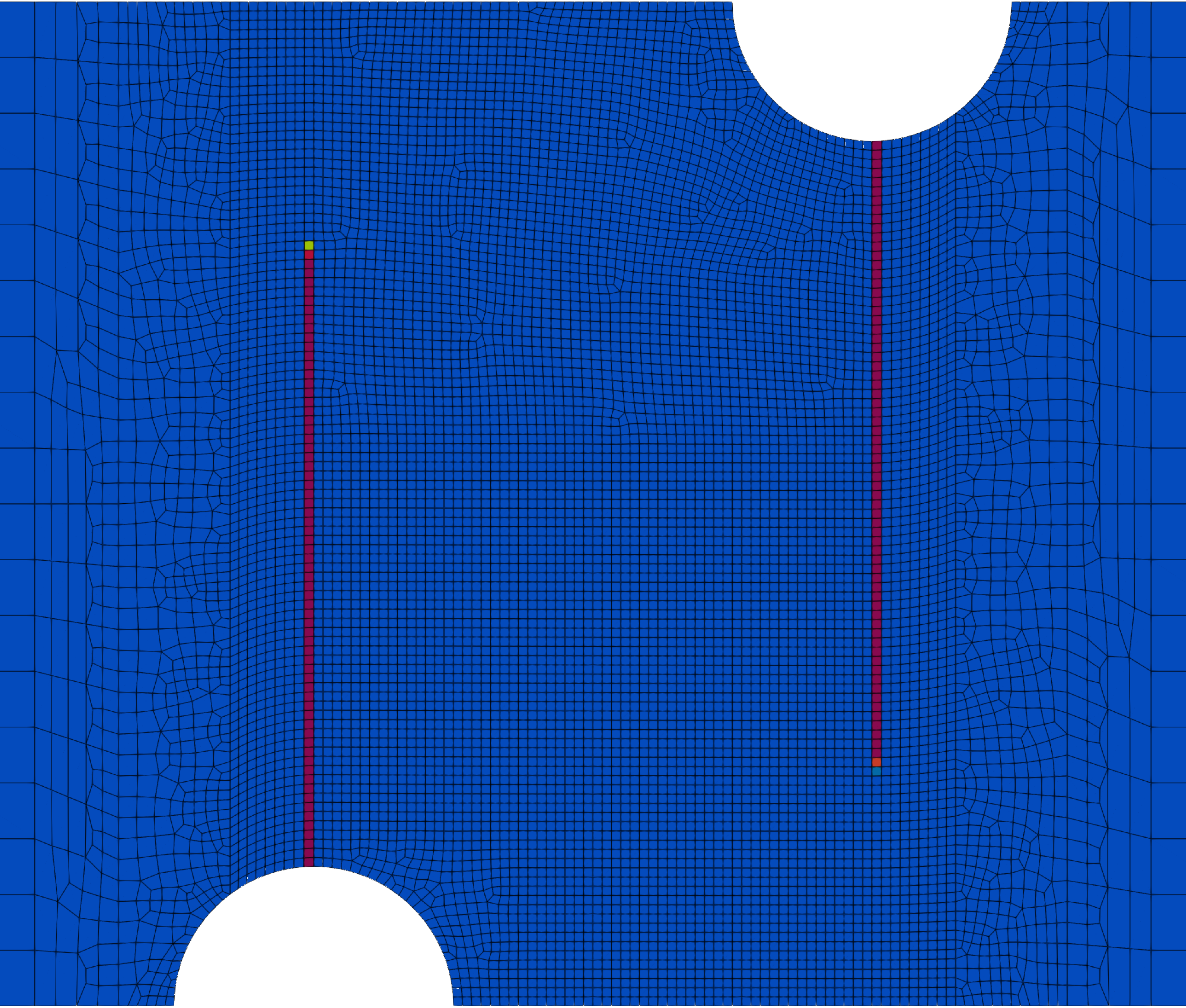}
    \caption{9667 elements}
  \end{subfigure}%
  \begin{subfigure}{.33\textwidth} 
    \centering 
    \includegraphics[height=50mm,angle=90]{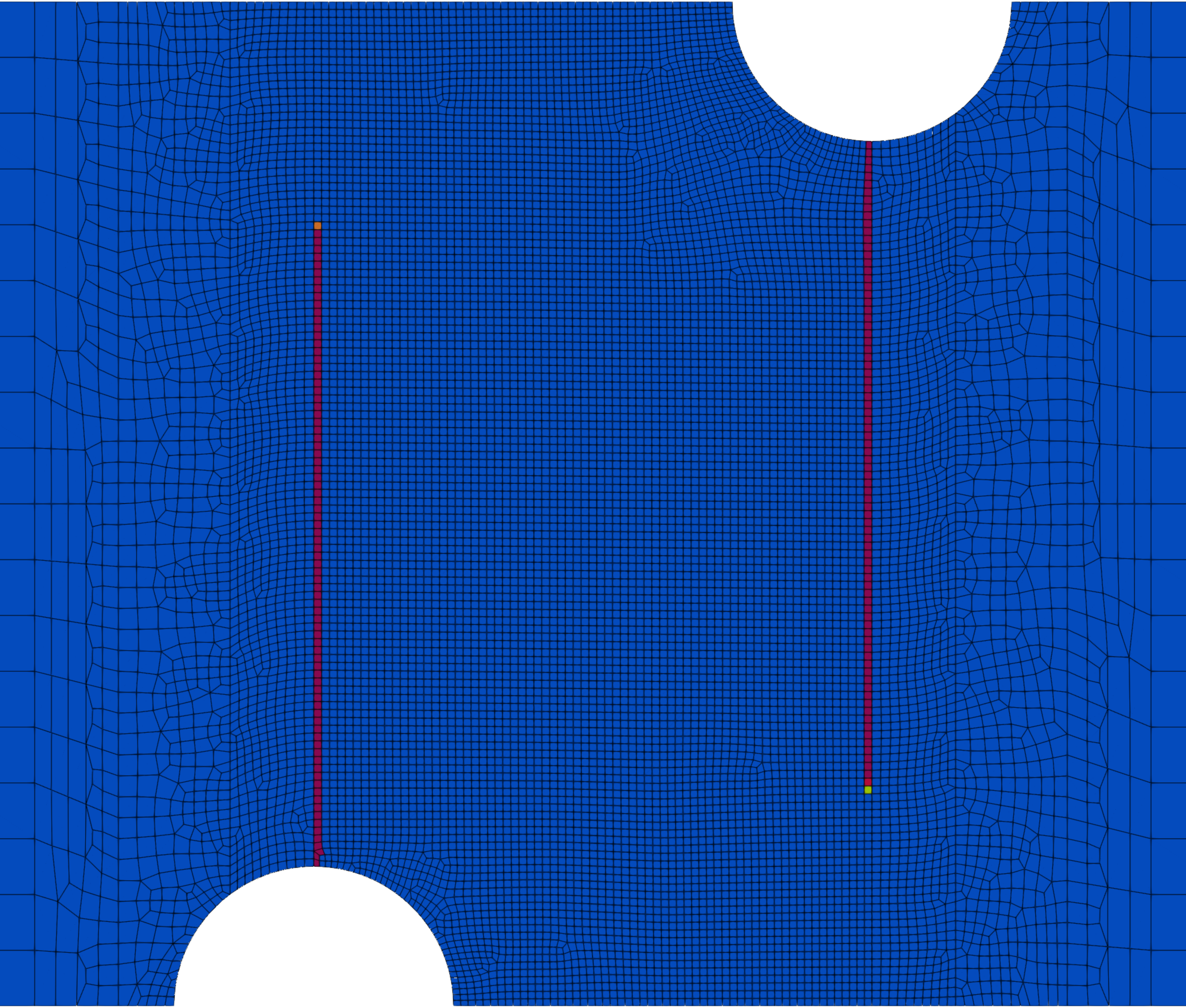}
    \caption{12704 elements}
  \end{subfigure}%
  \begin{subfigure}{.33\textwidth} 
    \centering 
    \includegraphics[height=50mm,angle=90]{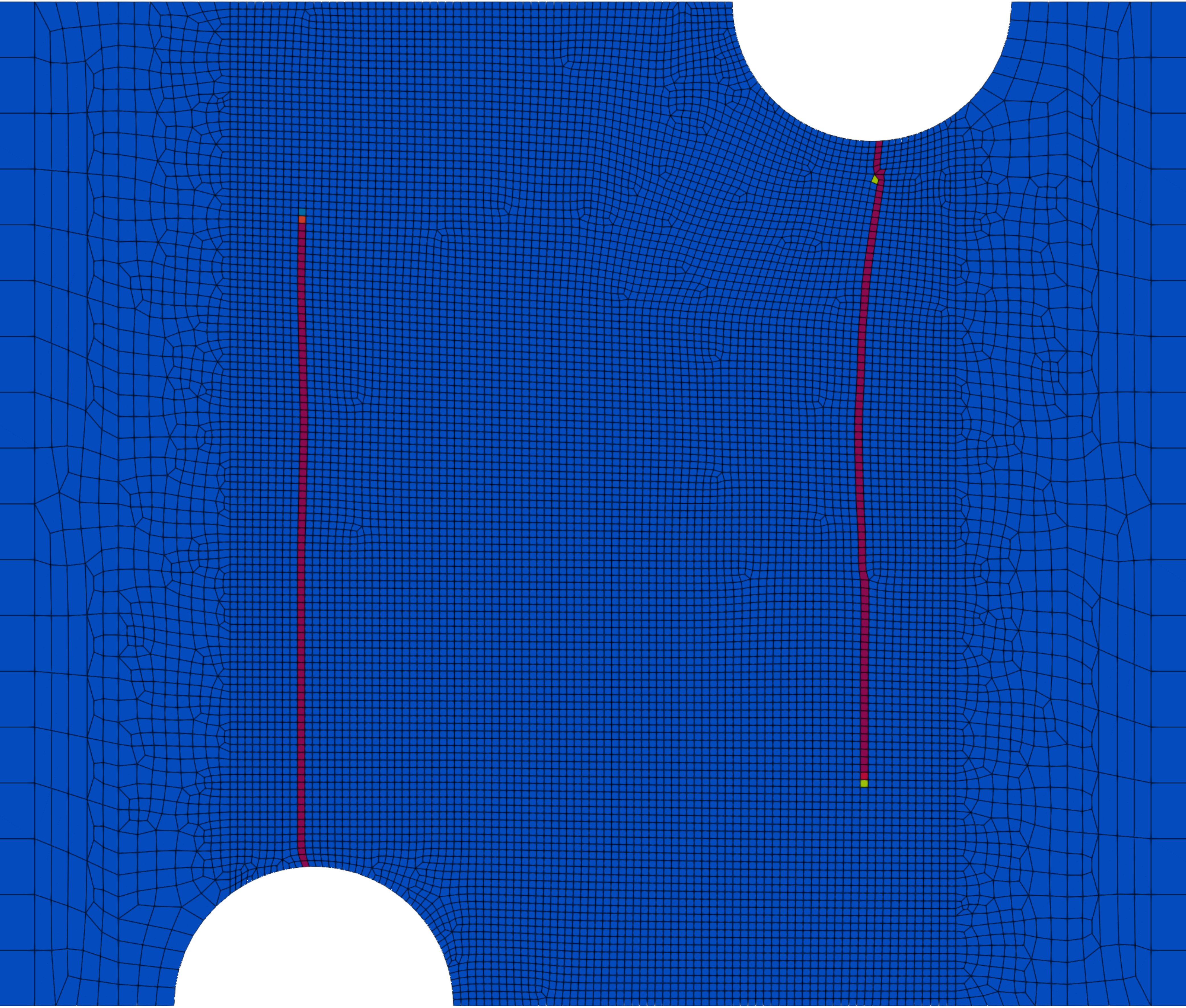}
    \caption{13955 elements}
    \label{fig:p4_AnotchedDlocal_zoom13955}     
  \end{subfigure}

  \caption{Damage contour plots with the local damage model for the asymmetrically notched specimen at $u/l \times \text{10}^\text{2}=0.4~\text{[--]}$. The full specimen is shown with the coarsest and finest mesh in Figs.~\ref{fig:p4_AnotchedDlocal_full1624} and \ref{fig:p4_AnotchedDlocal_full13955} and the center of the specimen is shown for all meshes in Figs.~\ref{fig:p4_AnotchedDlocal_zoom1624}-\ref{fig:p4_AnotchedDlocal_zoom13955}. For this study, the damage variables are averaged over all integration points per element.}
  \label{fig:p4_AnotchedDlocal}     
\end{figure}

Furthermore, the damage contour plots for the study of the local anisotropic damage model are presented in Fig.~\ref{fig:p4_AnotchedDlocal}. For all meshes, two cracks form at the notches and propagate horizontally through the specimen, where all cracks localize into a single row of elements. They do not exhibit a tendency of forming a shear crack and to coalesce, which contradicts the results of the gradient-extended solution in Figs.~\ref{fig:p4_ANotchedFu}-\ref{fig:p4_AnotchedDevolution} and the experimental investigations of e.g.~\cite{AmbatiKruseEtAl2016}.
Hence, the utilized artificial viscosity $\eta_v = 1~[\si{\MPa\s}]$ does not yield regularizing effects and the excellent mesh convergence in Figs.~\ref{fig:p4_ANotchedFuA}, \ref{fig:p4_ANotchedFuB}, and \ref{fig:p4_ANotchedFuC} is related to the gradient-extensions of models~A, B, and C.

\begin{figure}[htbp]
  \centering
  \includegraphics{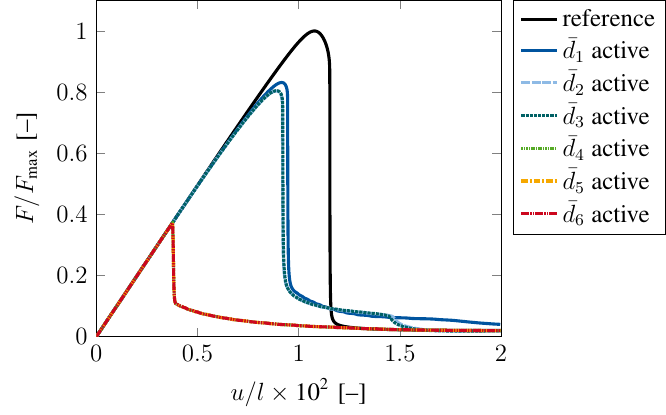}
  \caption{Force-displacement curves for the damage model with regularization of single components of the damage tensor using model~A for the asymmetrically notched specimen. The forces are normalized with respect to the maximum force of the reference regularization using the mesh with 1624 elements with $F_\text{max} = 4.4114 \times 10^4~[\si{\newton}]$.}
  \label{fig:p4_ANotchedFuStudyAi}
\end{figure}

Next, we investigate the possibility of a further reduction of the micromorphic tuple by using only a single degree of freedom. Therefore, model~A is again utilized for the simulation of the asymmetrically notched specimen, but each time only a single component of the micromorphic tuple is activated. For example, "$\bar{d}_1$ active" refers to the micromorphic gradient parameters $A_1^\text{A} = 1000~[\si{\MPa\mm\squared}]$, $A_2^\text{A} = A_3^\text{A} = A_4^\text{A} = A_5^\text{A} = A_6^\text{A} = 0~[\si{\MPa\mm\squared}]$, the micromorphic penalty parameters $H_1^\text{A} = 10^4~[\si{\MPa}]$, $H_2^\text{A} = H_3^\text{A} = H_4^\text{A} = H_5^\text{A} = H_6^\text{A} = 0~[\si{\MPa}]$, and the Dirichlet boundary conditions $\bar{d}_2 = \bar{d}_3 = \bar{d}_4 = \bar{d}_5 = \bar{d}_6 = 0~[\si{-}]$. For model~A, $\bar{d}_1$ is associated with the regularization of $D_{xx}$, $\bar{d}_2$ with $D_{yy}$, $\bar{d}_3$ with $D_{zz}$, $\bar{d}_4$ with $D_{xy}$, $\bar{d}_5$ with $D_{xz}$, and $\bar{d}_6$ with $D_{yz}$.

Fig.~\ref{fig:p4_ANotchedFuStudyAi} shows the corresponding force-displacement curves for the regularization of a single component of the damage tensor as well as the reference solution where all six independent components are regularized (from Fig.~\ref{fig:p4_ANotchedFuA}). The regularization of a single component does not suffice for the regularization of the shear crack, since all solutions obtained by a single-component regularization underestimate the maximum force carried by the specimen. The regularization of the normal component $D_{xx}$ by $\bar{d}_1$, i.e.~the damage tensor component in loading direction, yields an underestimation of $16.86~\si{[\percent]}$. The regularization of the normal components $D_{yy}$ and $D_{zz}$ by $\bar{d}_2$ and $\bar{d}_3$ yield an underestimation of $19.77~\si{[\percent]}$ and $19.55~\si{[\percent]}$ and the regularization of the shear components $D_{xy}$, $D_{xz}$, and $D_{yz}$ by $\bar{d}_4$, $\bar{d}_5$, and $\bar{d}_6$ each yields an underestimation of $62.80~\si{[\percent]}$.

\begin{figure}
  \centering 
  \begin{subfigure}{.18\textwidth} 
    \centering 
    \includegraphics[height=23mm,angle=90]{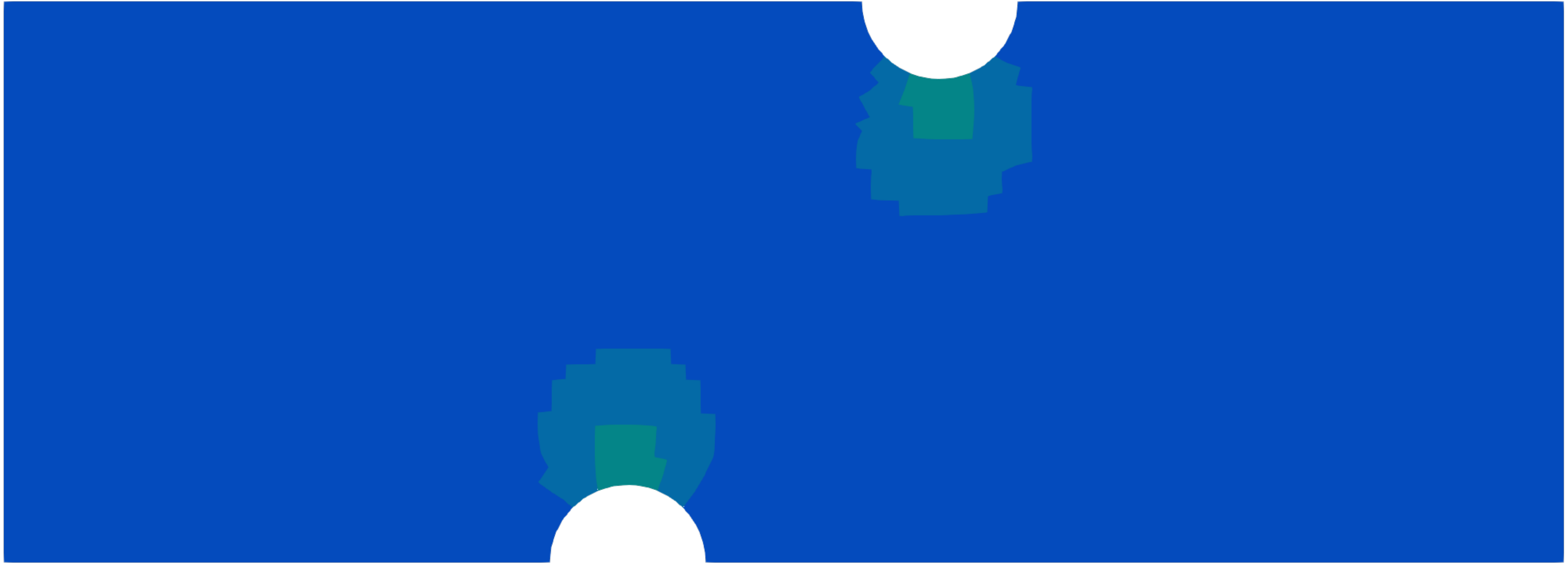}
  \end{subfigure}%
  \begin{subfigure}{.18\textwidth} 
    \centering 
    \includegraphics[height=23mm,angle=90]{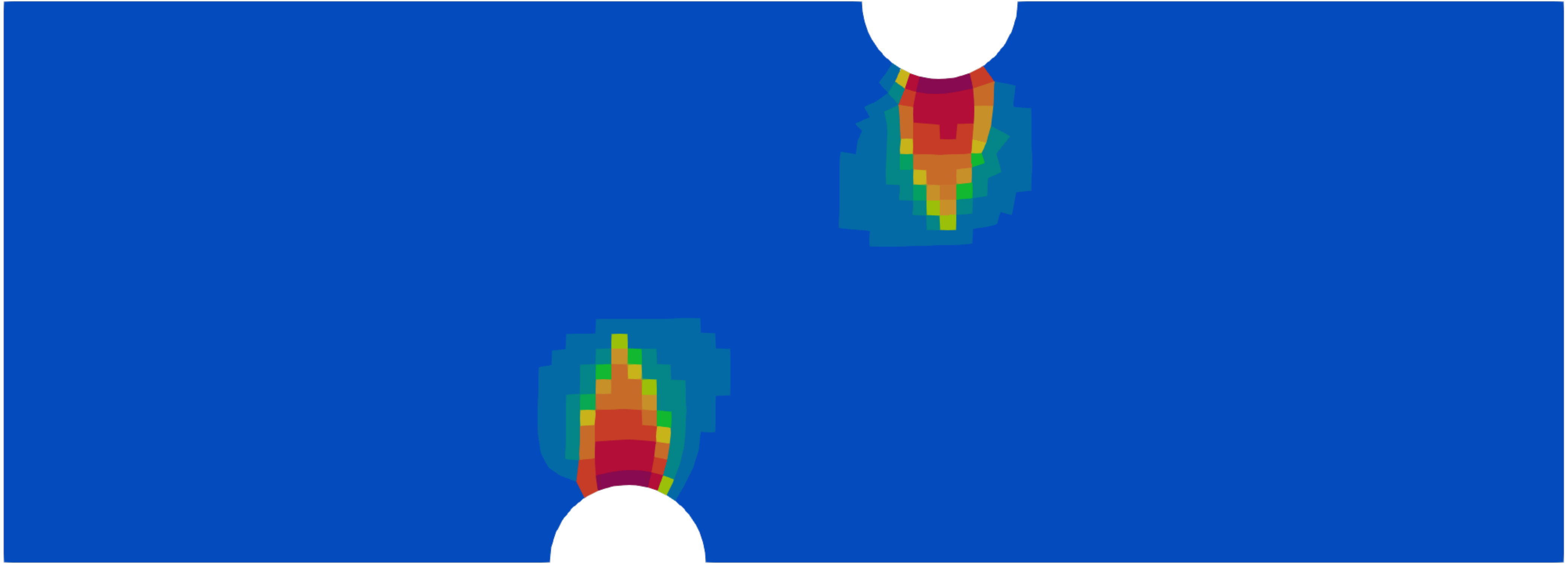}
  \end{subfigure}%
  \begin{subfigure}{.18\textwidth} 
    \centering 
    \includegraphics[height=23mm,angle=90]{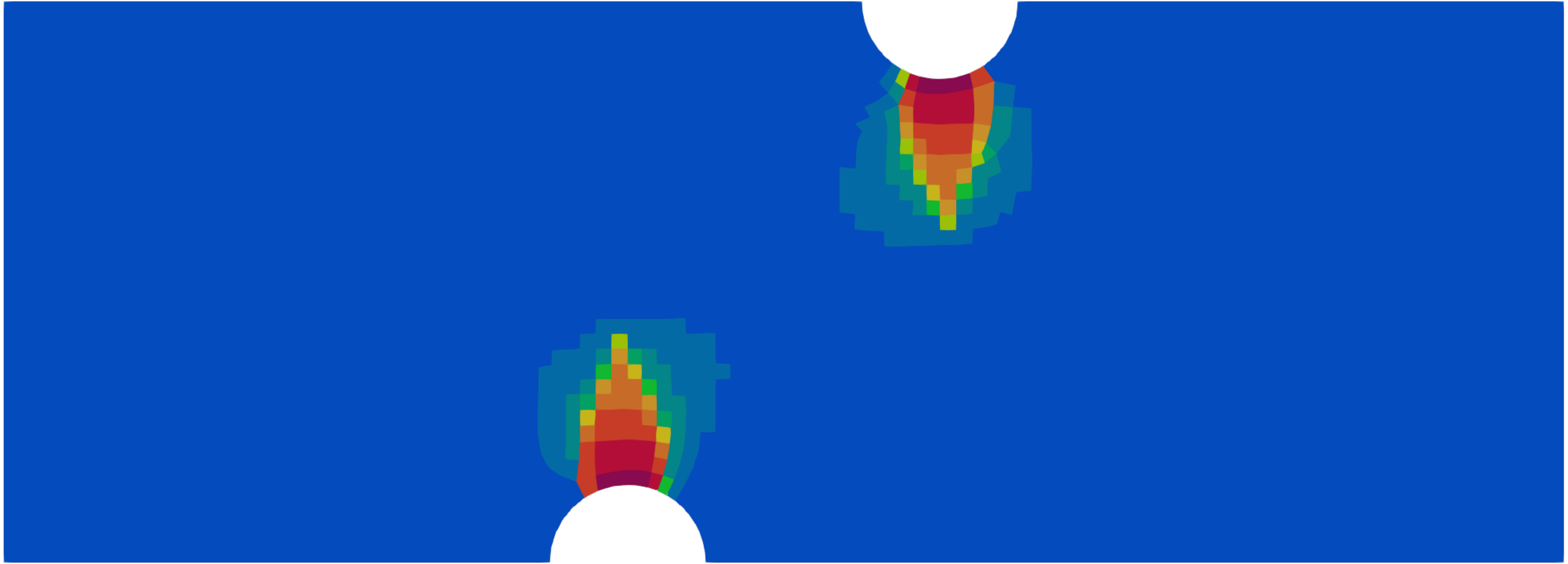}
  \end{subfigure}%
  \begin{subfigure}{.18\textwidth} 
    \centering 
    \includegraphics[height=23mm,angle=90]{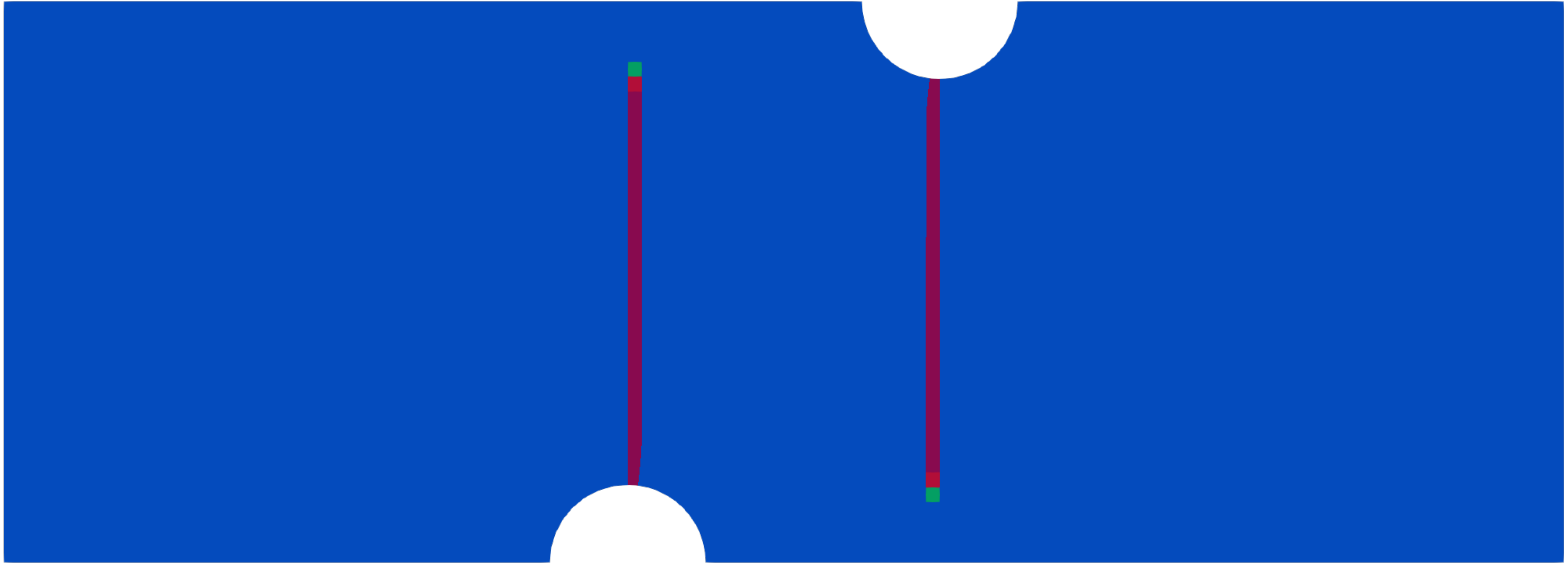}
  \end{subfigure}%
  \begin{subfigure}{.18\textwidth} 
    \centering 
    \includegraphics[height=23mm,angle=90]{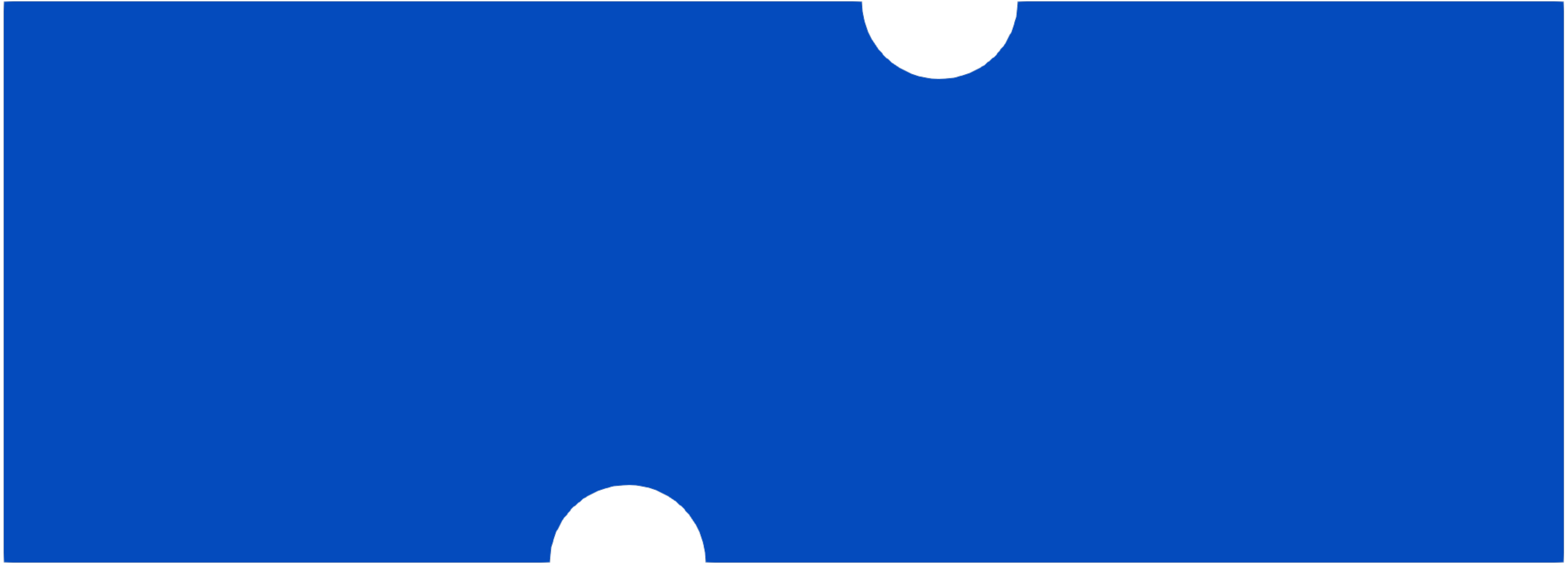}
  \end{subfigure}%
  \begin{subfigure}{.08\textwidth} 
    \centering 
    \begin{tikzpicture}
      \node[inner sep=0pt] (pic) at (0,0) {\includegraphics[height=40mm, width=5mm]
      {03_Contour/00_Color_Maps/Damage_Step_Vertical.pdf}};
      \node[inner sep=0pt] (0)   at ($(pic.south)+( 0.50, 0.15)$)  {$0$};
      \node[inner sep=0pt] (1)   at ($(pic.south)+( 0.50, 3.80)$)  {$1$};
      \node[inner sep=0pt] (d)   at ($(pic.south)+( 0.00, 4.35)$)  {$D_{xx}~\si{[-]}$};
    \end{tikzpicture} 
  \end{subfigure}

  \vspace{2mm}
  
  \begin{subfigure}{.18\textwidth} 
    \centering 
    \includegraphics[height=23mm,angle=90]{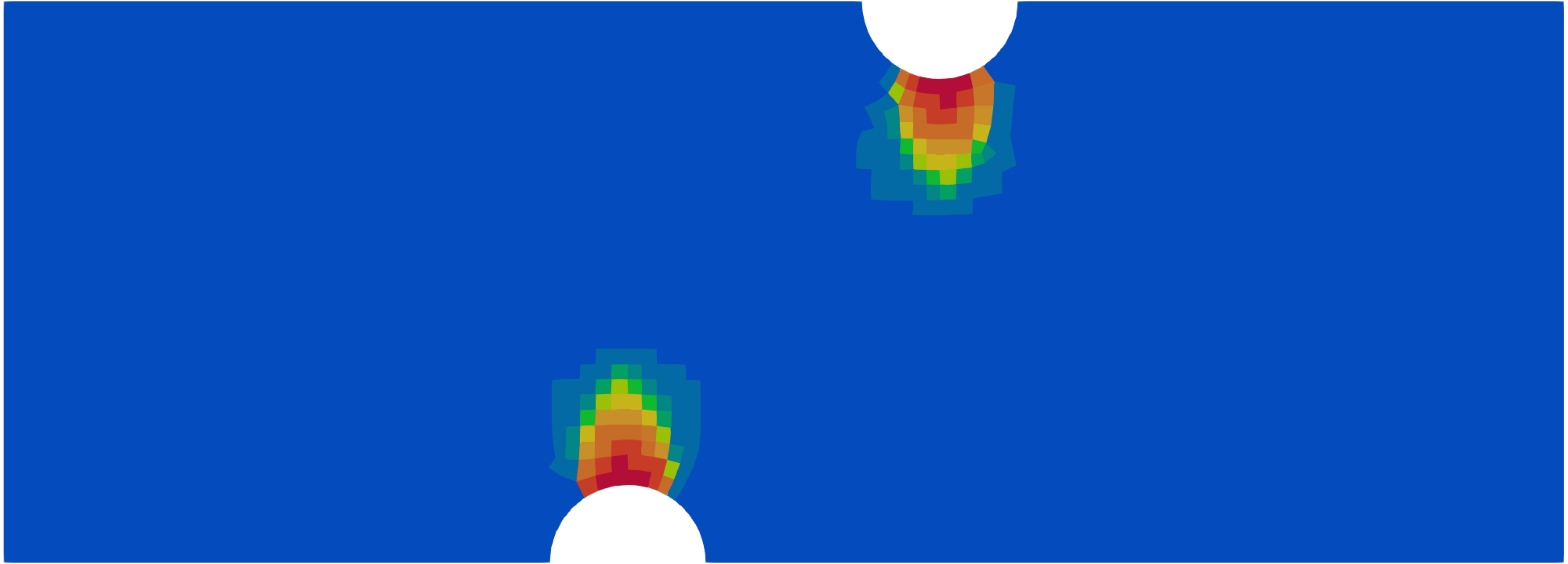}
  \end{subfigure}%
  \begin{subfigure}{.18\textwidth} 
    \centering 
    \includegraphics[height=23mm,angle=90]{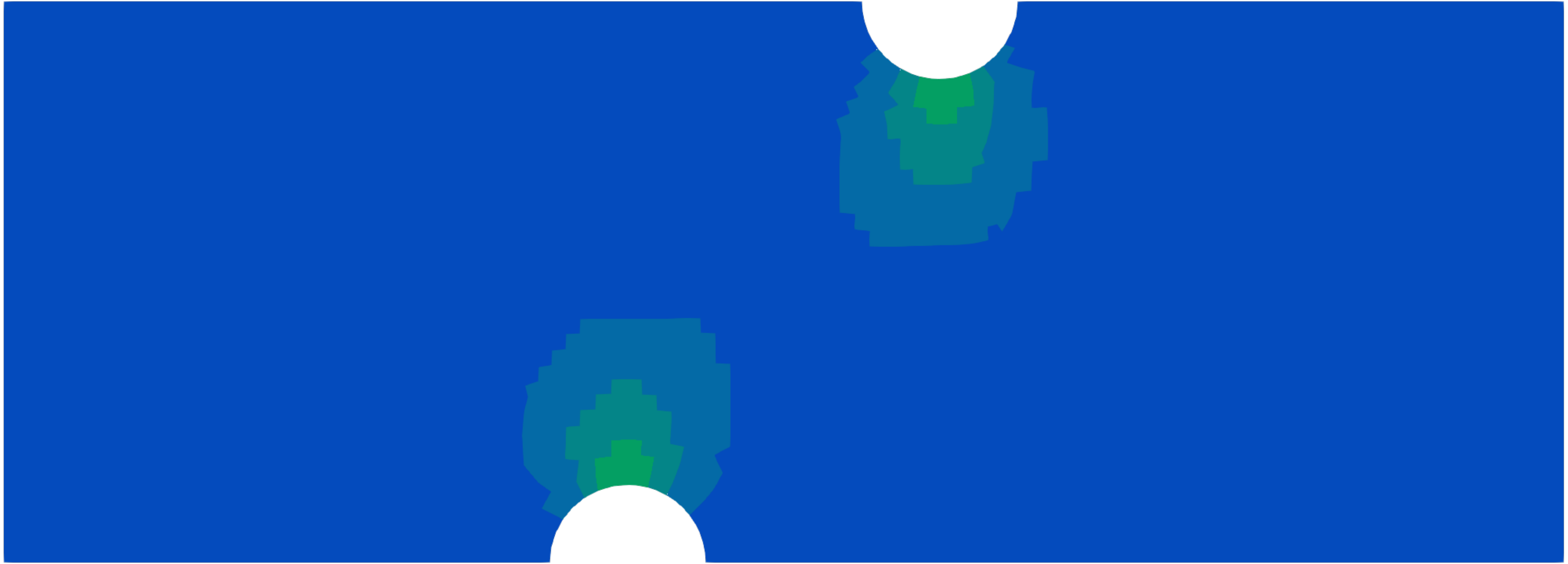}
  \end{subfigure}%
  \begin{subfigure}{.18\textwidth} 
    \centering 
    \includegraphics[height=23mm,angle=90]{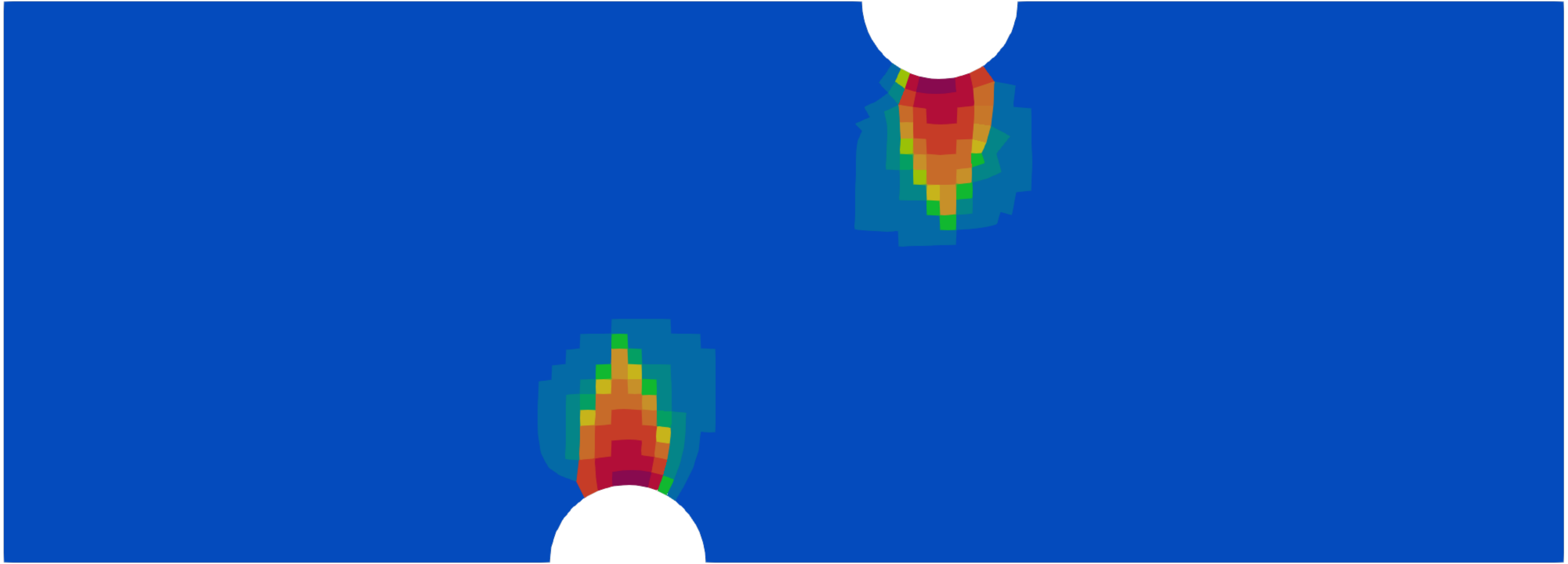}
  \end{subfigure}%
  \begin{subfigure}{.18\textwidth} 
    \centering 
    \includegraphics[height=23mm,angle=90]{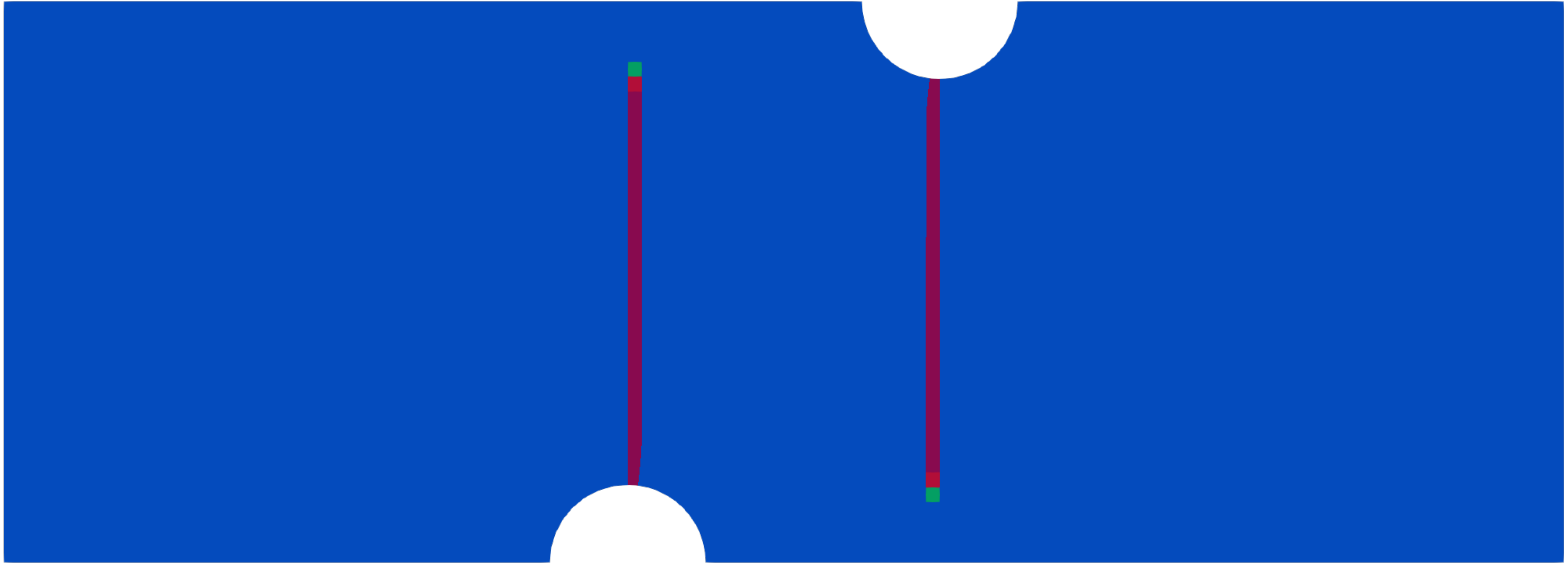}
  \end{subfigure}%
  \begin{subfigure}{.18\textwidth} 
    \centering 
    \includegraphics[height=23mm,angle=90]{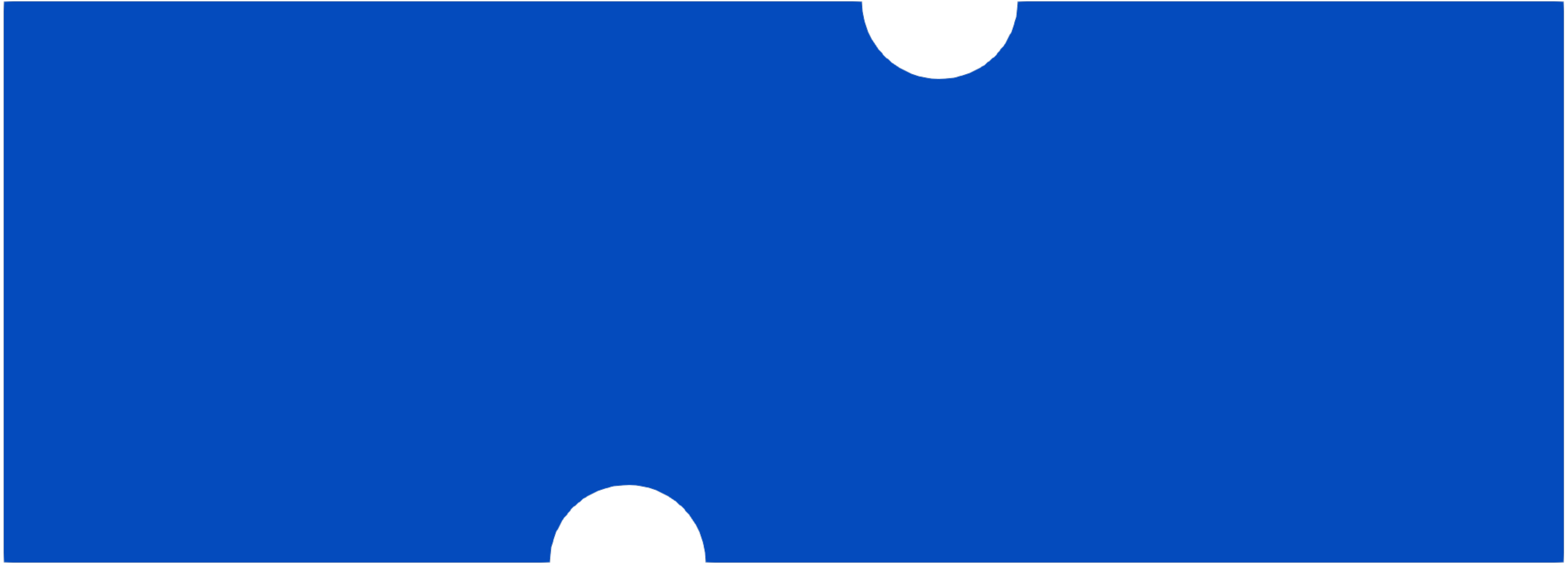}
  \end{subfigure}%
  \begin{subfigure}{.08\textwidth} 
    \centering 
    \begin{tikzpicture}
      \node[inner sep=0pt] (pic) at (0,0) {\includegraphics[height=40mm, width=5mm]
      {03_Contour/00_Color_Maps/Damage_Step_Vertical.pdf}};
      \node[inner sep=0pt] (0)   at ($(pic.south)+( 0.50, 0.15)$)  {$0$};
      \node[inner sep=0pt] (1)   at ($(pic.south)+( 0.50, 3.80)$)  {$1$};
      \node[inner sep=0pt] (d)   at ($(pic.south)+( 0.00, 4.35)$)  {$D_{yy}~\si{[-]}$};
    \end{tikzpicture} 
  \end{subfigure}
  
  \vspace{1mm}
  
  \begin{subfigure}{.18\textwidth} 
    \centering 
    \caption{$\bar{d}_1$ active}
    \label{fig:p4_AnotchedDstudyAi_u09_d1bar}
  \end{subfigure}%
  \begin{subfigure}{.18\textwidth} 
    \centering 
    \caption{$\bar{d}_2$ active}
    \label{fig:p4_AnotchedDstudyAi_u09_d2bar}
  \end{subfigure}%
  \begin{subfigure}{.18\textwidth} 
    \centering 
    \caption{$\bar{d}_3$ active}
    \label{fig:p4_AnotchedDstudyAi_u09_d3bar}
  \end{subfigure}%
  \begin{subfigure}{.18\textwidth} 
    \centering 
    \caption{$\bar{d}_4$ active}
    \label{fig:p4_AnotchedDstudyAi_u09_d4bar}
  \end{subfigure}%
  \begin{subfigure}{.18\textwidth} 
    \centering 
    \caption{reference}
    \label{fig:p4_AnotchedDstudyAi_u09_ref}
  \end{subfigure}%
  \begin{subfigure}{.08\textwidth} 
    \hphantom{Dxx}
  \end{subfigure}

  \caption{Damage contour plots for the damage model with regularization of single components of the damage tensor using model~A for the asymmetrically notched specimen (1624 elements) at $u/l \times \text{10}^\text{2}=0.9~\text{[--]}$ in Fig.~\ref{fig:p4_ANotchedFuStudyAi}. For this study, the damage variables are averaged over all integration points per element.}
  \label{fig:p4_AnotchedDstudyAi_u09}     
\end{figure}

\begin{figure}
  \centering 
  \begin{subfigure}{.18\textwidth} 
    \centering 
    \includegraphics[height=23mm,angle=90]{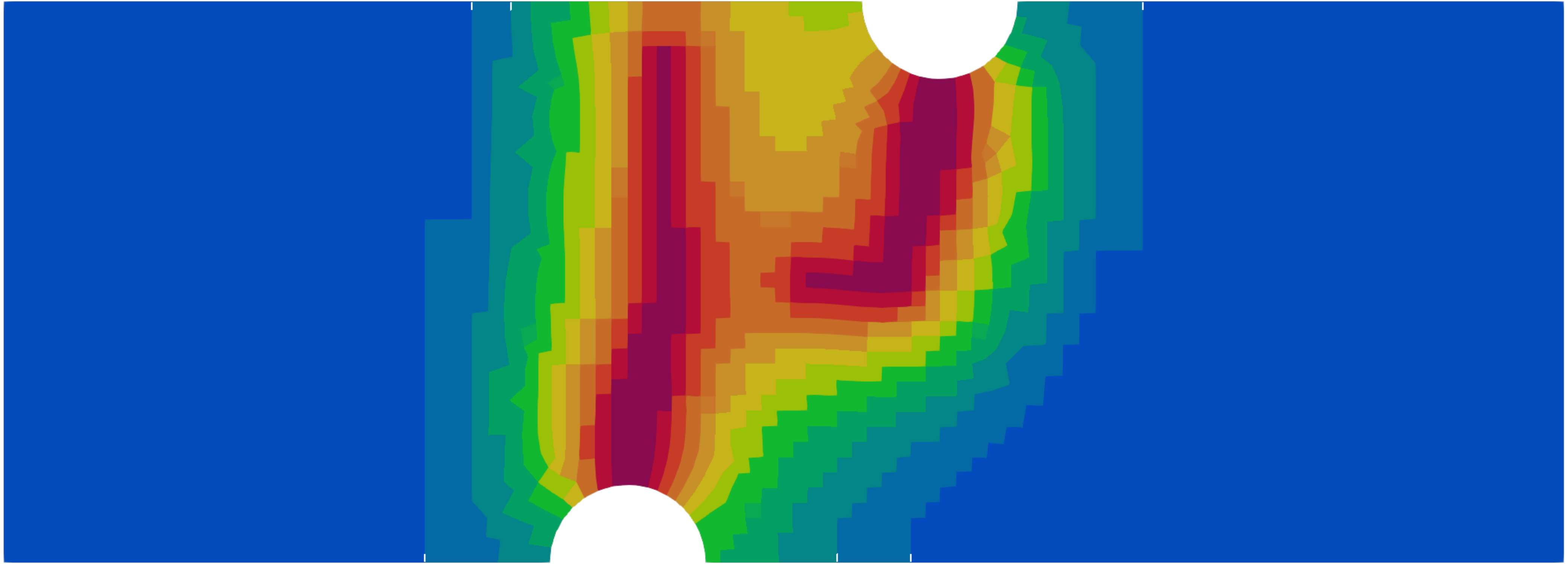}
  \end{subfigure}%
  \begin{subfigure}{.18\textwidth} 
    \centering 
    \includegraphics[height=23mm,angle=90]{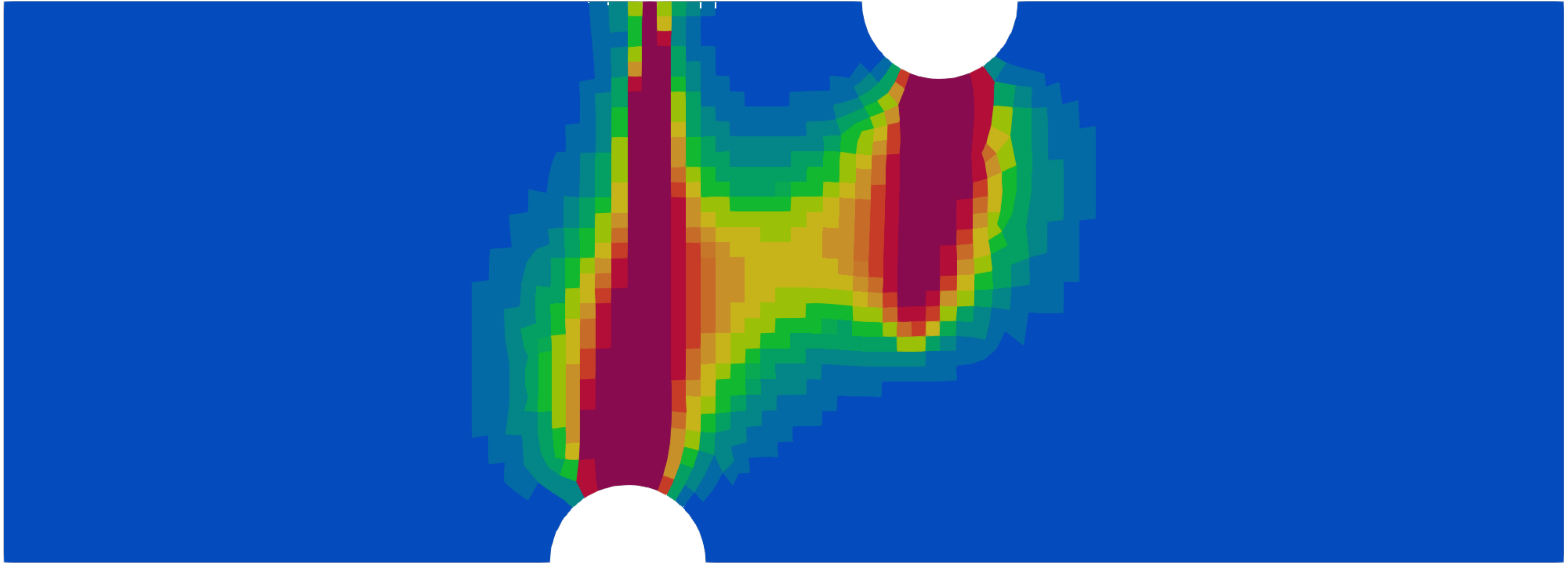}
  \end{subfigure}%
  \begin{subfigure}{.18\textwidth} 
    \centering 
    \includegraphics[height=23mm,angle=90]{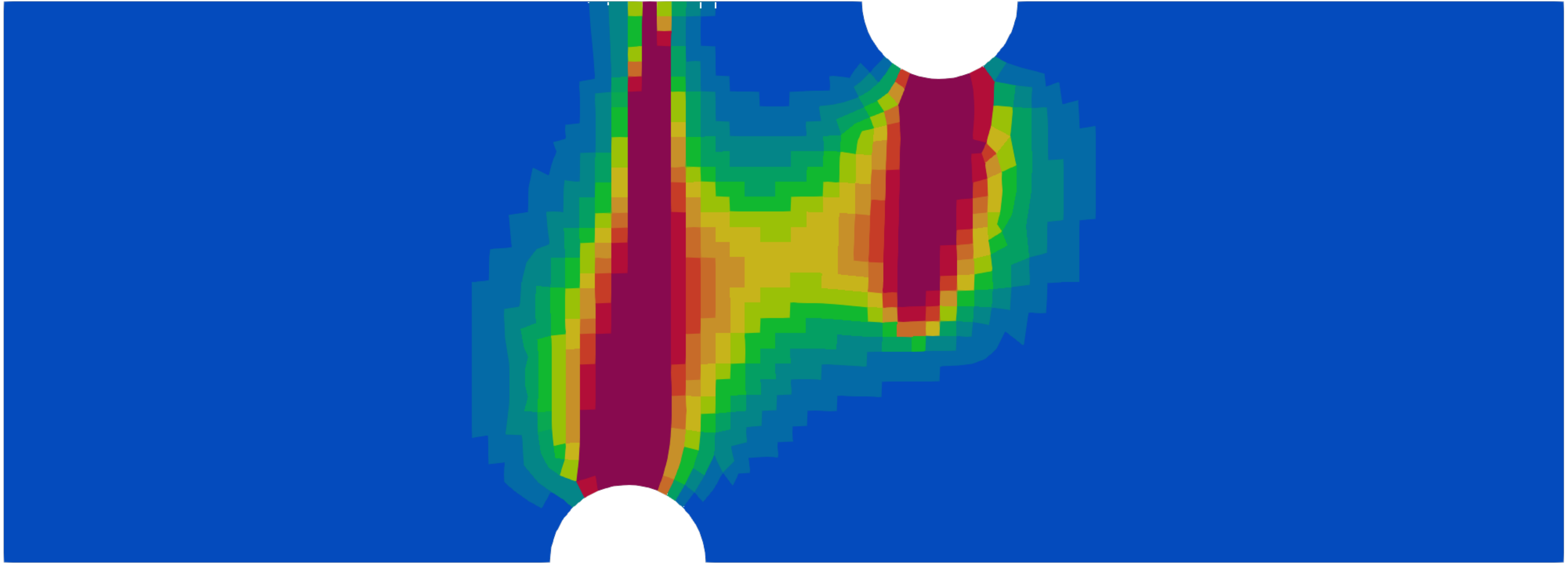}
  \end{subfigure}%
  \begin{subfigure}{.18\textwidth} 
    \centering 
    \includegraphics[height=23mm,angle=90]{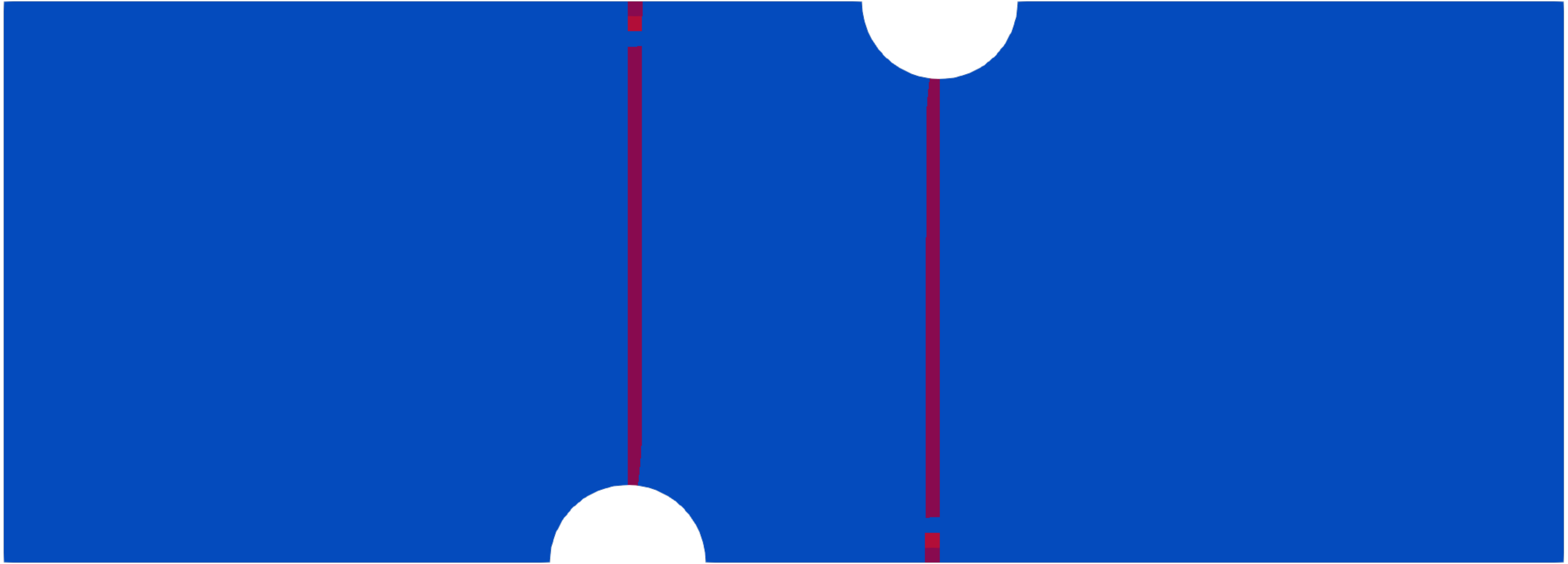}
  \end{subfigure}%
  \begin{subfigure}{.18\textwidth} 
    \centering 
    \includegraphics[height=23mm,angle=90]{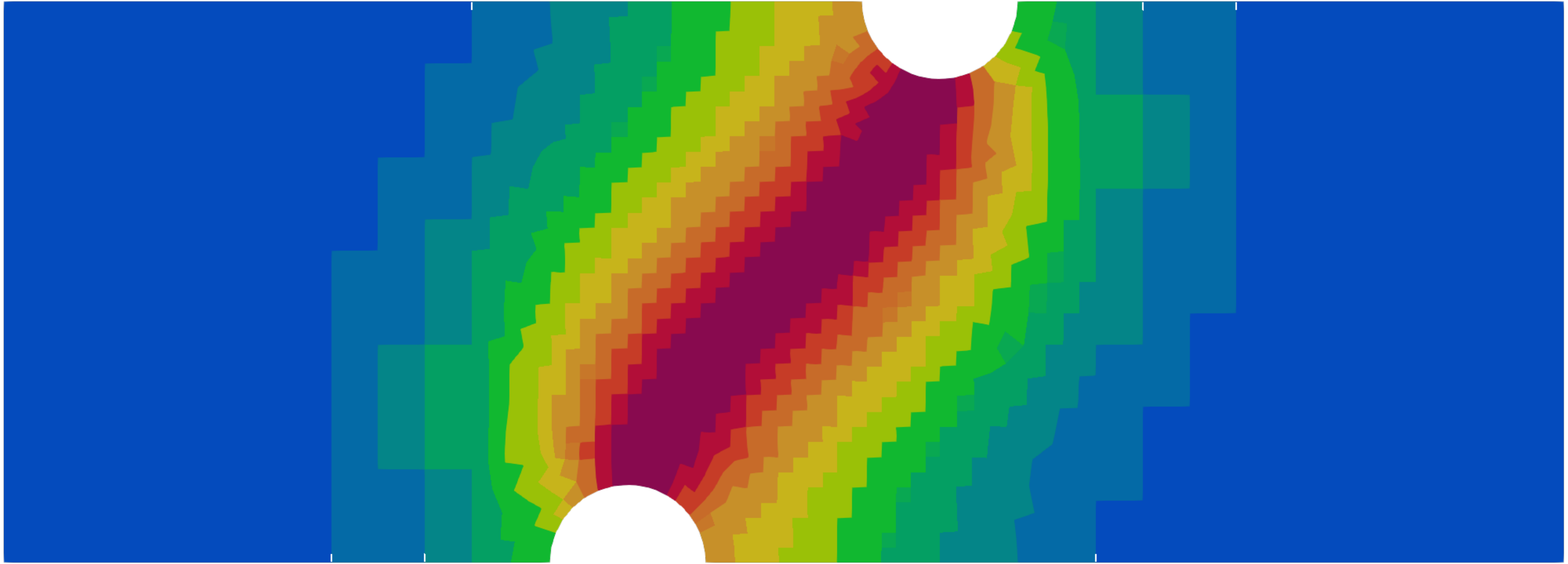}
  \end{subfigure}%
  \begin{subfigure}{.08\textwidth} 
    \centering 
    \begin{tikzpicture}
      \node[inner sep=0pt] (pic) at (0,0) {\includegraphics[height=40mm, width=5mm]
      {03_Contour/00_Color_Maps/Damage_Step_Vertical.pdf}};
      \node[inner sep=0pt] (0)   at ($(pic.south)+( 0.50, 0.15)$)  {$0$};
      \node[inner sep=0pt] (1)   at ($(pic.south)+( 0.50, 3.80)$)  {$1$};
      \node[inner sep=0pt] (d)   at ($(pic.south)+( 0.00, 4.35)$)  {$D_{xx}~\si{[-]}$};
    \end{tikzpicture} 
  \end{subfigure}

  \vspace{2mm}
  
  \begin{subfigure}{.18\textwidth} 
    \centering 
    \includegraphics[height=23mm,angle=90]{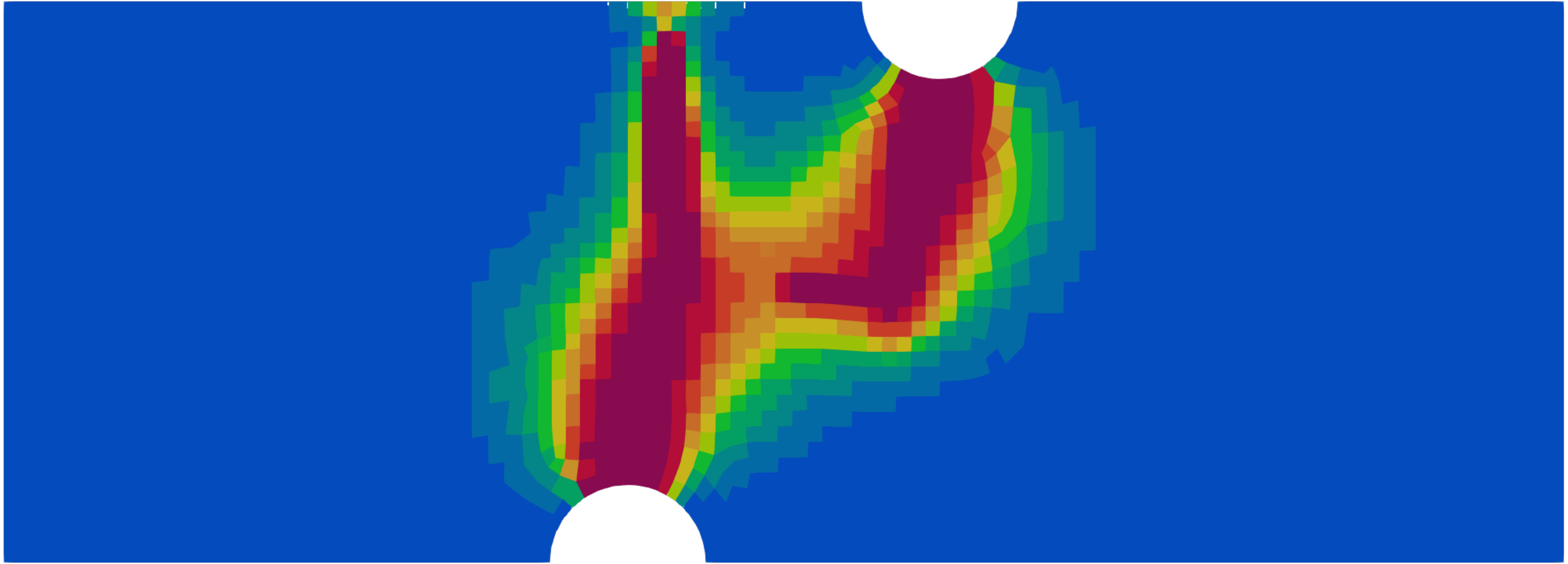}
  \end{subfigure}%
  \begin{subfigure}{.18\textwidth} 
    \centering 
    \includegraphics[height=23mm,angle=90]{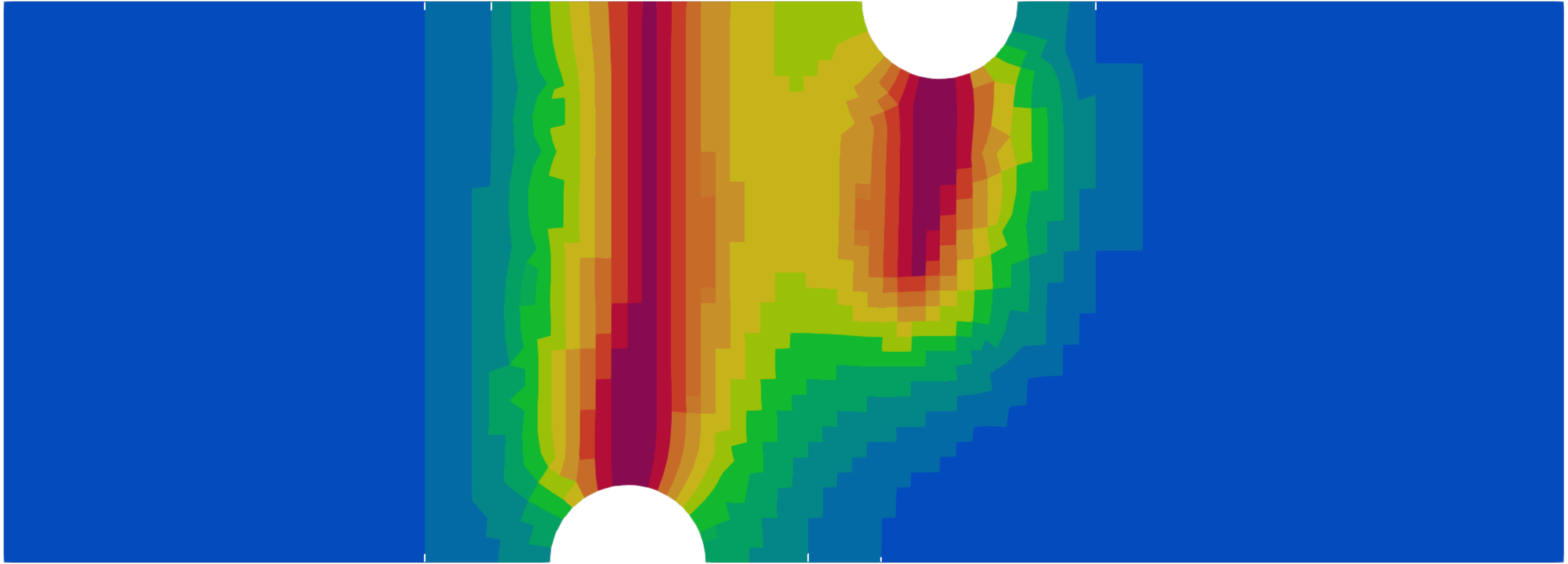}
  \end{subfigure}%
  \begin{subfigure}{.18\textwidth} 
    \centering 
    \includegraphics[height=23mm,angle=90]{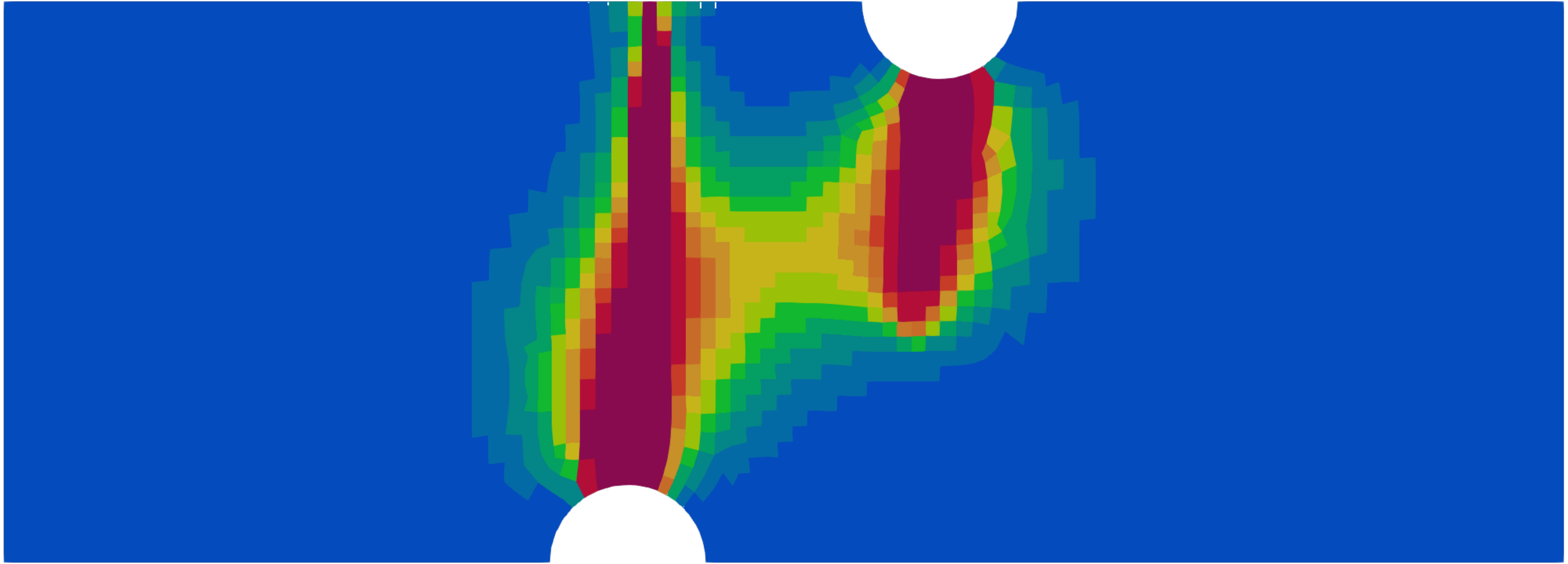}
  \end{subfigure}%
  \begin{subfigure}{.18\textwidth} 
    \centering 
    \includegraphics[height=23mm,angle=90]{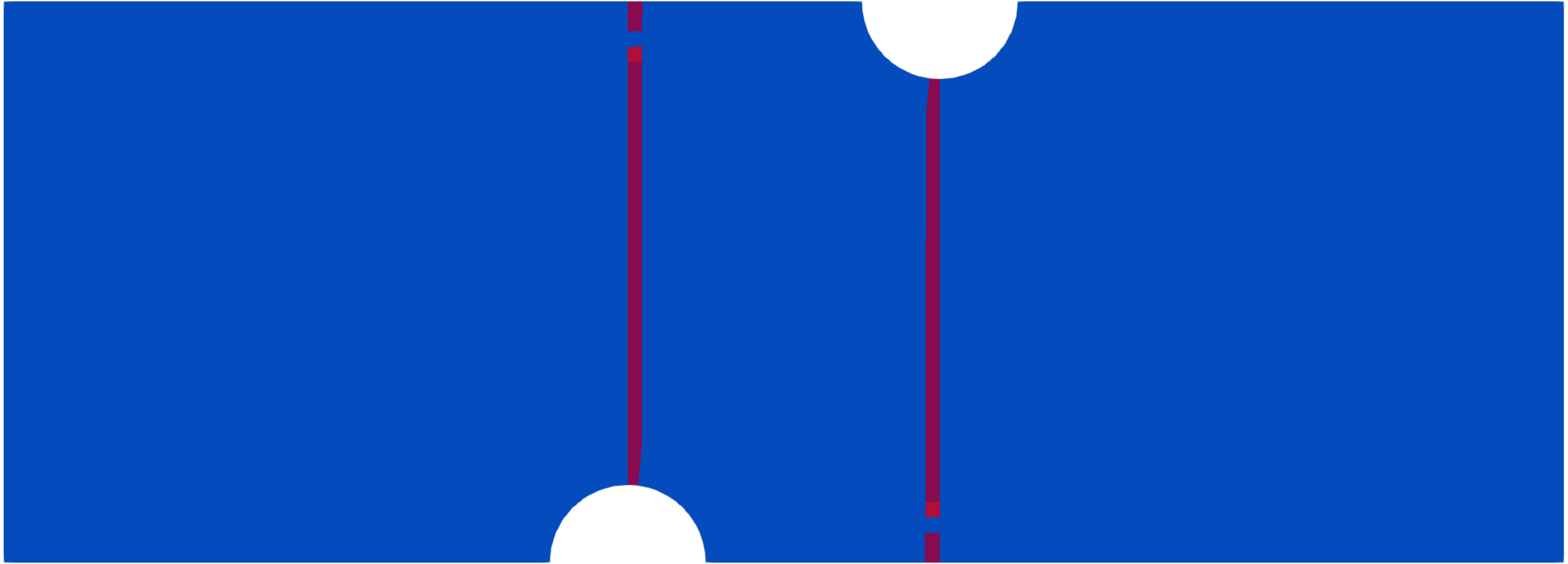}
  \end{subfigure}%
  \begin{subfigure}{.18\textwidth} 
    \centering 
    \includegraphics[height=23mm,angle=90]{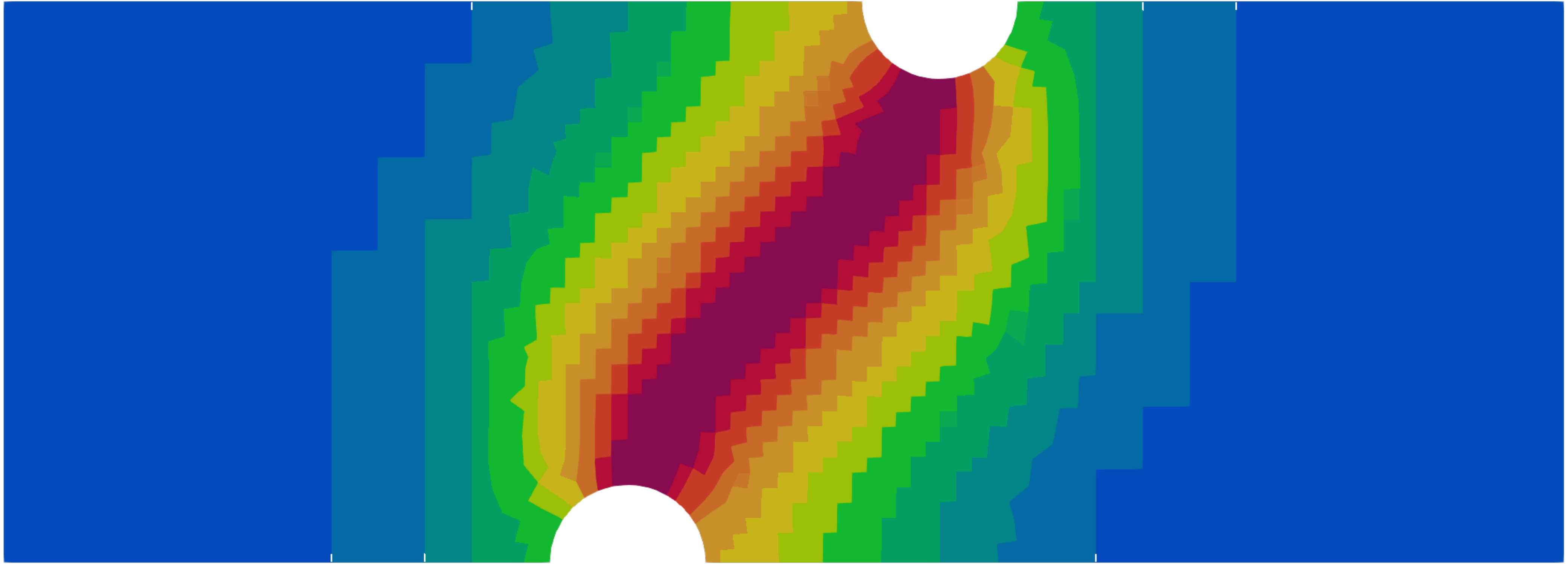}
  \end{subfigure}%
  \begin{subfigure}{.08\textwidth} 
    \centering 
    \begin{tikzpicture}
      \node[inner sep=0pt] (pic) at (0,0) {\includegraphics[height=40mm, width=5mm]
      {03_Contour/00_Color_Maps/Damage_Step_Vertical.pdf}};
      \node[inner sep=0pt] (0)   at ($(pic.south)+( 0.50, 0.15)$)  {$0$};
      \node[inner sep=0pt] (1)   at ($(pic.south)+( 0.50, 3.80)$)  {$1$};
      \node[inner sep=0pt] (d)   at ($(pic.south)+( 0.00, 4.35)$)  {$D_{yy}~\si{[-]}$};
    \end{tikzpicture} 
  \end{subfigure}
  
  \vspace{1mm}
  
  \begin{subfigure}{.18\textwidth} 
    \centering 
    \caption{$\bar{d}_1$ active}
    \label{fig:p4_AnotchedDstudyAi_u20_d1bar}
  \end{subfigure}%
  \begin{subfigure}{.18\textwidth} 
    \centering 
    \caption{$\bar{d}_2$ active}
    \label{fig:p4_AnotchedDstudyAi_u20_d2bar}
  \end{subfigure}%
  \begin{subfigure}{.18\textwidth} 
    \centering 
    \caption{$\bar{d}_3$ active}
    \label{fig:p4_AnotchedDstudyAi_u20_d3bar}
  \end{subfigure}%
  \begin{subfigure}{.18\textwidth} 
    \centering 
    \caption{$\bar{d}_4$ active}
    \label{fig:p4_AnotchedDstudyAi_u20_d4bar}
  \end{subfigure}%
  \begin{subfigure}{.18\textwidth} 
    \centering 
    \caption{reference}
    \label{fig:p4_AnotchedDstudyAi_u20_ref}
  \end{subfigure}%
  \begin{subfigure}{.08\textwidth} 
    \hphantom{Dxx}
  \end{subfigure}

  \caption{Damage contour plots for the damage model with regularization of single components of the damage tensor using model~A for the asymmetrically notched specimen (1624 elements) at $u/l \times \text{10}^\text{2}=2.0~\text{[--]}$ in Fig.~\ref{fig:p4_ANotchedFuStudyAi}. For this study, the damage variables are averaged over all integration points per element.}
  \label{fig:p4_AnotchedDstudyAi_u20}     
\end{figure}

The damage contour plots at $u/l \times \text{10}^\text{2}=0.9~\text{[--]}$ and $u/l \times \text{10}^\text{2}=2.0~\text{[--]}$ are provided in Figs.~\ref{fig:p4_AnotchedDstudyAi_u09} and \ref{fig:p4_AnotchedDstudyAi_u20}. According to the force-displacement curve of the reference solution in Fig.~\ref{fig:p4_ANotchedFuA}, an undamaged state exists at $u/l \times \text{10}^\text{2}=0.9~\text{[--]}$ and the completely damage state at $u/l \times \text{10}^\text{2}=2.0~\text{[--]}$. The regularization of single damage tensor components influences the specific component's evolution significantly, but overall fails to obtain the reference solution. If e.g.~$\bar{d}_1$ is active, it diminishes the damage initiation of $D_{xx}$ whereas $D_{yy}$ already shows signs of localization at the beginning of the failure process (Fig.~\ref{fig:p4_AnotchedDstudyAi_u09_d1bar}). Upon further loading, $D_{xx}$ still displays a diffuse damage zone at the edges of the crack, but eventually the crack localizes and propagates horizontally instead of slantingly (Fig.~\ref{fig:p4_AnotchedDstudyAi_u20_d1bar}). The evolution of $D_{yy}$ at further loading yields the same crack pattern, but exhibits a thicker fully damaged zone with little diffusive character at its edges. The activation of $\bar{d}_2$ instead of $\bar{d}_1$ yields a converse behavior of $D_{xx}$ and $D_{yy}$ (Figs.~\ref{fig:p4_AnotchedDstudyAi_u09_d2bar} and \ref{fig:p4_AnotchedDstudyAi_u20_d2bar}). The activation of $\bar{d}_3$ with a regularization of $D_{zz}$ yields a pronounced concentration behavior for $D_{xx}$ and $D_{yy}$ (Figs.~\ref{fig:p4_AnotchedDstudyAi_u09_d3bar} and \ref{fig:p4_AnotchedDstudyAi_u20_d3bar}) analogously to the previously unregularized quantities ($D_{yy}$ in Figs.~\ref{fig:p4_AnotchedDstudyAi_u09_d1bar} and \ref{fig:p4_AnotchedDstudyAi_u20_d1bar} and $D_{xx}$ in Figs.~\ref{fig:p4_AnotchedDstudyAi_u09_d2bar} and \ref{fig:p4_AnotchedDstudyAi_u20_d2bar}). The activation of $\bar{d}_4$, $\bar{d}_5$, and $\bar{d}_6$ with a regularization of $D_{xy}$, $D_{xz}$, and $D_{yz}$ yields a localization of the normal components $D_{xx}$ and $D_{yy}$ into a single row of elements (Figs.~\ref{fig:p4_AnotchedDstudyAi_u09_d4bar} and \ref{fig:p4_AnotchedDstudyAi_u20_d4bar}).

\subsection{Rotor blade}
\label{sec:p4_RB}

After confirming the accuracy of the reduced volumetric-deviatoric regularization of model~C in Section \ref{sec:p4_ANotched}, it is applied for the three-dimensional simulation of a rotor blade specimen. This aims at studying the performance of the universal anisotropic damage framework, here specified for a Neo-Hookean material regularized by model~C, to predict damage evolution on a complex structural example. The number of nodes increases significantly in three-dimensional simulations and, thus, only model~C is employed for the simulation with two nonlocal degrees of freedom.

\begin{figure}
  \centering 
  \begin{subfigure}{.5\textwidth} 
    \includegraphics[width=0.48\textwidth]{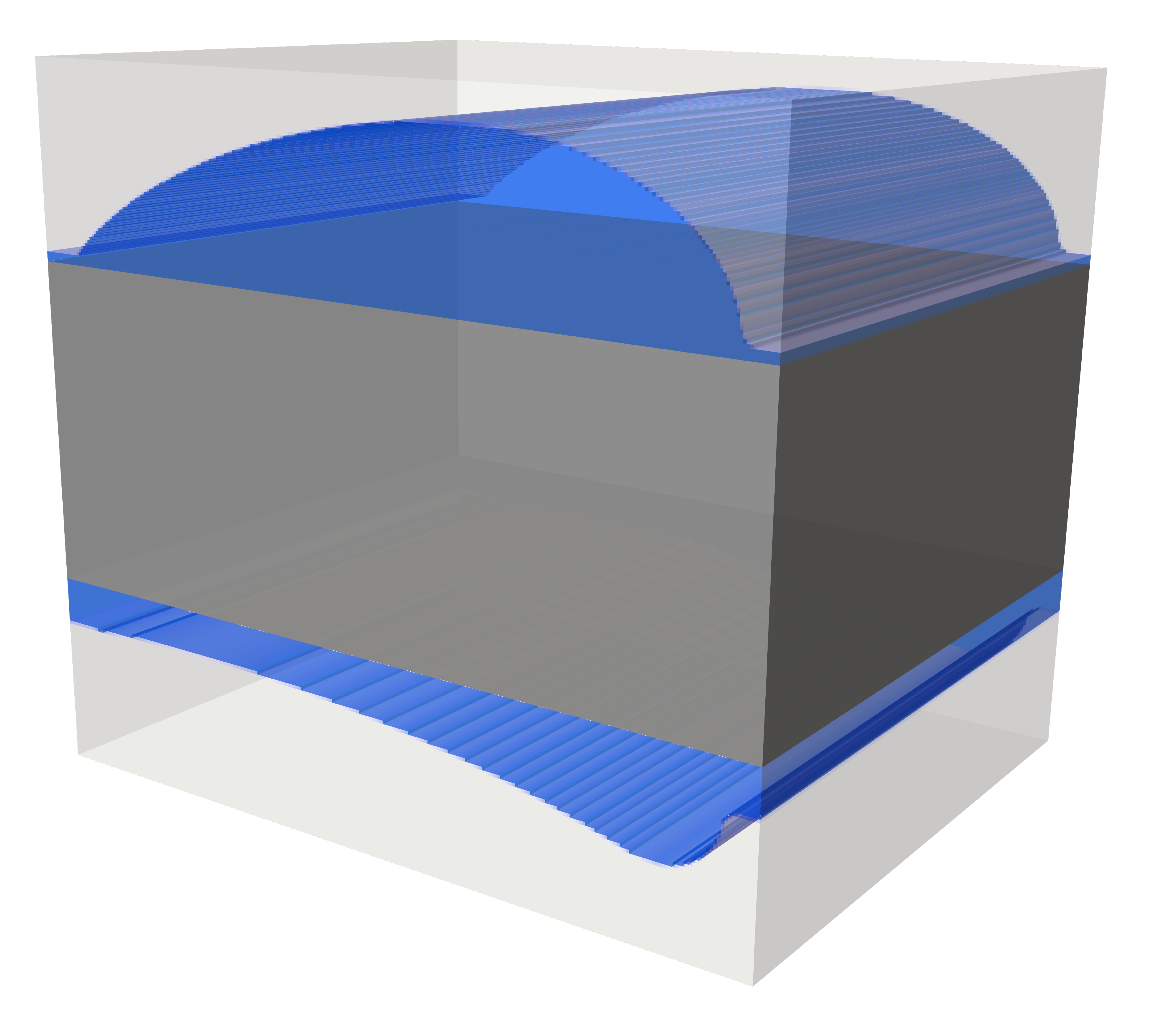}%
    \,\,\,%
    \includegraphics[width=0.48\textwidth]{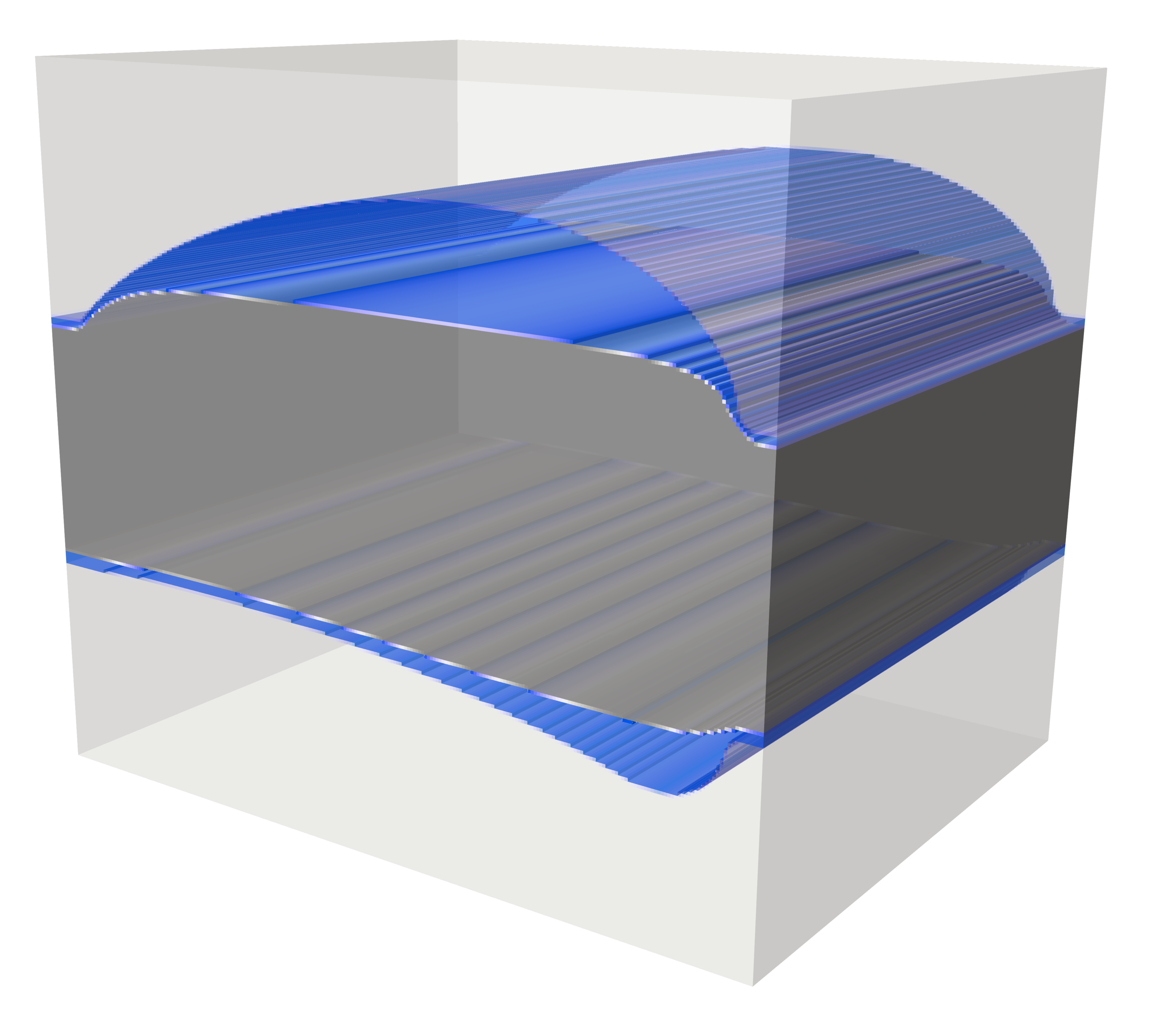}
    \includegraphics[width=0.48\textwidth]{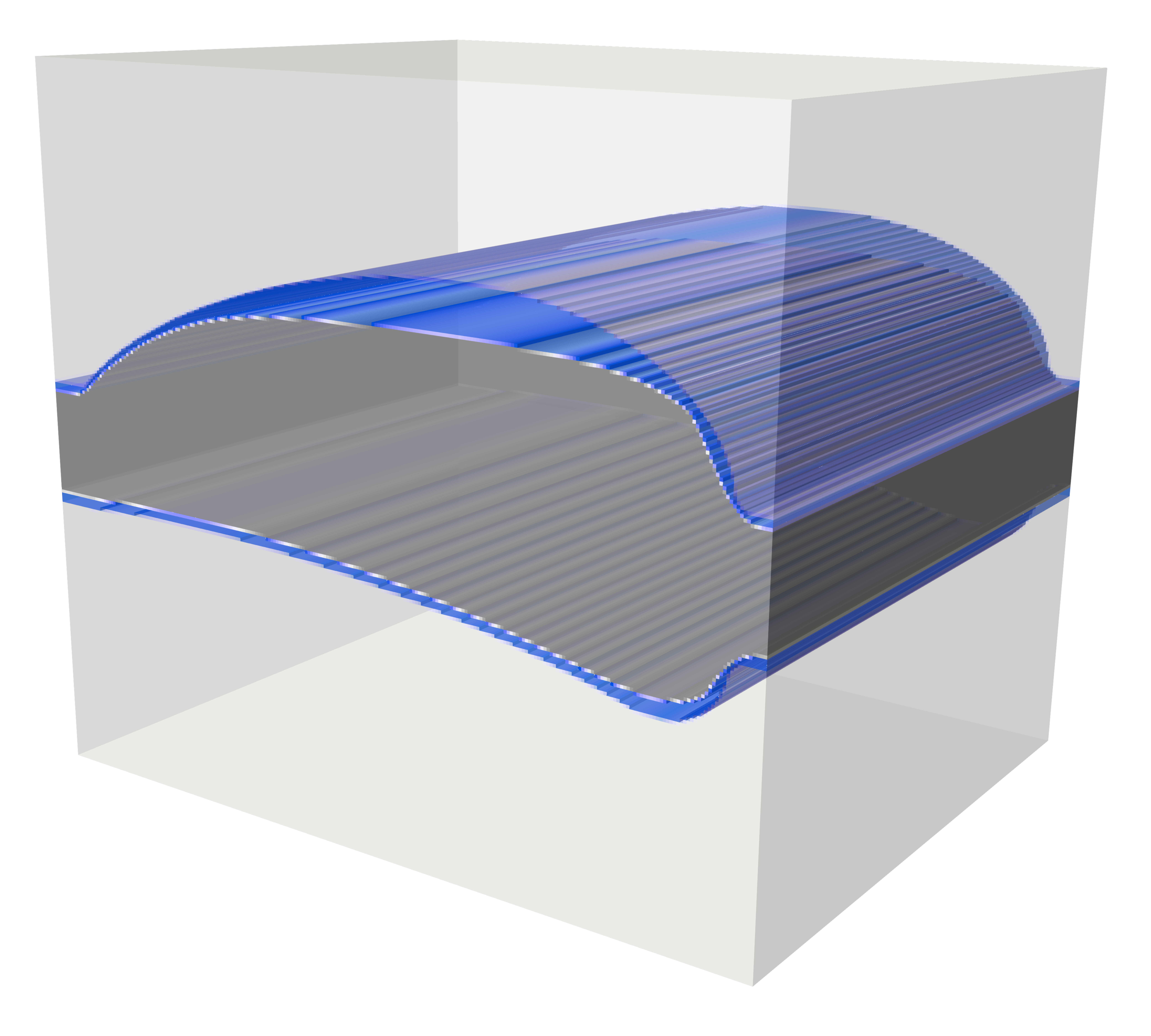}%
    \,\,\,%
    \includegraphics[width=0.48\textwidth]{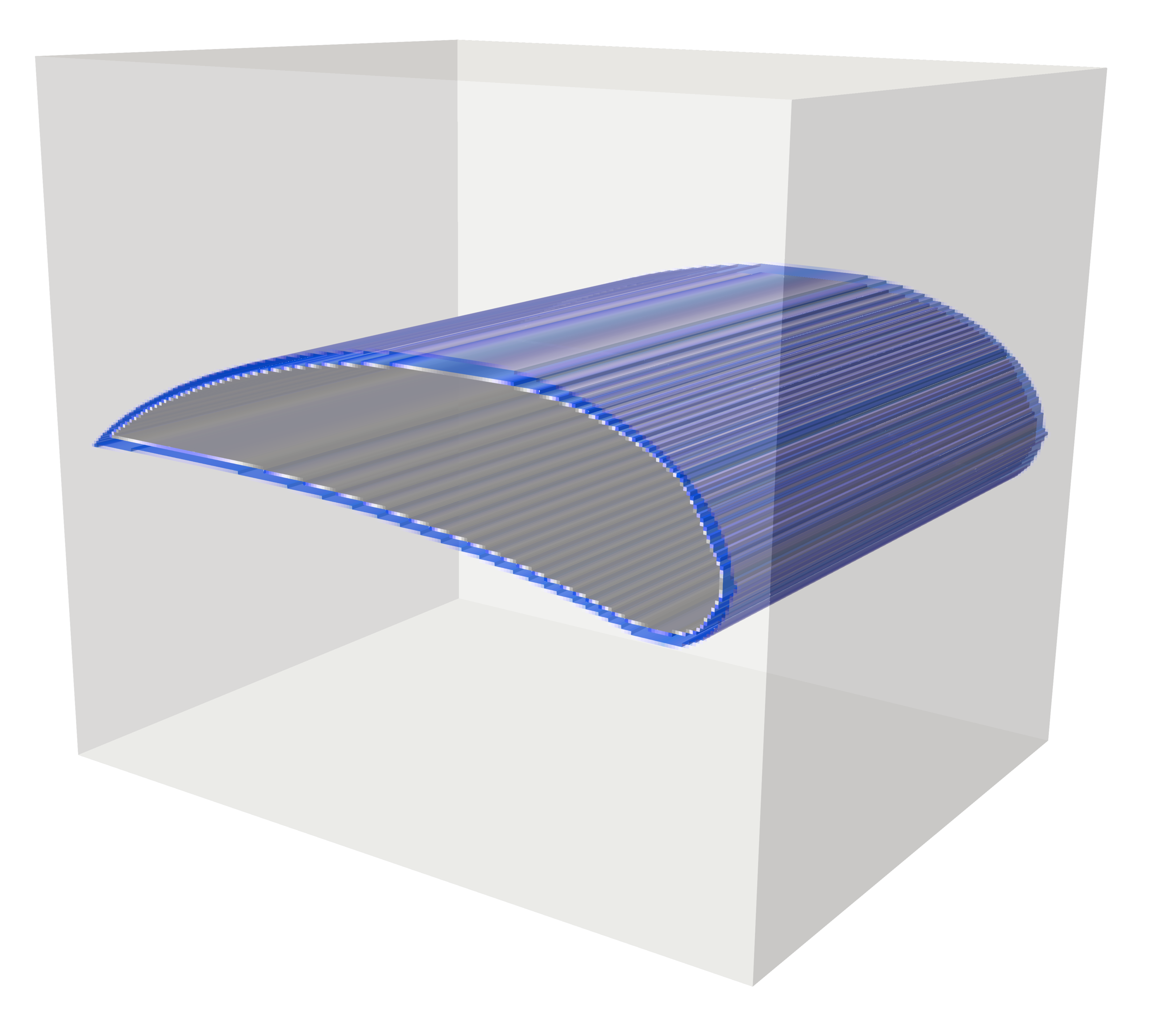}
    \caption{blade production (\cite{vanderVeldenRitzertEtAl2023})}
    \label{fig:p4_RB_motivation_production}
  \end{subfigure}%
  \begin{subfigure}{.5\textwidth} 
    \centering
    \includegraphics[width=0.85\textwidth]{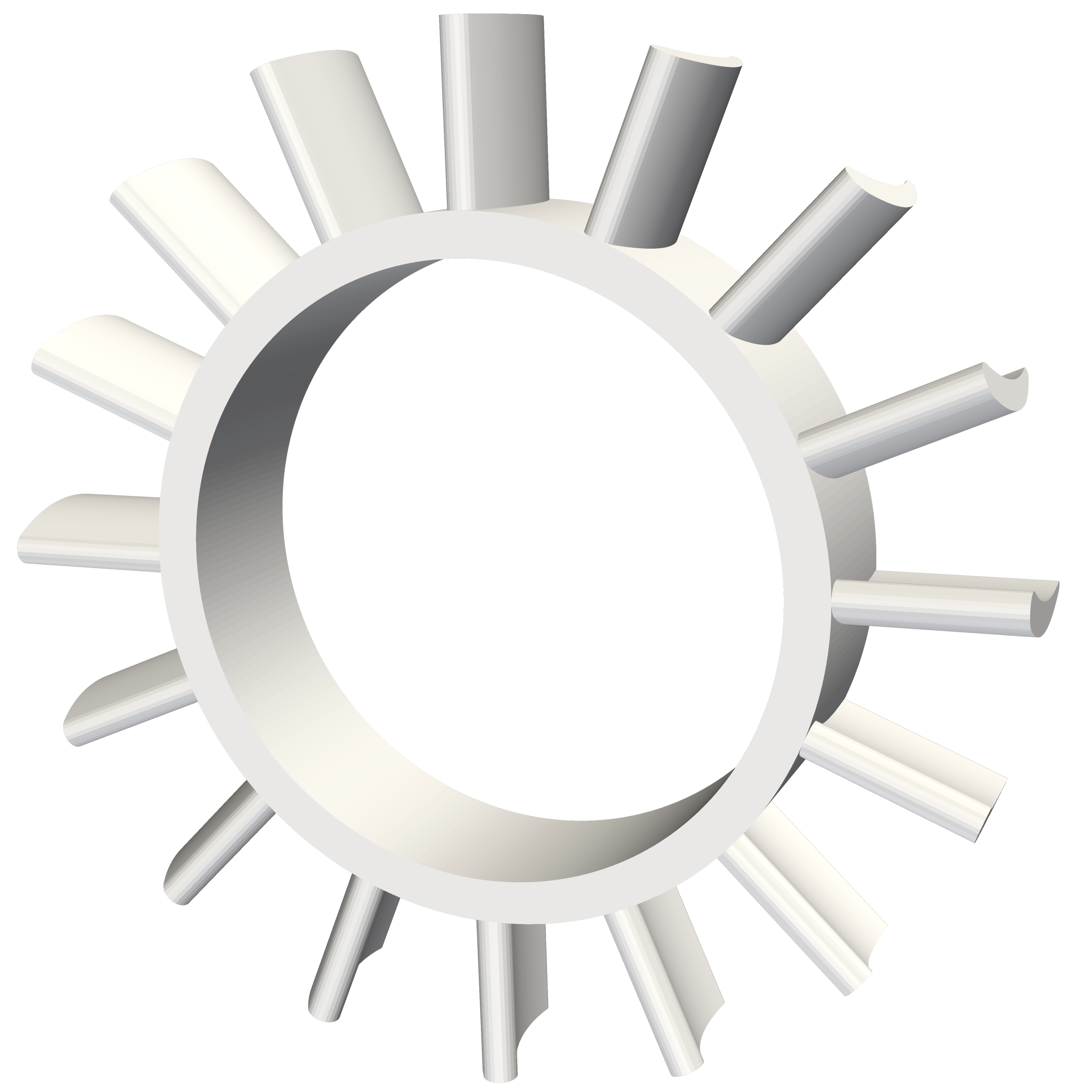}
    \caption{rotor}
    \label{fig:p4_RB_motivation_rotor}
  \end{subfigure}%
  \caption{Motivation for the rotor blade specimen. In Fig.~\ref{fig:p4_RB_motivation_production}, a single blade is manufactured by electrochemical machining from a solid metal workpiece (dark gray). The tool (light gray) defines the shape and the electrolyte (blue) serves as an electrical conductor for a direct current.}
  \label{fig:p4_motivation}
\end{figure}

The geometry is inspired by previous works in \cite{vanderVeldenRitzertEtAl2023}, where the electrochemical machining process is simulated for the manufacturing of a rotor blade (Fig.~\ref{fig:p4_RB_motivation_production}) that can be assembled to an entire rotor (Fig.~\ref{fig:p4_RB_motivation_rotor}).

\begin{figure}
    \centering 
    \beginpgfgraphicnamed{p4_ANgeombvp}    
    \begin{tikzpicture}
      \node[inner sep=0pt] (pic) at (0,0) {\includegraphics[width=.75\textwidth]{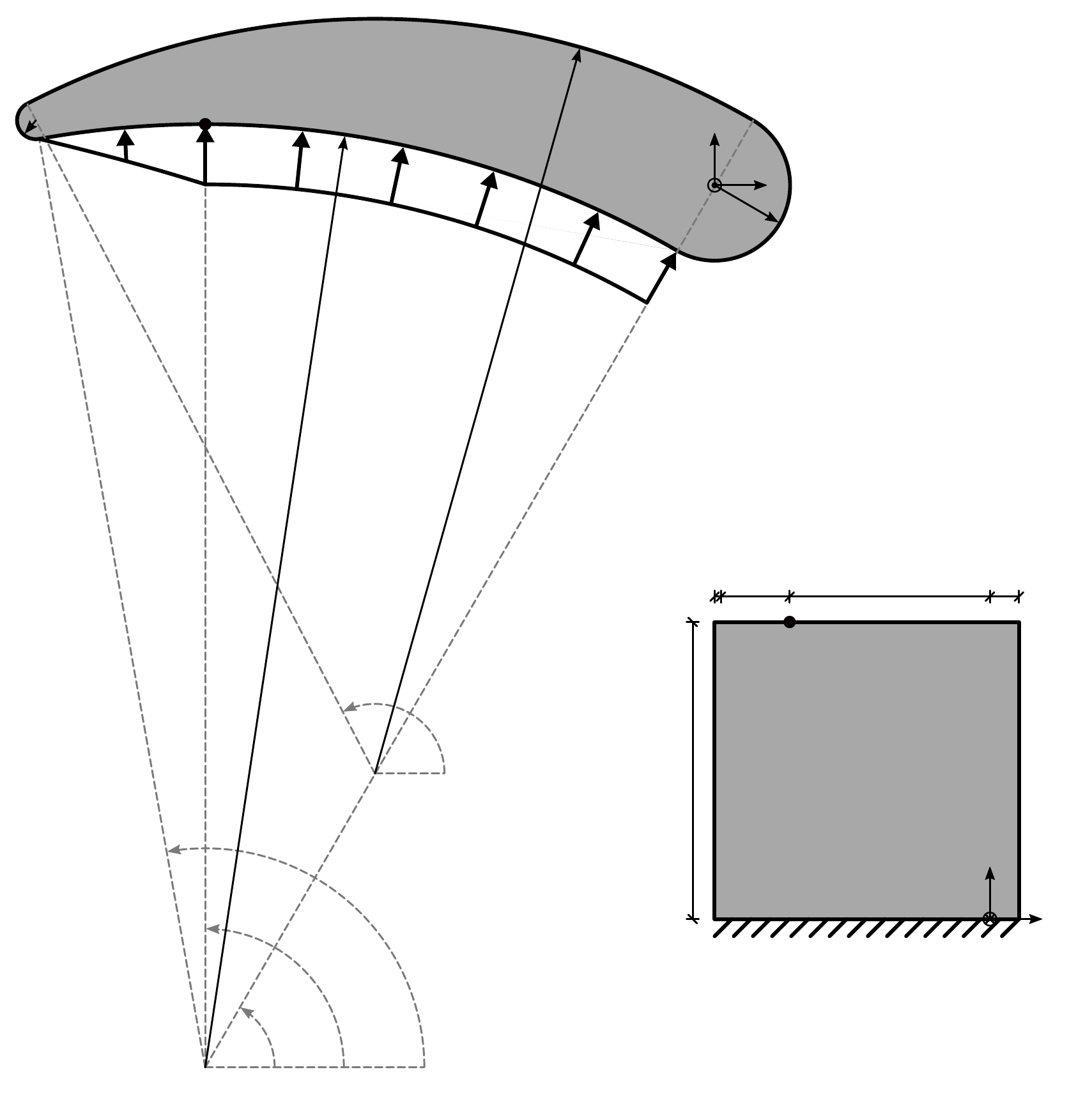}};
      \node[inner sep=0pt] at ($(pic.south) +(+2.35,10.25)$)  {$x$};
      \node[inner sep=0pt] at ($(pic.south) +(+1.60,10.50)$)  {$y$};
      \node[inner sep=0pt] at ($(pic.south) +(+1.60,10.00)$)  {$z$};
      \node[inner sep=0pt] at ($(pic.south) +(+5.55,02.15)$)  {$x$};
      \node[inner sep=0pt] at ($(pic.south) +(+4.65,02.12)$)  {$y$};
      \node[inner sep=0pt] at ($(pic.south) +(+4.65,02.54)$)  {$z$};
      \node[inner sep=0pt] at ($(pic.south) +(-3.75,10.95)$)  {$A$};
      \node[inner sep=0pt] at ($(pic.south) +(+2.65,04.90)$)  {$A$};
      \node[inner sep=0pt] at ($(pic.south) +(+2.85,09.35)$)  {$r_1$};
      \node[inner sep=0pt] at ($(pic.south) +(+0.50,11.75)$)  {$r_2$};
      \node[inner sep=0pt] at ($(pic.south) +(-2.10,10.80)$)  {$r_3$};
      \node[inner sep=0pt] at ($(pic.south) +(-5.80,10.30)$)  {$r_4$};
      \node[inner sep=0pt] at ($(pic.south) +(+1.40,08.60)$)  {$p$};
      \node[inner sep=0pt] at ($(pic.south) +(-2.30,03.60)$)  {$P_1$};
      \node[inner sep=0pt] at ($(pic.south) +(-4.10,00.30)$)  {$P_2$};
      \node[inner sep=0pt] at ($(pic.south) +(-2.90,01.00)$)  {$\varphi_1$};
      \node[inner sep=0pt] at ($(pic.south) +(-3.15,02.00)$)  {$\varphi_2$};
      \node[inner sep=0pt] at ($(pic.south) +(-3.97,03.00)$)  {$\varphi_3$};
      \node[inner sep=0pt] at ($(pic.south) +(-1.95,04.55)$)  {$\varphi_4$};
      \node[inner sep=0pt] at ($(pic.south) +(+1.40,03.50)$)  {$l$};
      \node[inner sep=0pt] at ($(pic.south) +(+5.05,05.70)$)  {$r_1$};
      \node[inner sep=0pt] at ($(pic.south) +(+1.90,05.70)$)  {$r_4$};
      \node[inner sep=0pt] at ($(pic.south) +(+3.85,05.70)$)  {$w_1$};
      \node[inner sep=0pt] at ($(pic.south) +(+2.35,05.70)$)  {$w_2$};
    \end{tikzpicture} 
    \endpgfgraphicnamed
    \caption{Geometry and boundary value problem for the rotor blade specimen. Side view ($x$-$y$) and bottom view ($x$-$z$).}
    \label{fig:p4_RBgeombvp}
\end{figure}

Fig.~\ref{fig:p4_RBgeombvp} provides the side and the bottom view of the geometry and boundary value problem where the geometrical dimensions, angles, and positions read $l = 20~[\si{\mm}]$, $r_1 = 2~[\si{\mm}]$, $r_2 = 20~[\si{\mm}]$, $r_3 = 25~[\si{\mm}]$, $r_4 = 0.5~[\si{\mm}]$, $w_1 = 13.5~[\si{\mm}]$, $w_2 = 4.882~[\si{\mm}]$, $\varphi_1 = 60~[\si{\degree}]$, $\varphi_2 = 90~[\si{\degree}]$, $\varphi_3 = 100.137~[\si{\degree}]$, $\varphi_4 = 117.444~[\si{\degree}]$, $P_1 = (-9.0~|~-15.5884~|~0.0)~[\si{\mm}]$, $P_2 = (-13.5~|~-23.3827~|~0.0)~[\si{\mm}]$. The rear edge of the rotor blade is clamped at $z = 0~[\si{\mm}]$ and the loading case of a pressure load applied from the bottom side is considered.

\begin{figure}
  \centering 
  \begin{subfigure}{.5\textwidth} 
    \includegraphics[width=\textwidth]{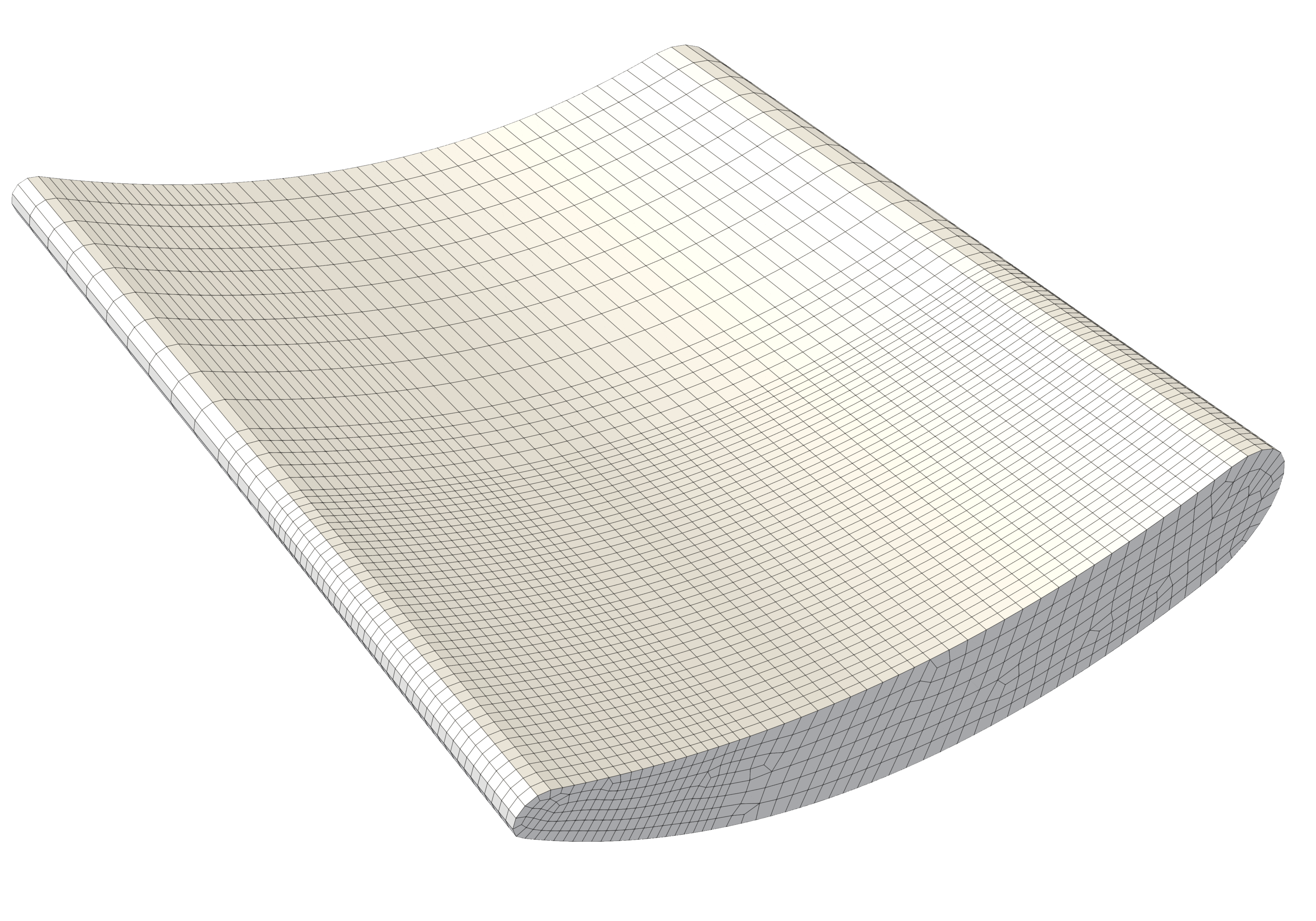}
    \caption{coarsest mesh (21560 elements)}
  \end{subfigure}%
  \begin{subfigure}{.5\textwidth} 
    \includegraphics[width=\textwidth]{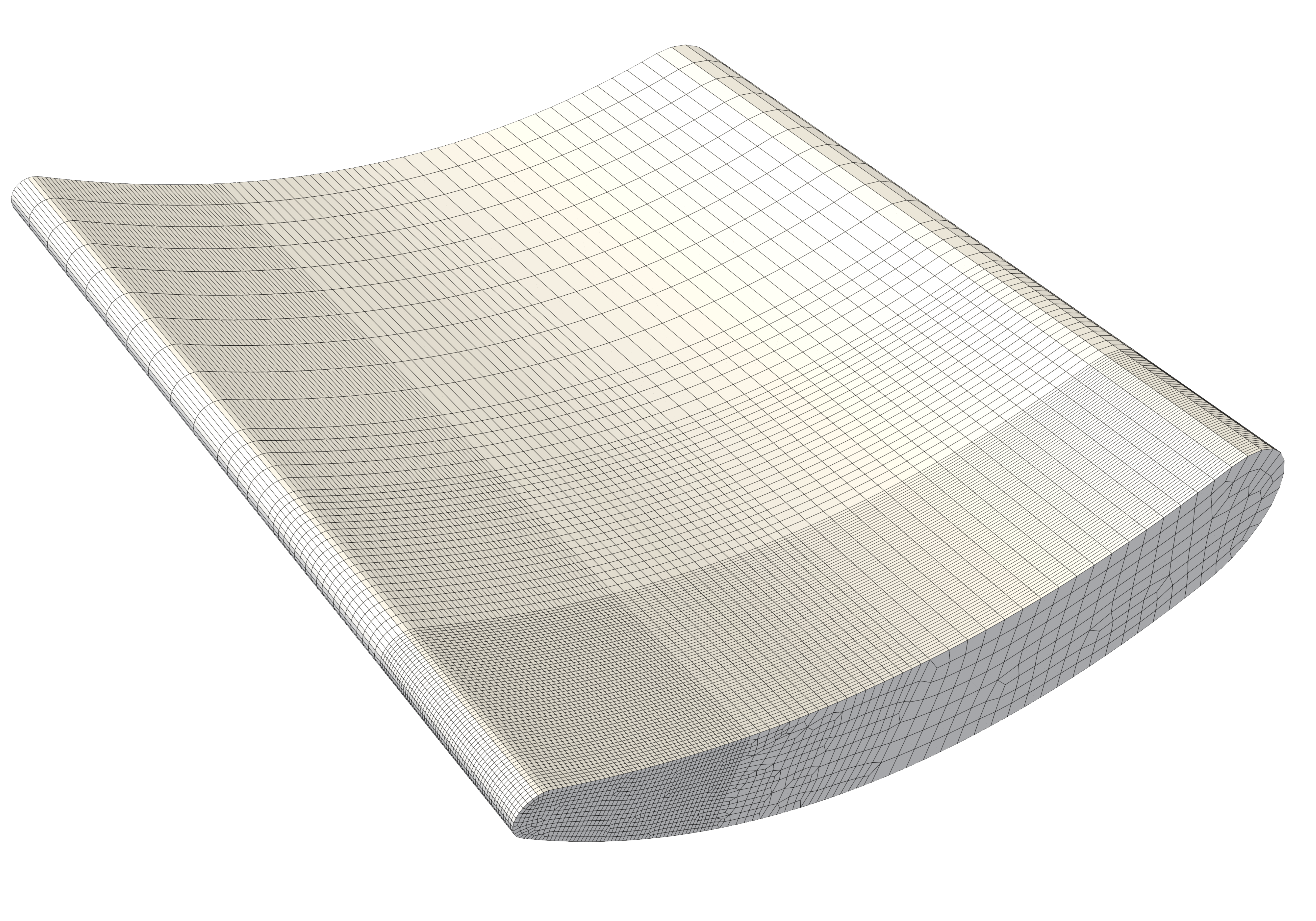}
    \caption{finest mesh (116565 elements)}
  \end{subfigure}%
  \caption{Finite element meshes for the rotor blade specimen (bottom view).}
  \label{fig:p4_RBmeshes}
\end{figure}

In Fig.~\ref{fig:p4_RBmeshes}, the finite element meshes are shown, which are refined towards the clamped edge and towards the side with the smaller radius due to results of preliminary studies that revealed damage initiation in these regions.

\begin{figure}[htbp]
  \centering
  \includegraphics{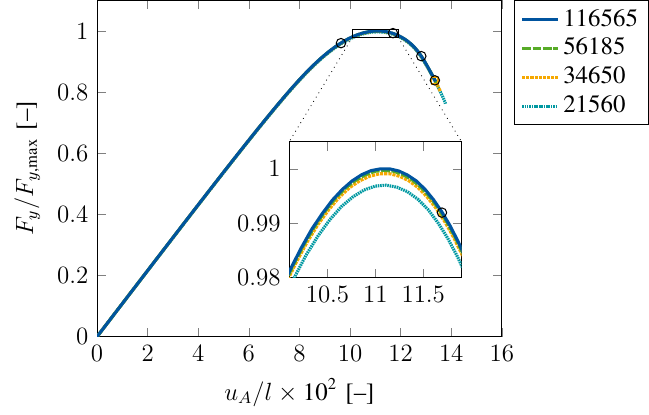}
  \caption{Force-displacement curves for the mesh convergence study of the rotor blade specimen using model~C. The forces are normalized with respect to the maximum force of the finest mesh (116565 elements) with $F_\text{max} = 1.5738 \times 10^4~[\si{\newton}]$. The orange boxes indicate the points of comparison for mesh convergence in Figs.~\ref{fig:p4_Dmeshconbottom} and \ref{fig:p4_Dmeshcontop}. The black circles indicate the points of evaluation for the damage evolution in Figs.~\ref{fig:p4_Dnormal_evolution} and \ref{fig:p4_Dshear_evolution}.}
  \label{fig:p4_RBFu}
\end{figure}

The normalized force-displacement curves in Fig.~\ref{fig:p4_RBFu} depict the sum of the forces in $y$-direction at the clamped edge over the deflection in $y$-direction of point~$A$ (see Fig.~\ref{fig:p4_RBgeombvp}) that is located at position $P_A = (-13.5~|~1.6173~|~20.0)~[\si{\mm}]$. The mesh convergence study yields a close agreement in the force-displacement curves between the results of all meshes. Moreover, the coarsest mesh with 21560~elements underestimates the structural load bearing capacity of the rotor blade that is obtained with the finest mesh with 116565~elements by just $0.30~\si{[\percent]}$.

\begin{figure}
  \centering 
  \begin{subfigure}{.45\textwidth} 
    \includegraphics[width=\textwidth]{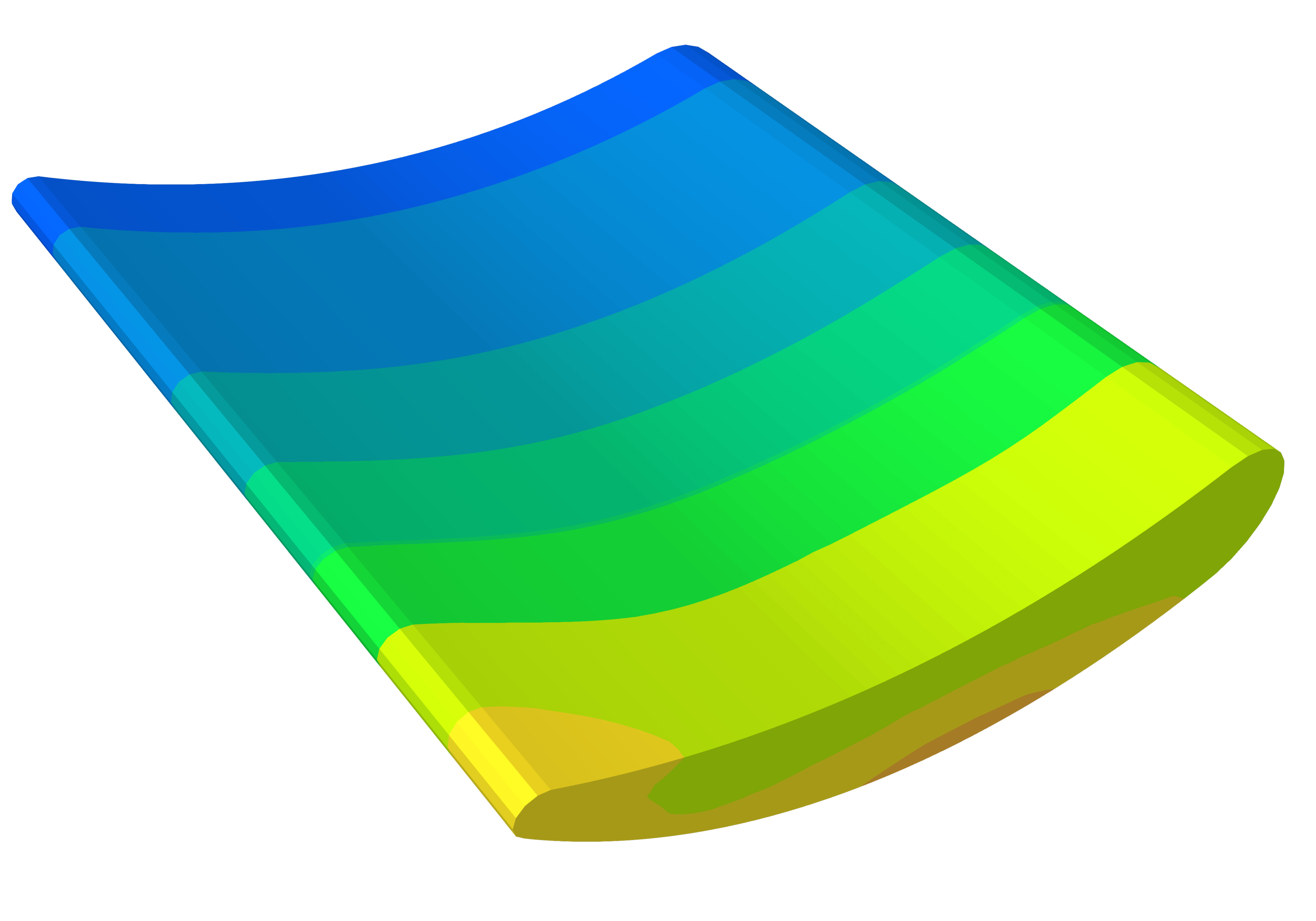}
  \end{subfigure}%
  \begin{subfigure}{.45\textwidth} 
    \includegraphics[width=\textwidth]{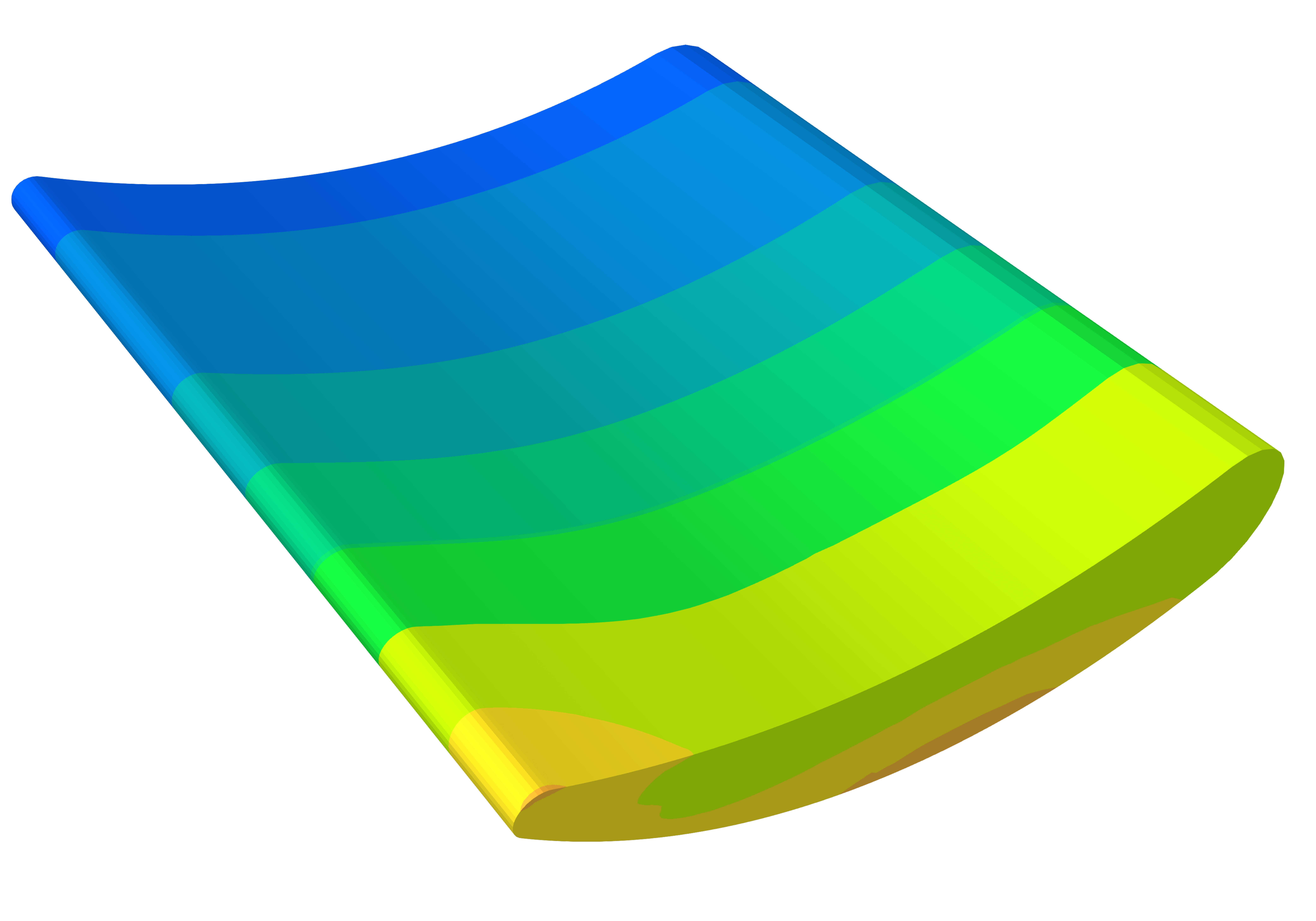}
  \end{subfigure}%
  \begin{subfigure}{.08\textwidth} 
    \centering 
    \begin{tikzpicture}
      \node[inner sep=0pt] (pic) at (0,0) {\includegraphics[height=40mm, width=5mm]
      {03_Contour/00_Color_Maps/Damage_Step_Vertical.pdf}};
      \node[inner sep=0pt] (0)   at ($(pic.south)+( 0.50, 0.15)$)  {$0$};
      \node[inner sep=0pt] (1)   at ($(pic.south)+( 0.50, 3.80)$)  {$1$};
      \node[inner sep=0pt] (d)   at ($(pic.south)+( 0.00, 4.35)$)  {$D_{xx}~\si{[-]}$};
    \end{tikzpicture} 
  \end{subfigure}
  
  \vspace{2mm}
  
  \begin{subfigure}{.45\textwidth} 
    \includegraphics[width=\textwidth]{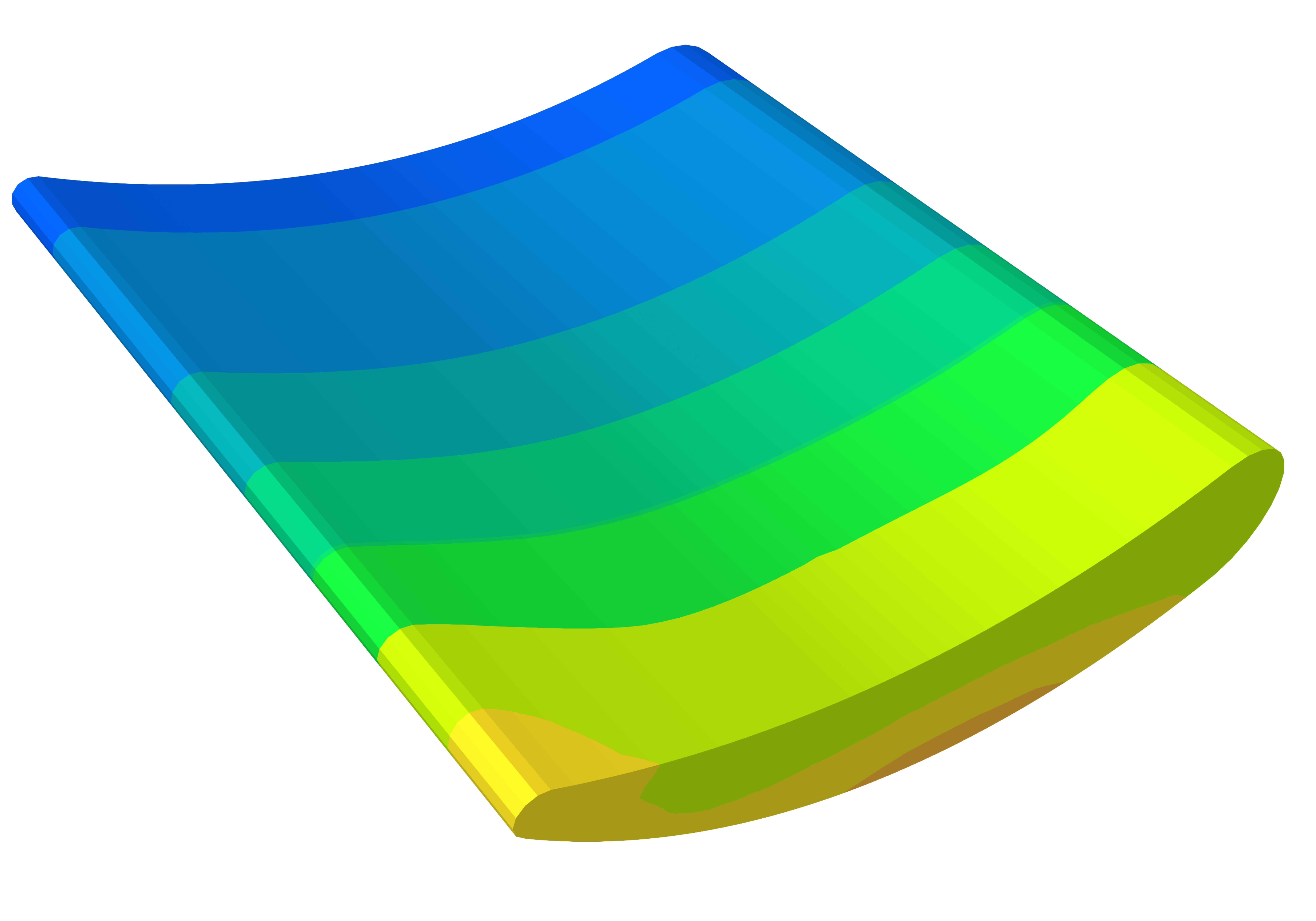}
  \end{subfigure}%
  \begin{subfigure}{.45\textwidth} 
    \includegraphics[width=\textwidth]{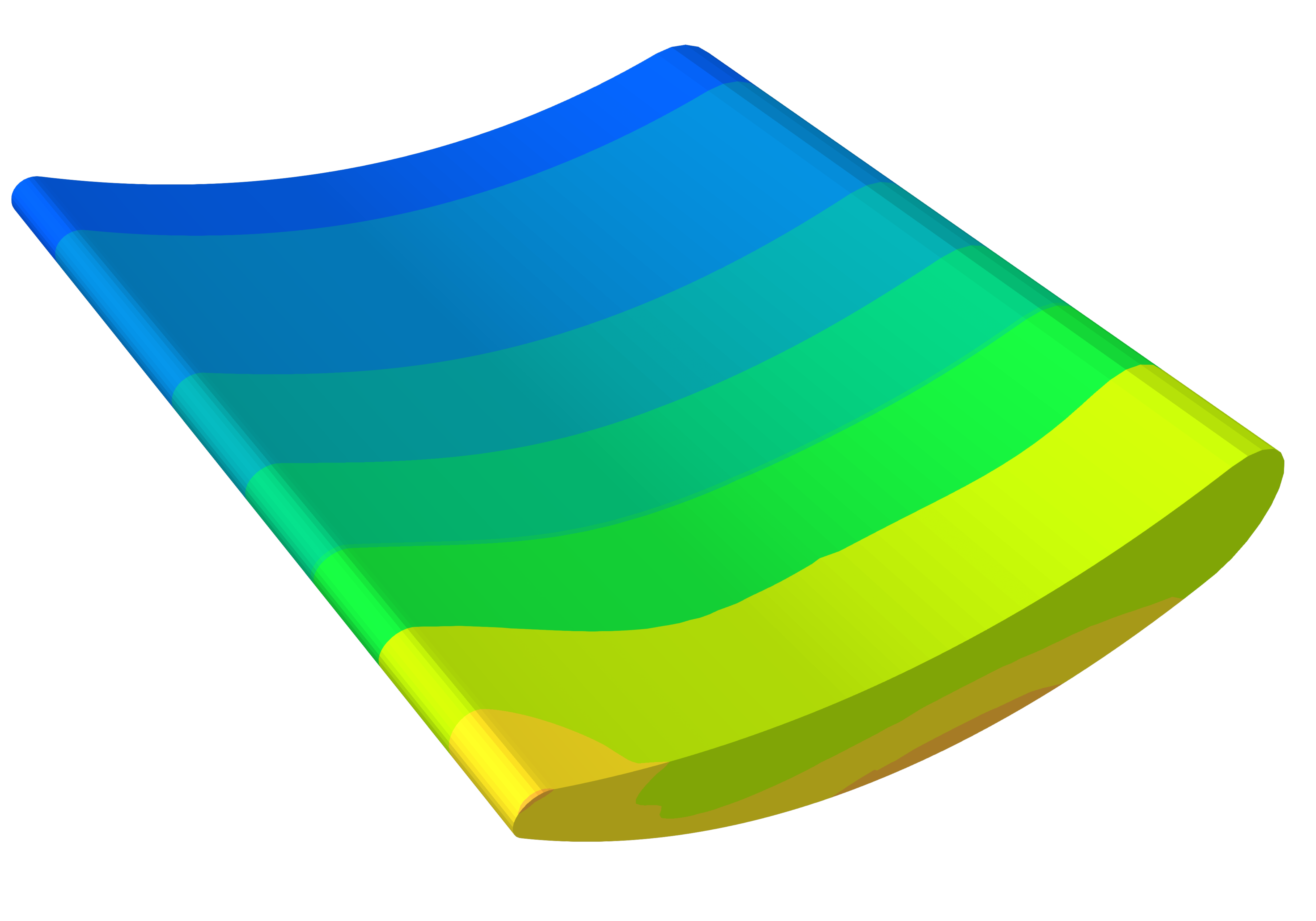}
  \end{subfigure}%
  \begin{subfigure}{.08\textwidth} 
    \centering 
    \begin{tikzpicture}
      \node[inner sep=0pt] (pic) at (0,0) {\includegraphics[height=40mm, width=5mm]
      {03_Contour/00_Color_Maps/Damage_Step_Vertical.pdf}};
      \node[inner sep=0pt] (0)   at ($(pic.south)+( 0.50, 0.15)$)  {$0$};
      \node[inner sep=0pt] (1)   at ($(pic.south)+( 0.50, 3.80)$)  {$1$};
      \node[inner sep=0pt] (d)   at ($(pic.south)+( 0.00, 4.35)$)  {$D_{yy}~\si{[-]}$};
    \end{tikzpicture} 
  \end{subfigure}
  
  \vspace{2mm}
  
  \begin{subfigure}{.45\textwidth} 
    \includegraphics[width=\textwidth]{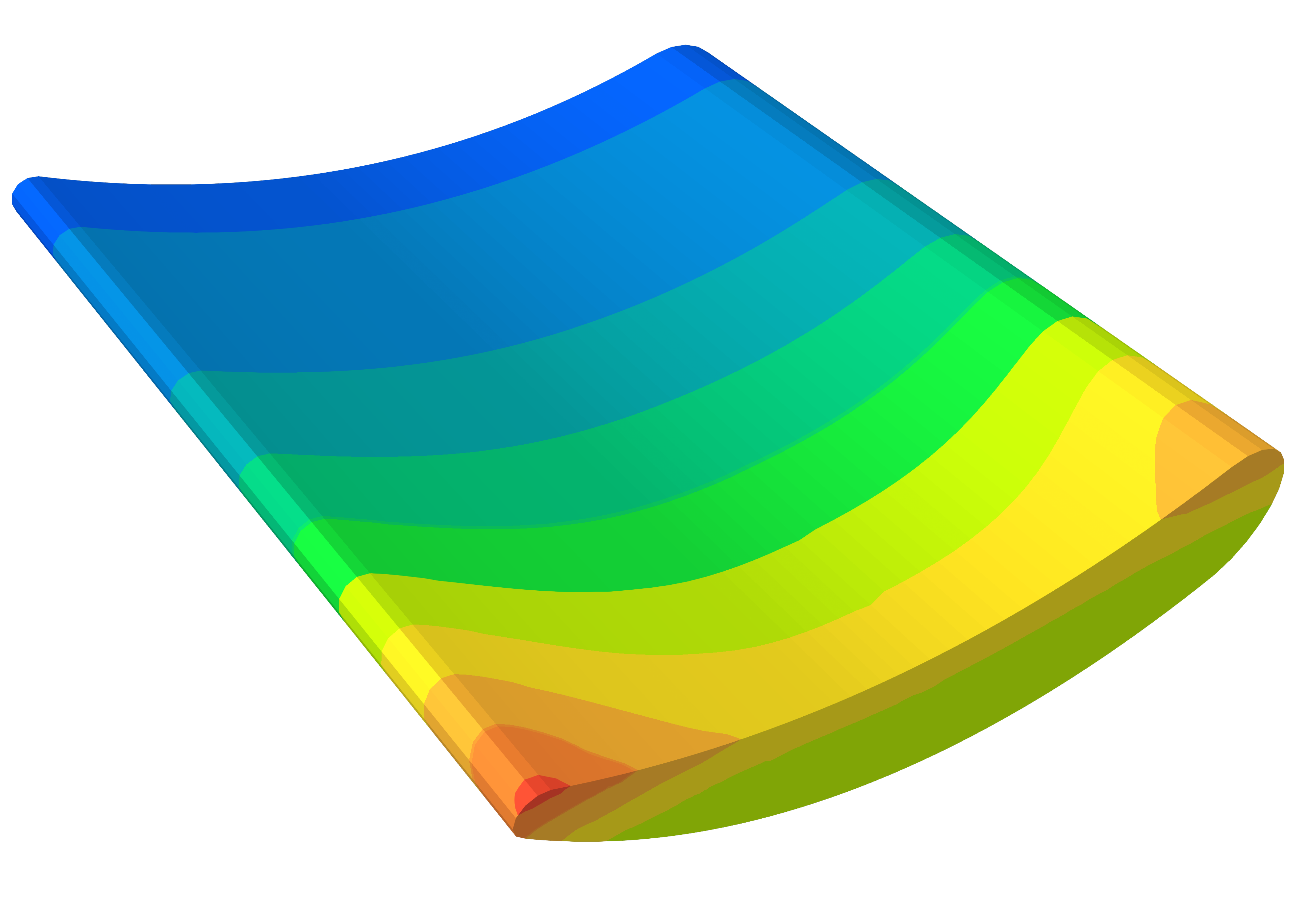}
    \caption{21560 elements}
    \label{fig:p4_Dmeshconbottom21560}
  \end{subfigure}%
  \begin{subfigure}{.45\textwidth} 
    \includegraphics[width=\textwidth]{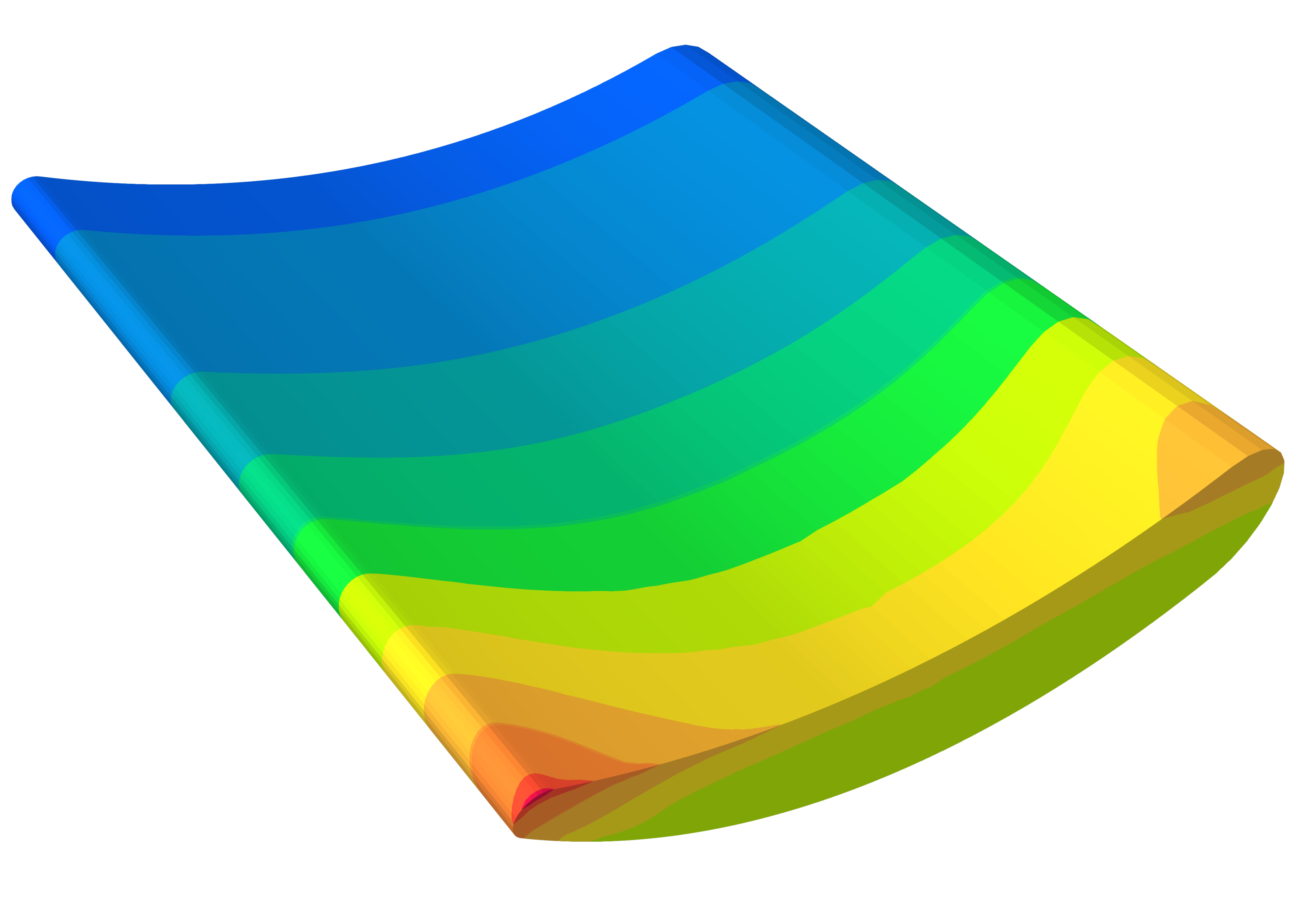}
    \caption{116565 elements}
    \label{fig:p4_Dmeshconbottom116565}
  \end{subfigure}%
  \begin{subfigure}{.08\textwidth} 
    \centering 
    \begin{tikzpicture}
      \node[inner sep=0pt] (pic) at (0,0) {\includegraphics[height=40mm, width=5mm]
      {03_Contour/00_Color_Maps/Damage_Step_Vertical.pdf}};
      \node[inner sep=0pt] (0)   at ($(pic.south)+( 0.50, 0.15)$)  {$0$};
      \node[inner sep=0pt] (1)   at ($(pic.south)+( 0.50, 3.80)$)  {$1$};
      \node[inner sep=0pt] (d)   at ($(pic.south)+( 0.00, 4.35)$)  {$D_{zz}~\si{[-]}$};
    \end{tikzpicture} 
  \end{subfigure}
  
  \caption{Contour plots for the mesh convergence of the normal components of the damage tensor for the rotor blade specimen (bottom view).}
  \label{fig:p4_Dmeshconbottom}
\end{figure}

\begin{figure}
  \centering 
  \begin{subfigure}{.45\textwidth} 
    \includegraphics[width=\textwidth]{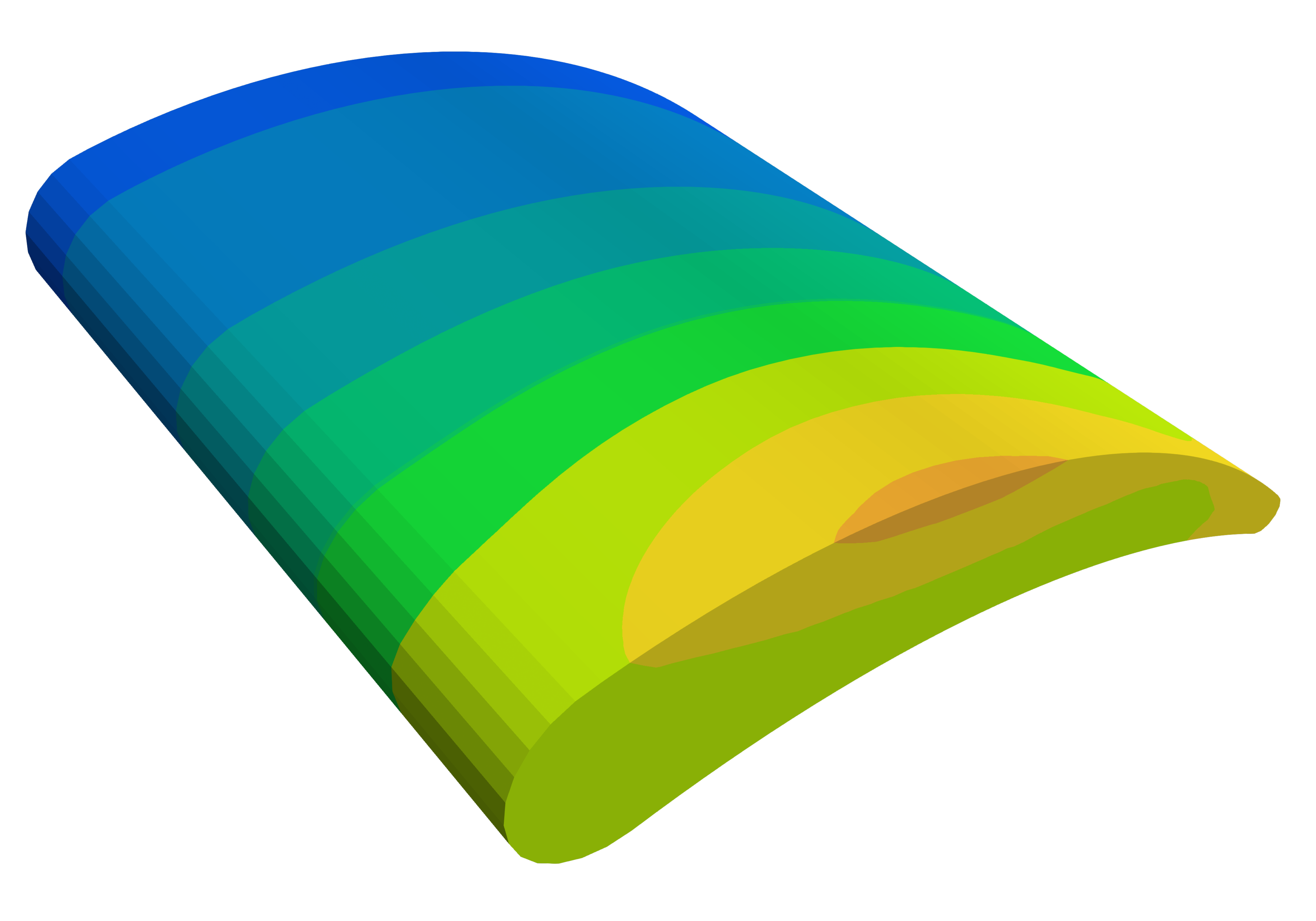}
  \end{subfigure}%
  \begin{subfigure}{.45\textwidth} 
    \includegraphics[width=\textwidth]{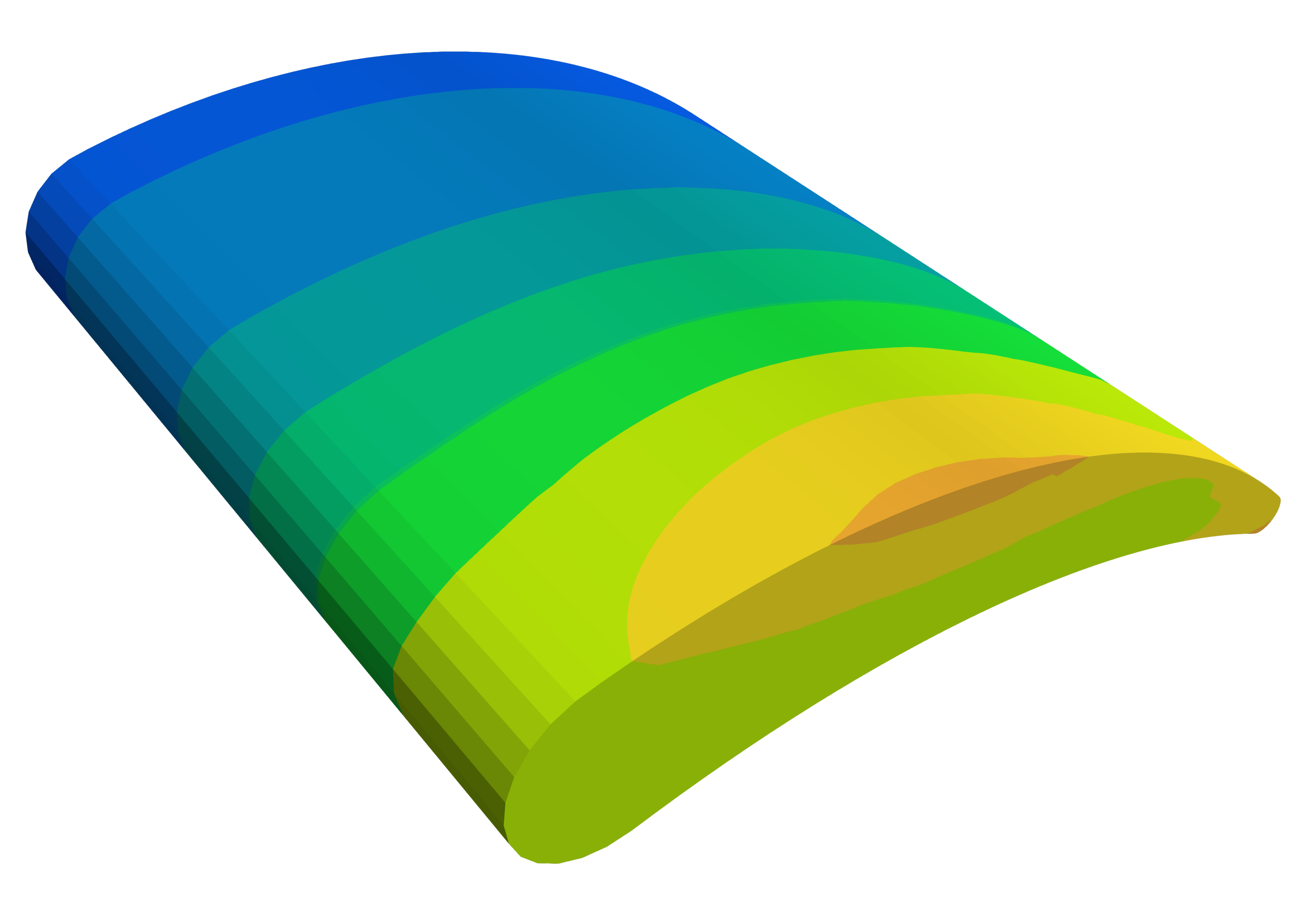}
  \end{subfigure}%
  \begin{subfigure}{.08\textwidth} 
    \centering 
    \begin{tikzpicture}
      \node[inner sep=0pt] (pic) at (0,0) {\includegraphics[height=40mm, width=5mm]
      {03_Contour/00_Color_Maps/Damage_Step_Vertical.pdf}};
      \node[inner sep=0pt] (0)   at ($(pic.south)+( 0.50, 0.15)$)  {$0$};
      \node[inner sep=0pt] (1)   at ($(pic.south)+( 0.50, 3.80)$)  {$1$};
      \node[inner sep=0pt] (d)   at ($(pic.south)+( 0.00, 4.35)$)  {$D_{xx}~\si{[-]}$};
    \end{tikzpicture} 
  \end{subfigure}
  
  \vspace{2mm}
  
  \begin{subfigure}{.45\textwidth} 
    \includegraphics[width=\textwidth]{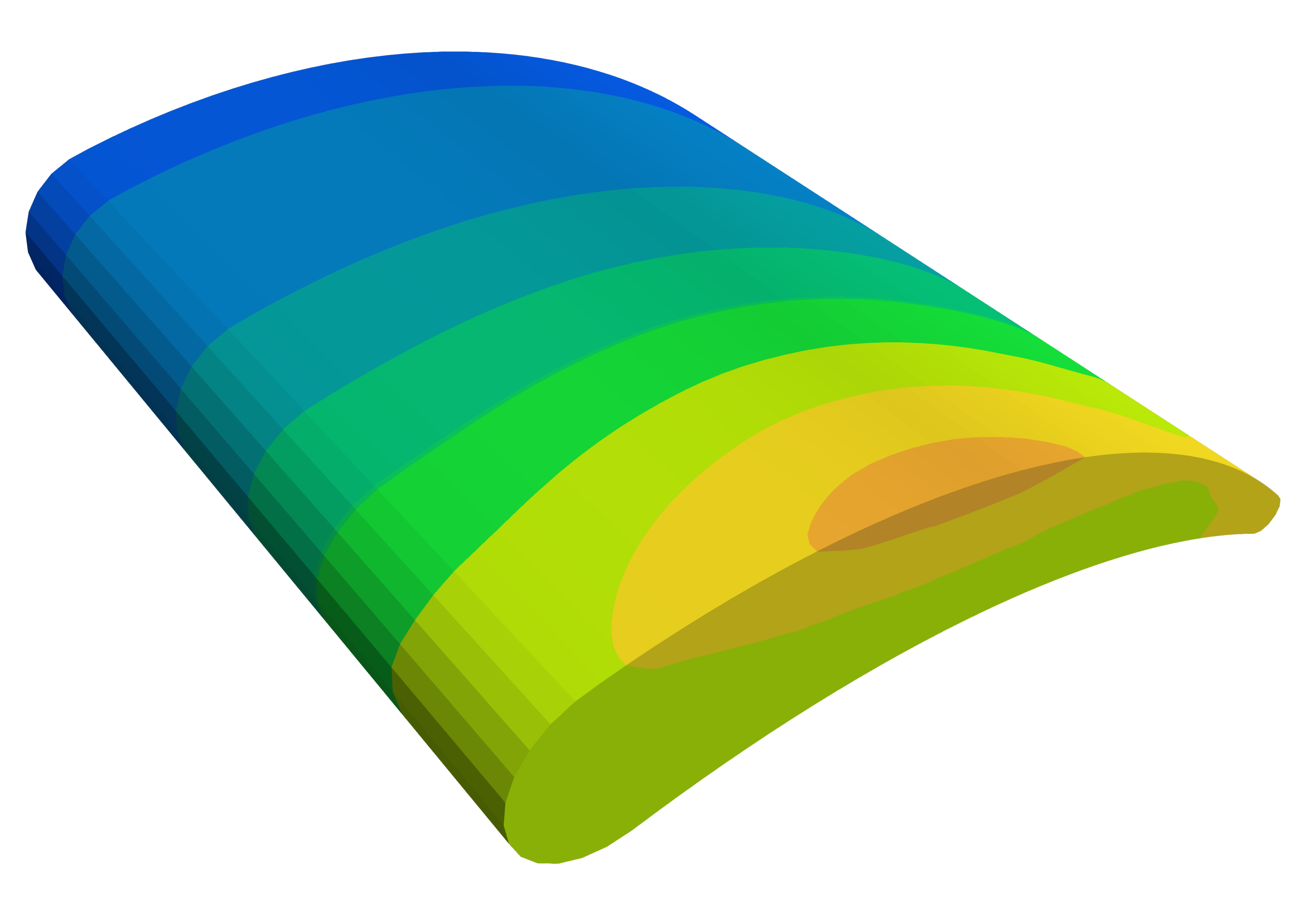}
  \end{subfigure}%
  \begin{subfigure}{.45\textwidth} 
    \includegraphics[width=\textwidth]{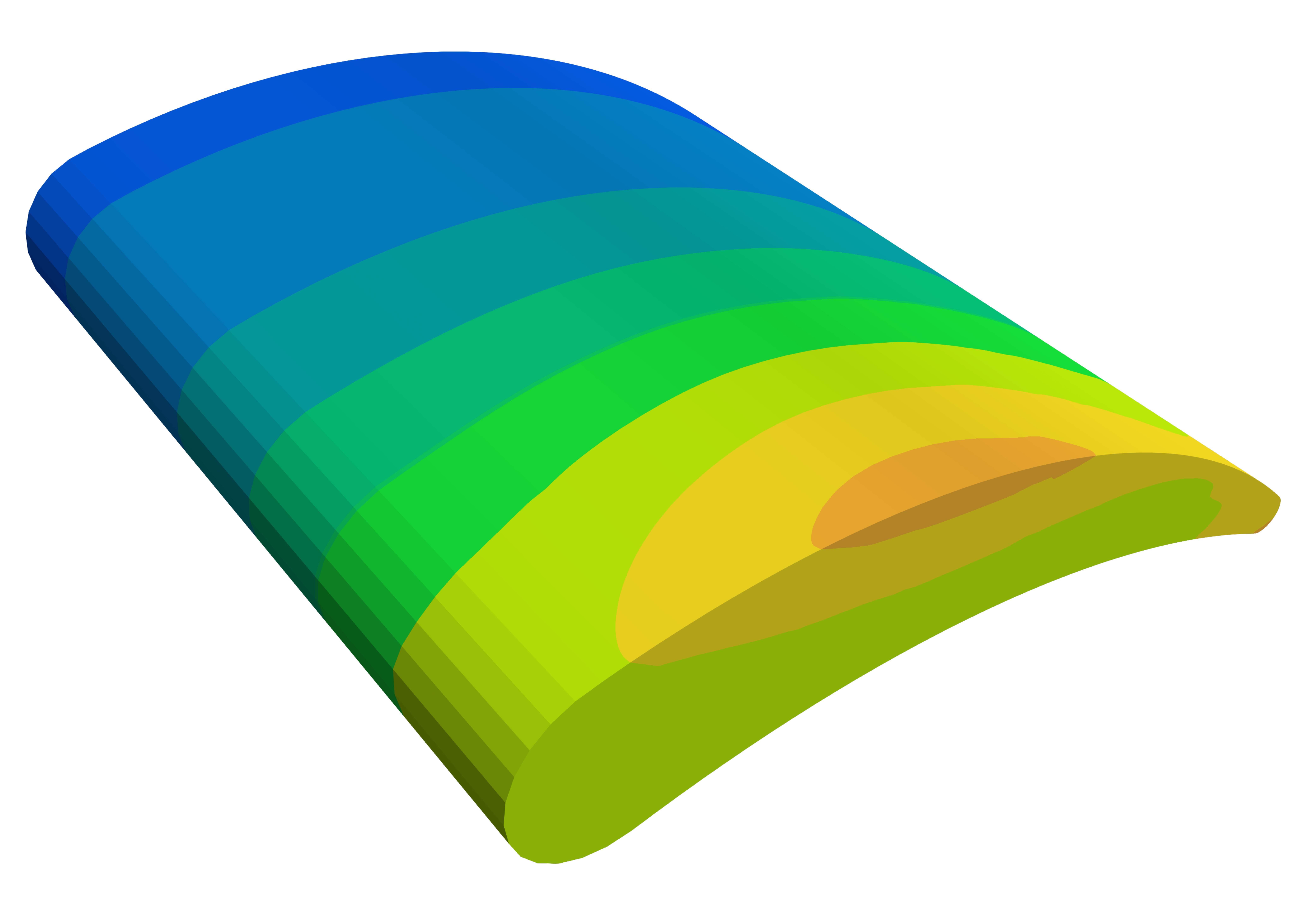}
  \end{subfigure}%
  \begin{subfigure}{.08\textwidth} 
    \centering 
    \begin{tikzpicture}
      \node[inner sep=0pt] (pic) at (0,0) {\includegraphics[height=40mm, width=5mm]
      {03_Contour/00_Color_Maps/Damage_Step_Vertical.pdf}};
      \node[inner sep=0pt] (0)   at ($(pic.south)+( 0.50, 0.15)$)  {$0$};
      \node[inner sep=0pt] (1)   at ($(pic.south)+( 0.50, 3.80)$)  {$1$};
      \node[inner sep=0pt] (d)   at ($(pic.south)+( 0.00, 4.35)$)  {$D_{yy}~\si{[-]}$};
    \end{tikzpicture} 
  \end{subfigure}
  
  \vspace{2mm}
  
  \begin{subfigure}{.45\textwidth} 
    \includegraphics[width=\textwidth]{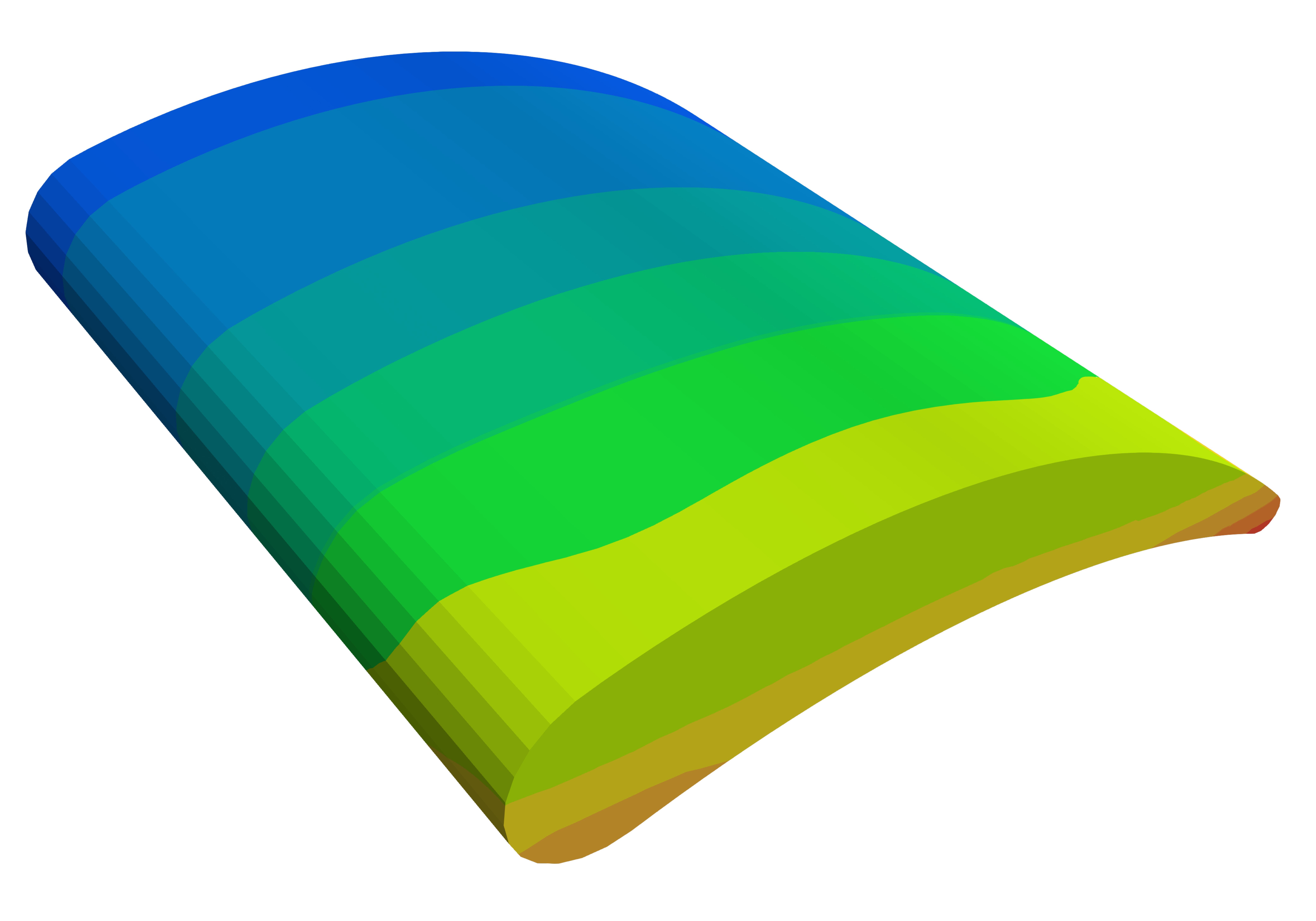}
    \caption{21560 elements}
    \label{fig:p4_Dmeshcontop21560}
  \end{subfigure}%
  \begin{subfigure}{.45\textwidth} 
    \includegraphics[width=\textwidth]{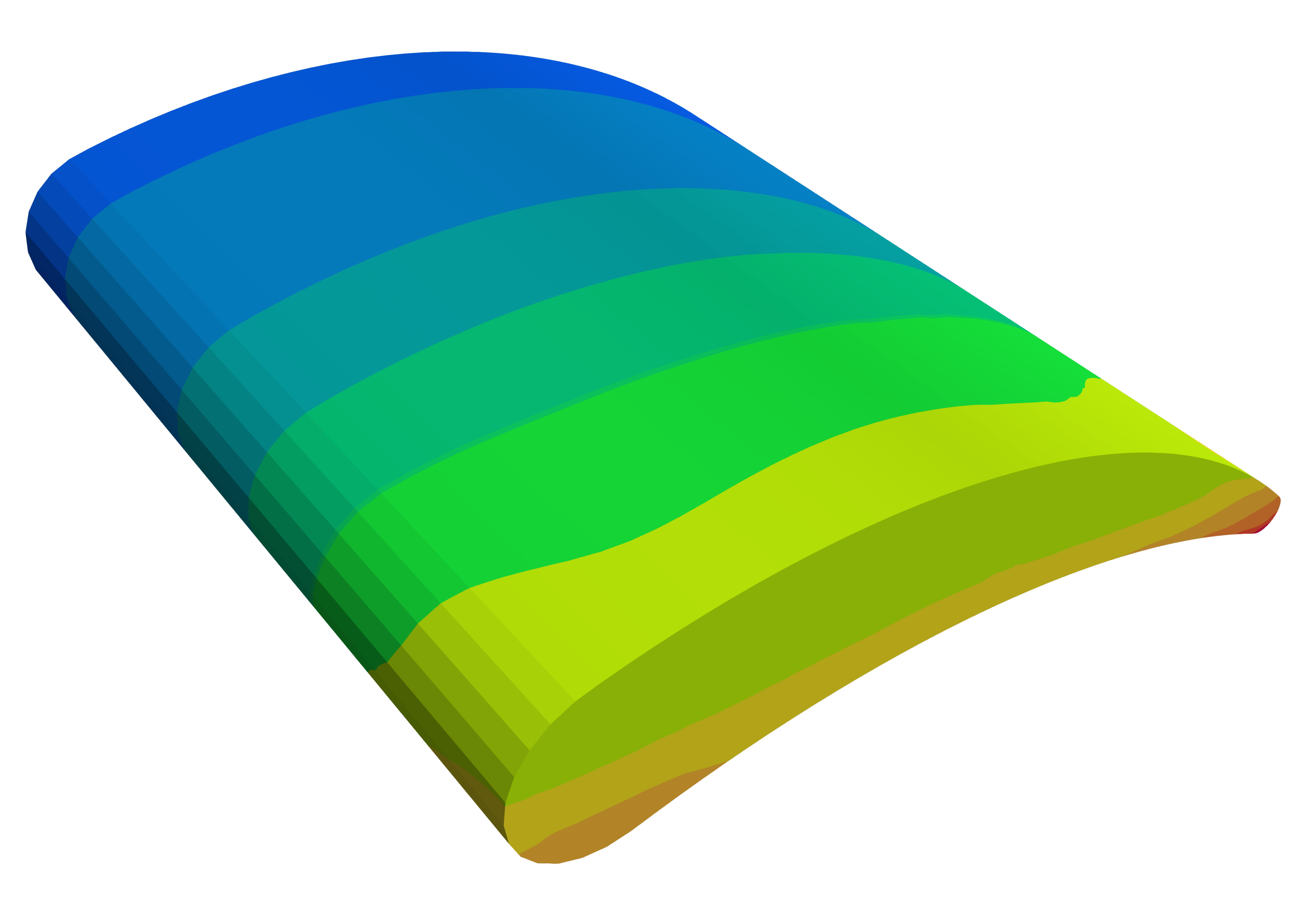}
    \caption{116565 elements}
    \label{fig:p4_Dmeshcontop116565}
  \end{subfigure}%
  \begin{subfigure}{.08\textwidth} 
    \centering 
    \begin{tikzpicture}
      \node[inner sep=0pt] (pic) at (0,0) {\includegraphics[height=40mm, width=5mm]
      {03_Contour/00_Color_Maps/Damage_Step_Vertical.pdf}};
      \node[inner sep=0pt] (0)   at ($(pic.south)+( 0.50, 0.15)$)  {$0$};
      \node[inner sep=0pt] (1)   at ($(pic.south)+( 0.50, 3.80)$)  {$1$};
      \node[inner sep=0pt] (d)   at ($(pic.south)+( 0.00, 4.35)$)  {$D_{zz}~\si{[-]}$};
    \end{tikzpicture} 
  \end{subfigure}
  
  \caption{Contour plots for the mesh convergence of the normal components of the damage tensor for the rotor blade specimen (top view).}
  \label{fig:p4_Dmeshcontop}
\end{figure}

The excellent coarse mesh accuracy, which has been observed with respect to the load bearing capacity, is also reflected in the comparison of the damage contour plots in Figs.~\ref{fig:p4_Dmeshconbottom} (bottom view) and \ref{fig:p4_Dmeshcontop} (top view), where high agreement between the normal components of the damage tensor obtained with the coarsest and finest mesh is observed.
Fig.~\ref{fig:p4_Dmeshcontop} reveals the concentrated evolution of the damage tensor components $D_{xx}$ and $D_{yy}$ in the middle of the upper side of the blade at the clamped end. In Fig.~\ref{fig:p4_Dmeshconbottom}, the component $D_{zz}$ predominantly evolves at the edges of the lower side at the clamped end. The coarse mesh, in Figs.~\ref{fig:p4_Dmeshconbottom21560} and \ref{fig:p4_Dmeshcontop21560}, predicts these damage patterns accurately and the fine mesh, in Figs.~\ref{fig:p4_Dmeshconbottom116565} and \ref{fig:p4_Dmeshcontop116565}, corroborates these results. 

\begin{figure}
  \centering 
  
  \begin{subfigure}{.22\textwidth} 
    \includegraphics[width=\textwidth]{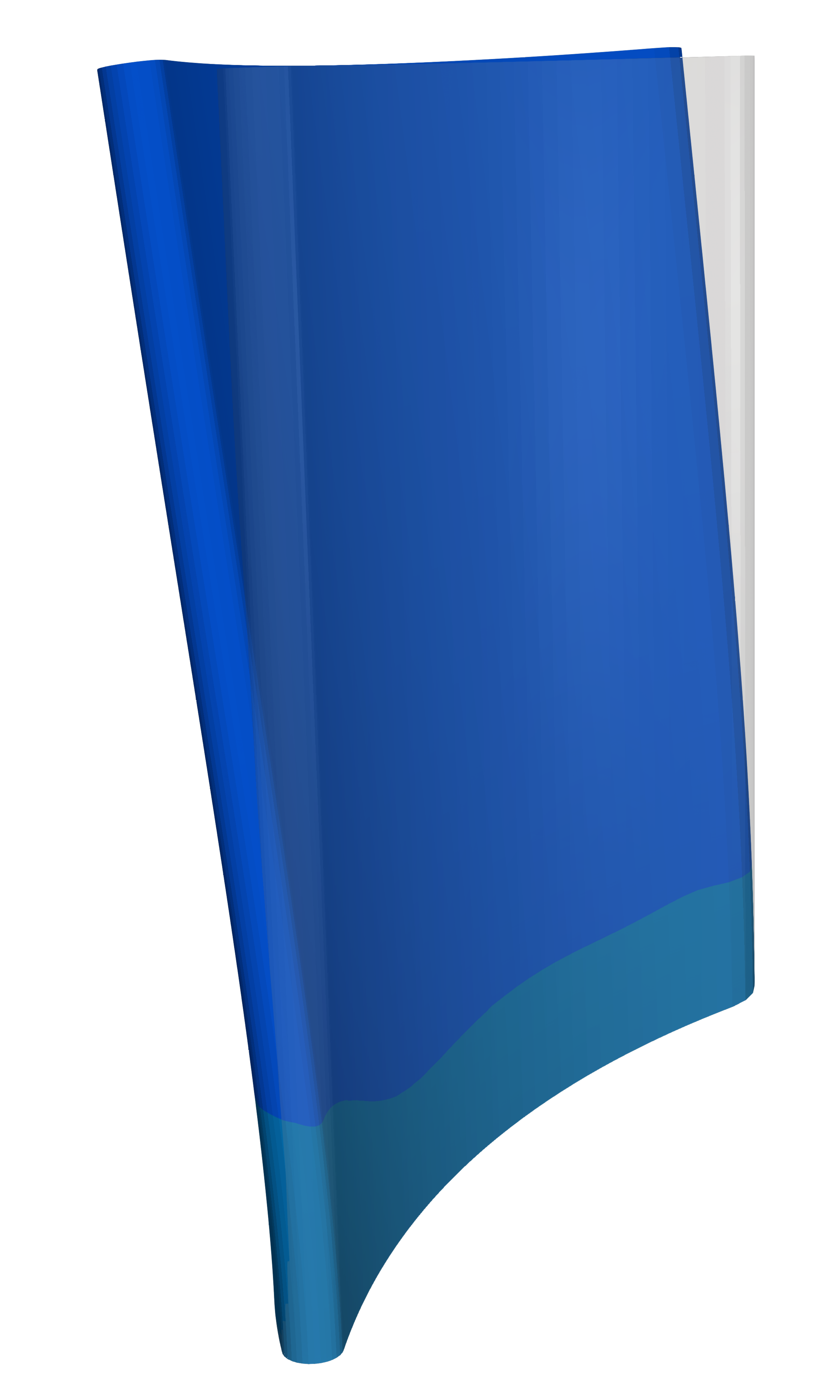}
  \end{subfigure}%
  \begin{subfigure}{.22\textwidth} 
    \includegraphics[width=\textwidth]{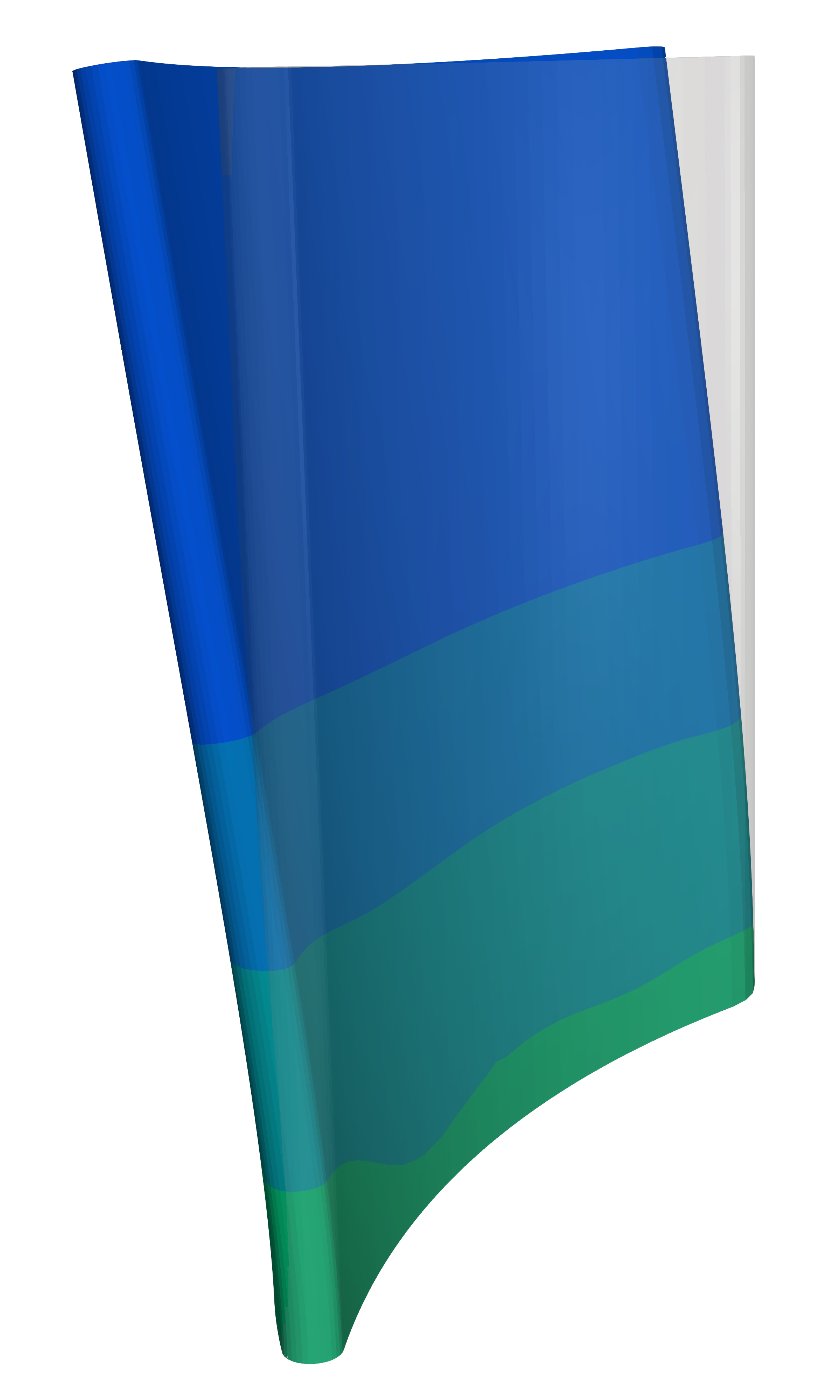}
  \end{subfigure}%
  \begin{subfigure}{.22\textwidth} 
    \includegraphics[width=\textwidth]{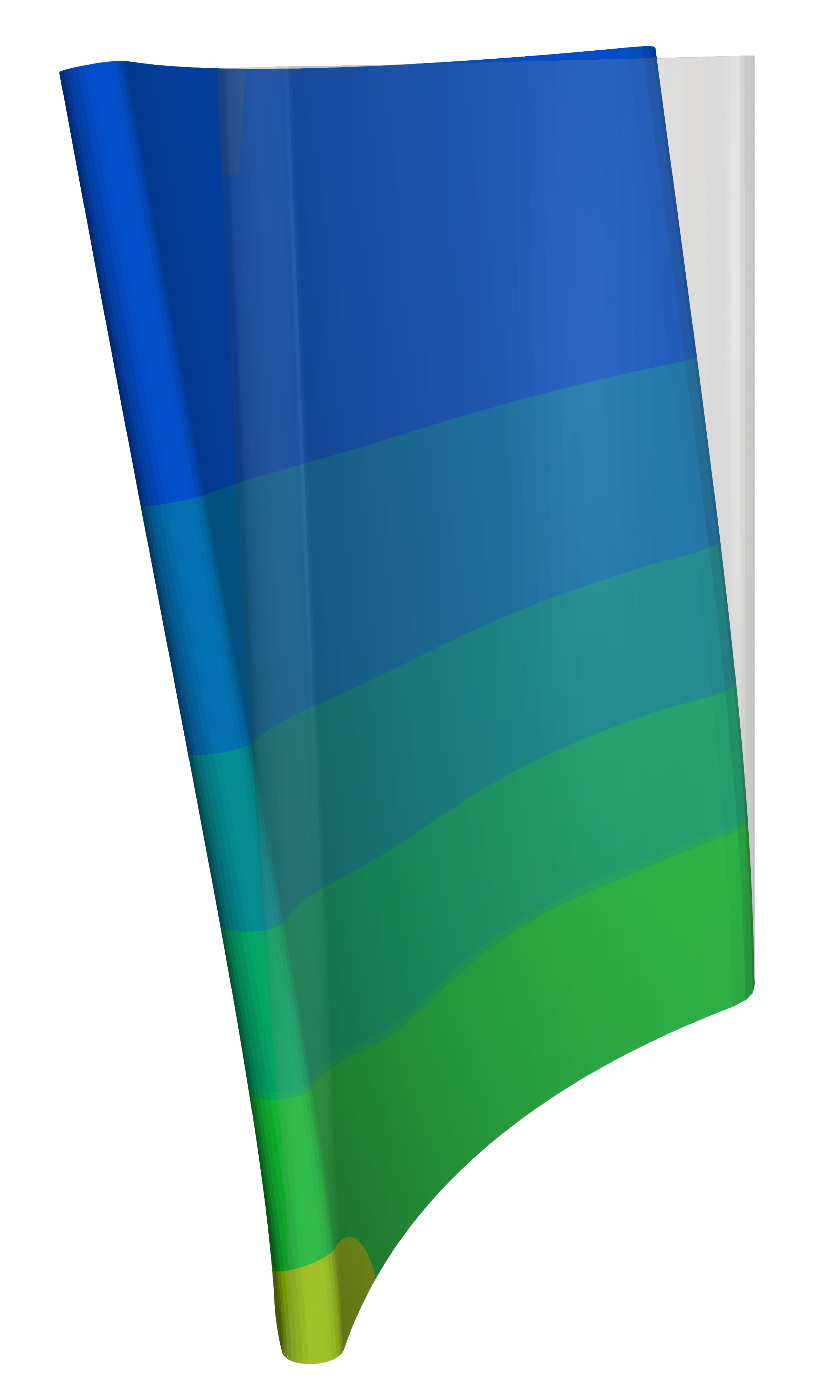}
  \end{subfigure}%
  \begin{subfigure}{.22\textwidth} 
    \includegraphics[width=\textwidth]{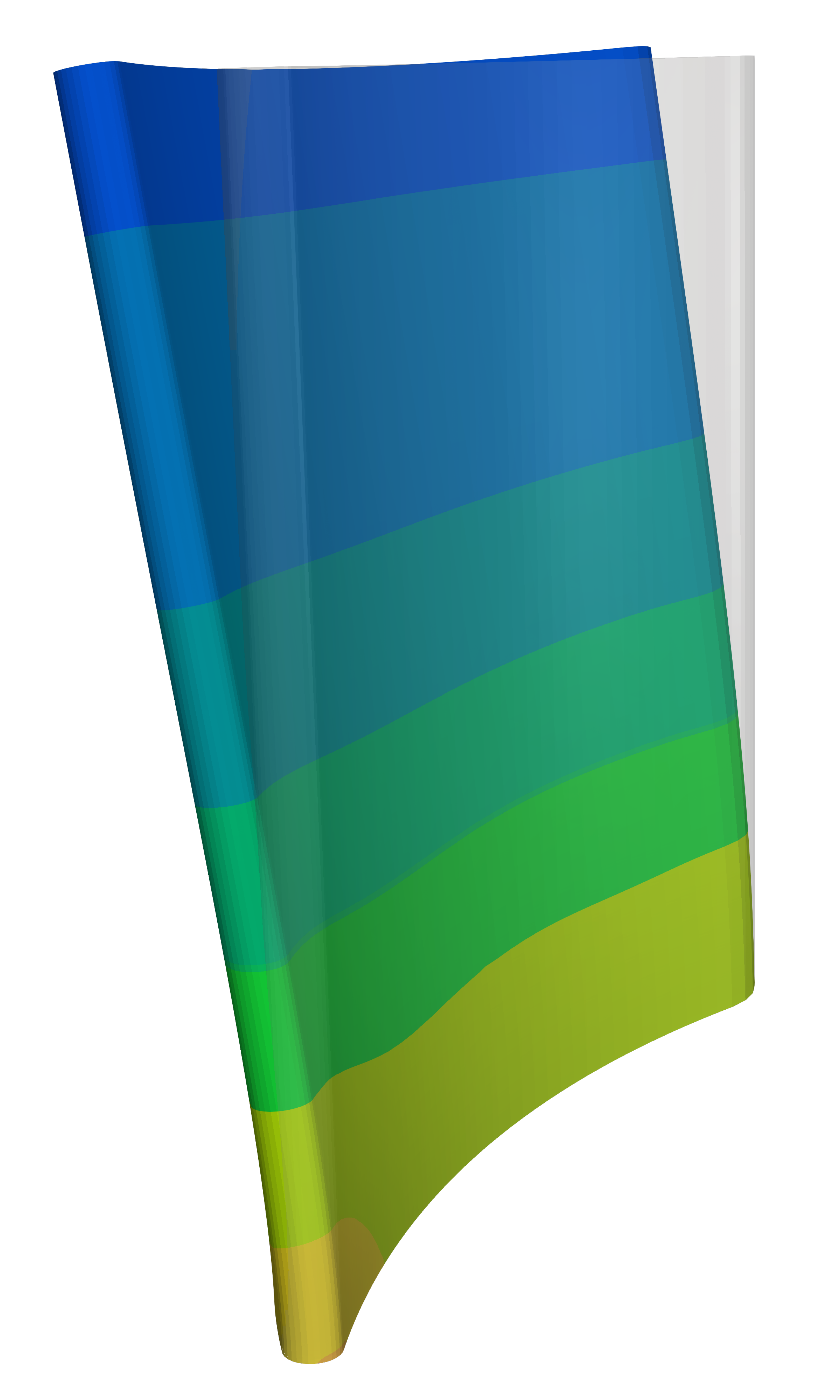}
  \end{subfigure}
  \quad
  \begin{subfigure}{.08\textwidth} 
    \centering 
    \begin{tikzpicture}
      \node[inner sep=0pt] (pic) at (0,0) {\includegraphics[height=40mm, width=5mm]
      {03_Contour/00_Color_Maps/Damage_Step_Vertical.pdf}};
      \node[inner sep=0pt] (0)   at ($(pic.south)+( 0.50, 0.15)$)  {$0$};
      \node[inner sep=0pt] (1)   at ($(pic.south)+( 0.50, 3.80)$)  {$1$};
      \node[inner sep=0pt] (d)   at ($(pic.south)+( 0.00, 4.35)$)  {$D_{xx}~\si{[-]}$};
    \end{tikzpicture} 
  \end{subfigure}
  
  \vspace{2mm}
  
  \begin{subfigure}{.22\textwidth} 
    \includegraphics[width=\textwidth]{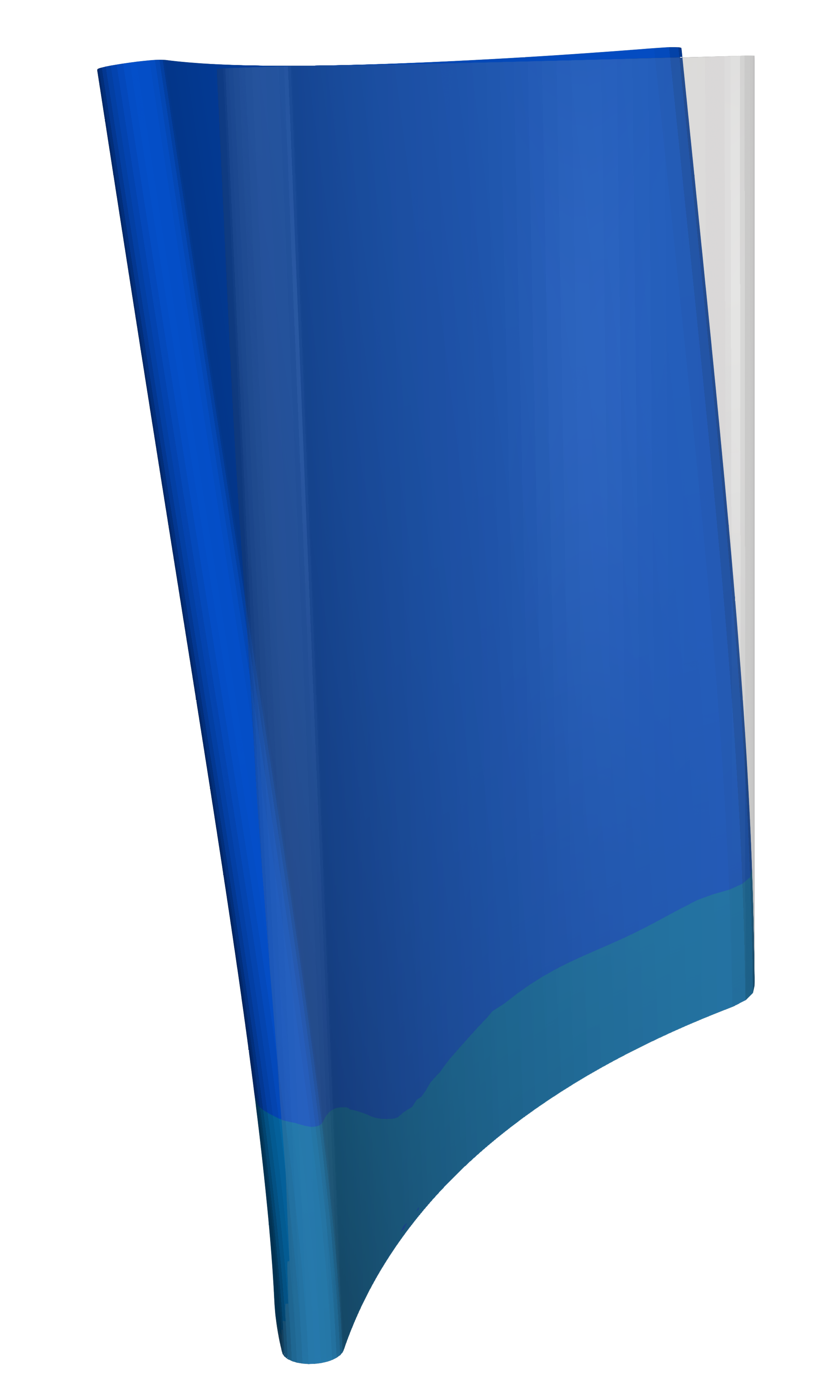}
  \end{subfigure}%
  \begin{subfigure}{.22\textwidth} 
    \includegraphics[width=\textwidth]{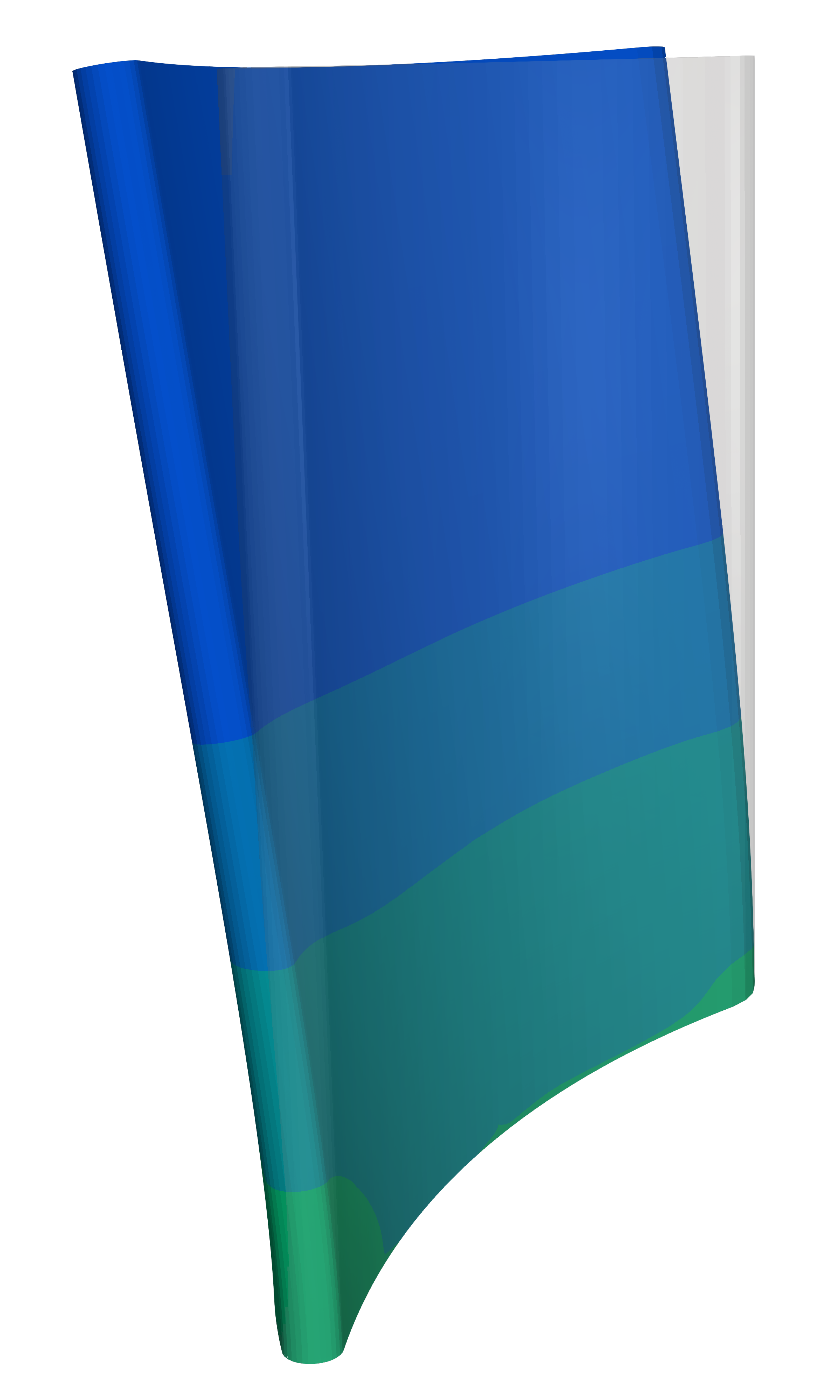}
  \end{subfigure}%
  \begin{subfigure}{.22\textwidth} 
    \includegraphics[width=\textwidth]{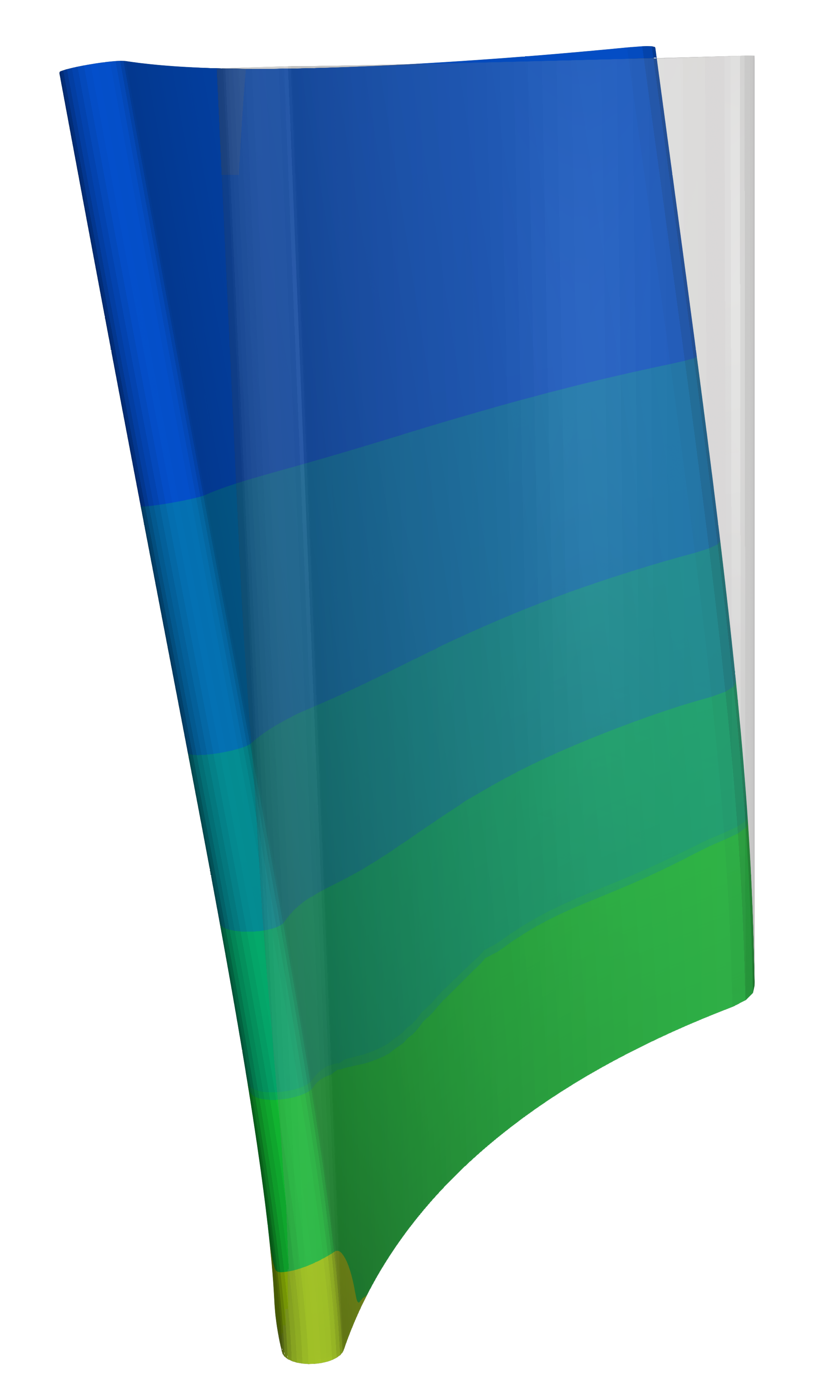}
  \end{subfigure}%
  \begin{subfigure}{.22\textwidth} 
    \includegraphics[width=\textwidth]{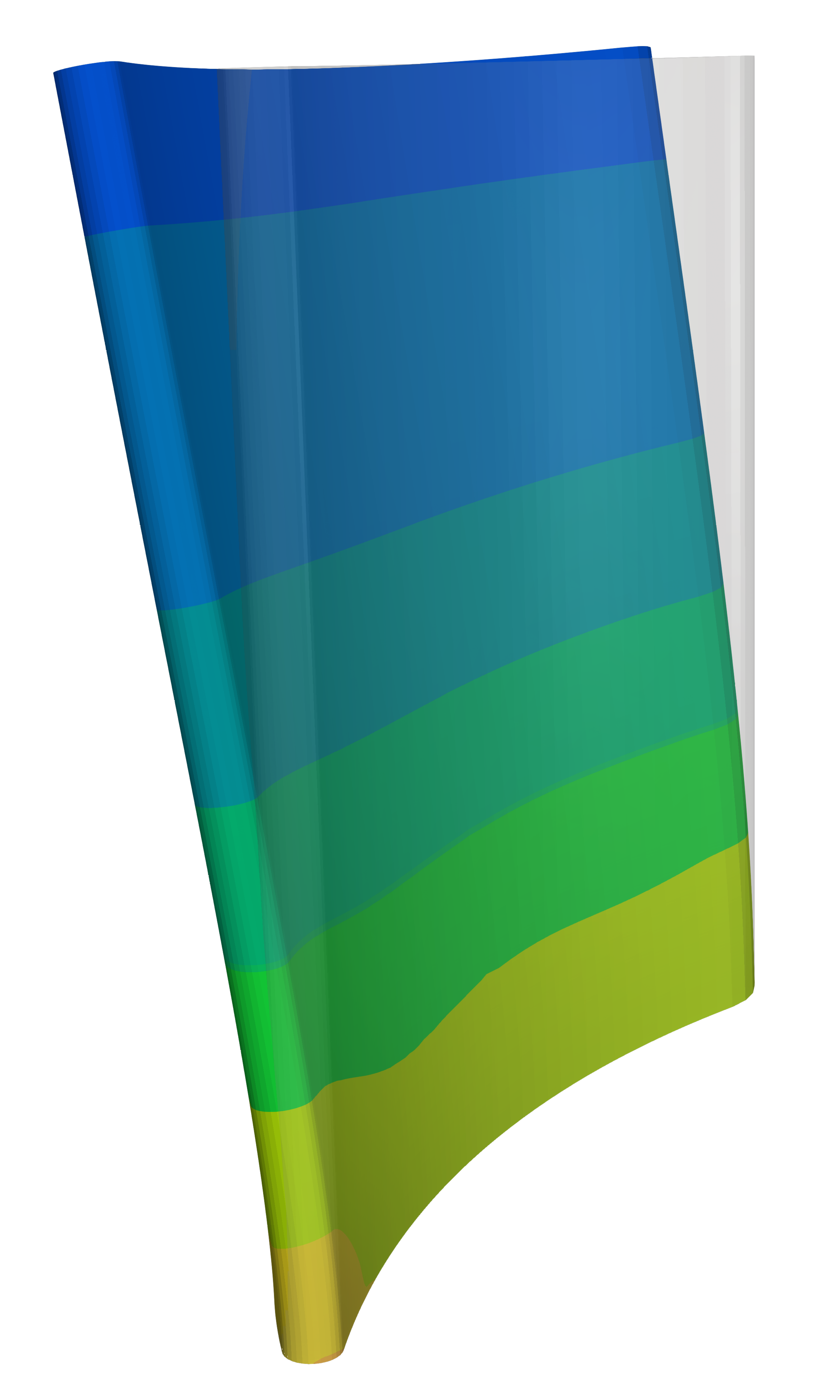}
  \end{subfigure}
  \quad
  \begin{subfigure}{.08\textwidth} 
    \centering 
    \begin{tikzpicture}
      \node[inner sep=0pt] (pic) at (0,0) {\includegraphics[height=40mm, width=5mm]
      {03_Contour/00_Color_Maps/Damage_Step_Vertical.pdf}};
      \node[inner sep=0pt] (0)   at ($(pic.south)+( 0.50, 0.15)$)  {$0$};
      \node[inner sep=0pt] (1)   at ($(pic.south)+( 0.50, 3.80)$)  {$1$};
      \node[inner sep=0pt] (d)   at ($(pic.south)+( 0.00, 4.35)$)  {$D_{yy}~\si{[-]}$};
    \end{tikzpicture} 
  \end{subfigure}
  
  \vspace{2mm}
  
  \begin{subfigure}{.22\textwidth} 
    \includegraphics[width=\textwidth]{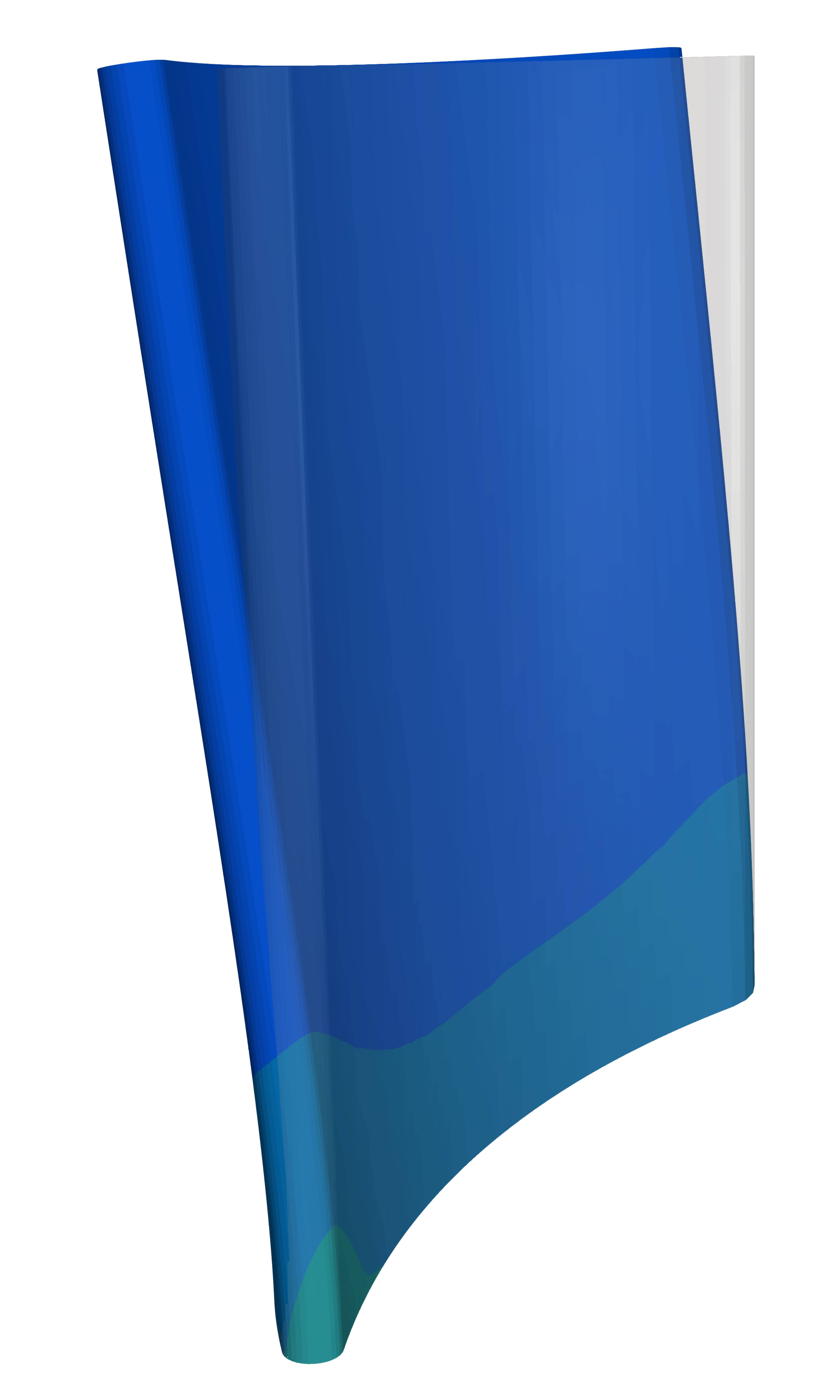}
    \centering
    \caption{damage initiation}
    \label{fig:p4_Dnormal_evolution_damage_initiation}
  \end{subfigure}%
  \begin{subfigure}{.22\textwidth} 
    \includegraphics[width=\textwidth]{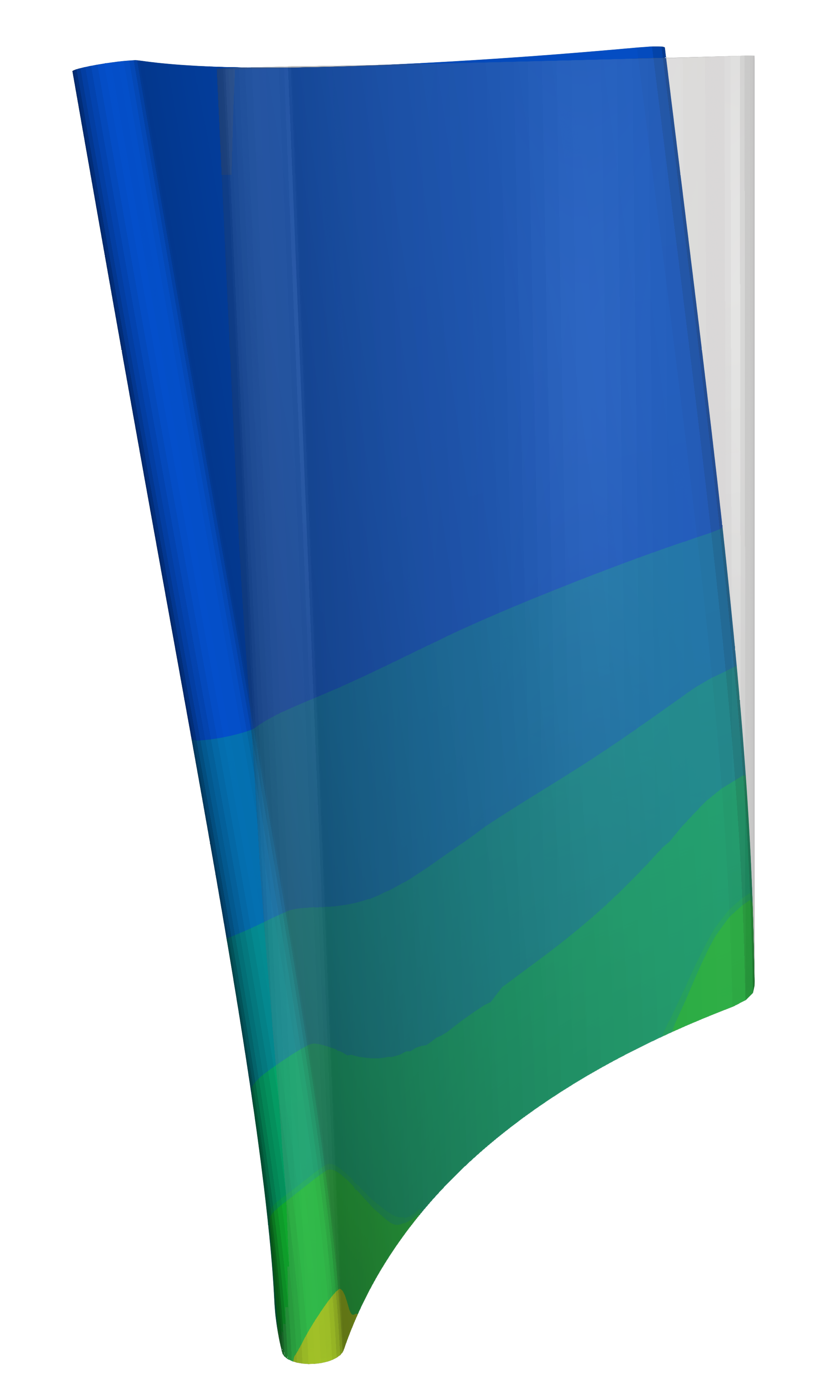}
    \centering
    \caption{intermediate~1}
    \label{fig:p4_Dnormal_evolution_intermediate_1}
  \end{subfigure}%
  \begin{subfigure}{.22\textwidth} 
    \includegraphics[width=\textwidth]{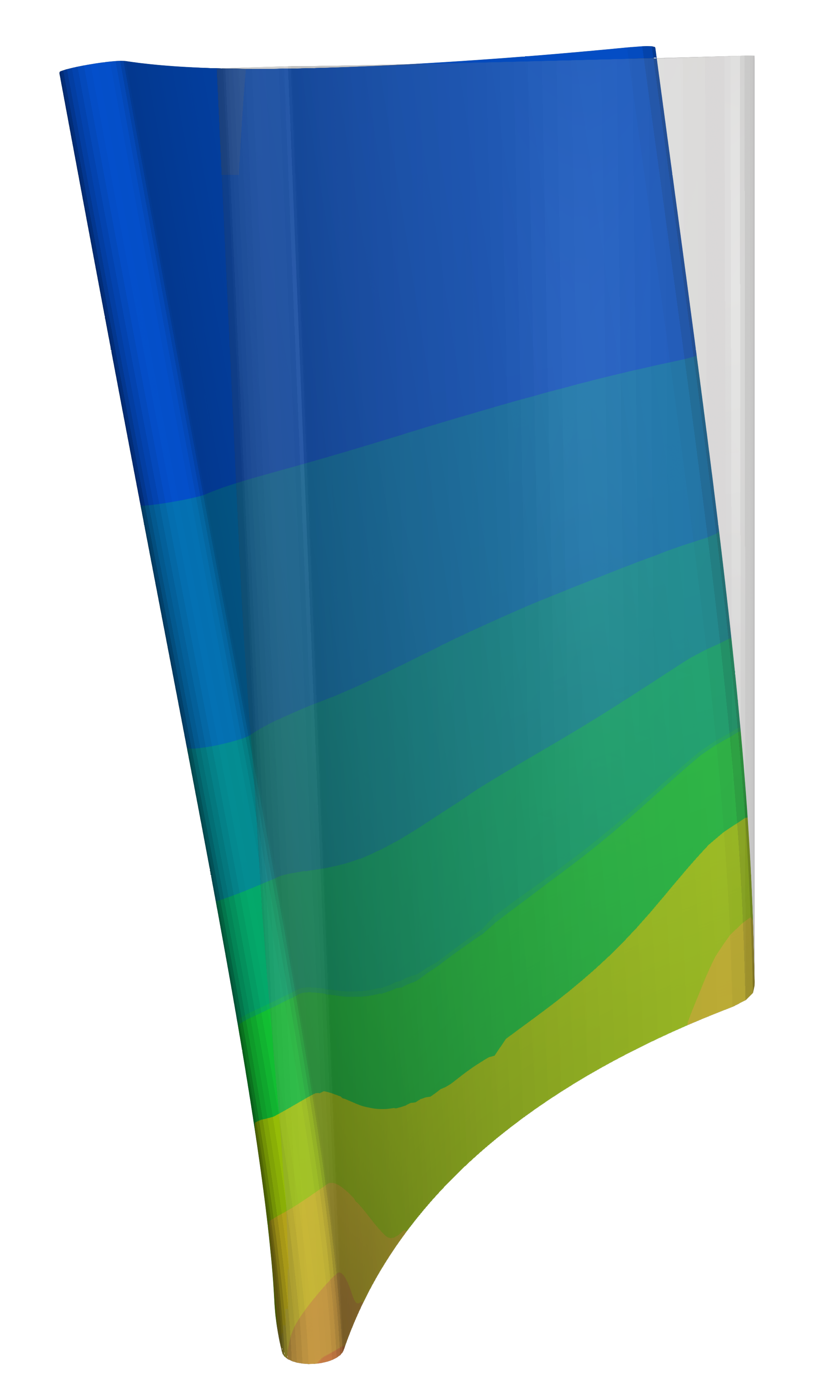}
    \centering
    \caption{intermediate~2}
    \label{fig:p4_Dnormal_evolution_intermediate_2}
  \end{subfigure}%
  \begin{subfigure}{.22\textwidth} 
    \includegraphics[width=\textwidth]{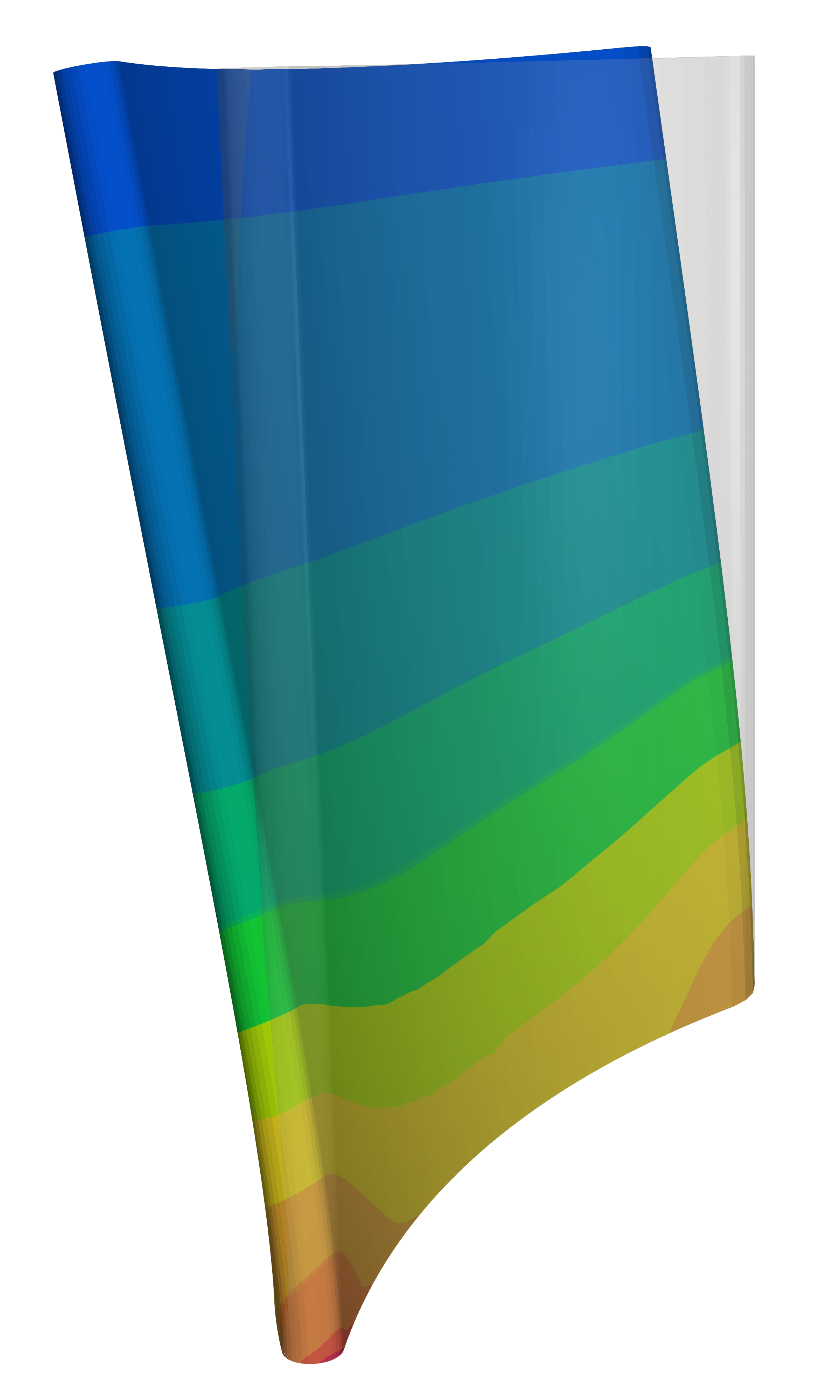}
    \centering
    \caption{crack opening}
    \label{fig:p4_Dnormal_evolution_crack_opening}
  \end{subfigure}
  \quad
  \begin{subfigure}{.08\textwidth} 
    \centering 
    \begin{tikzpicture}
      \node[inner sep=0pt] (pic) at (0,0) {\includegraphics[height=40mm, width=5mm]
      {03_Contour/00_Color_Maps/Damage_Step_Vertical.pdf}};
      \node[inner sep=0pt] (0)   at ($(pic.south)+( 0.50, 0.15)$)  {$0$};
      \node[inner sep=0pt] (1)   at ($(pic.south)+( 0.50, 3.80)$)  {$1$};
      \node[inner sep=0pt] (d)   at ($(pic.south)+( 0.00, 4.35)$)  {$D_{zz}~\si{[-]}$};
    \end{tikzpicture} 
    \vphantom{dp}
  \end{subfigure}

  \caption{Contour plots for the evolution of the normal components of the damage tensor for the rotor blade specimen (side view). The contours are plotted on the deformed configuration and the opaque solid shapes indicate the reference configuration.}
  \label{fig:p4_Dnormal_evolution}
\end{figure}

\begin{figure}
  \centering 
  
  \begin{subfigure}{.22\textwidth} 
    \includegraphics[width=\textwidth]{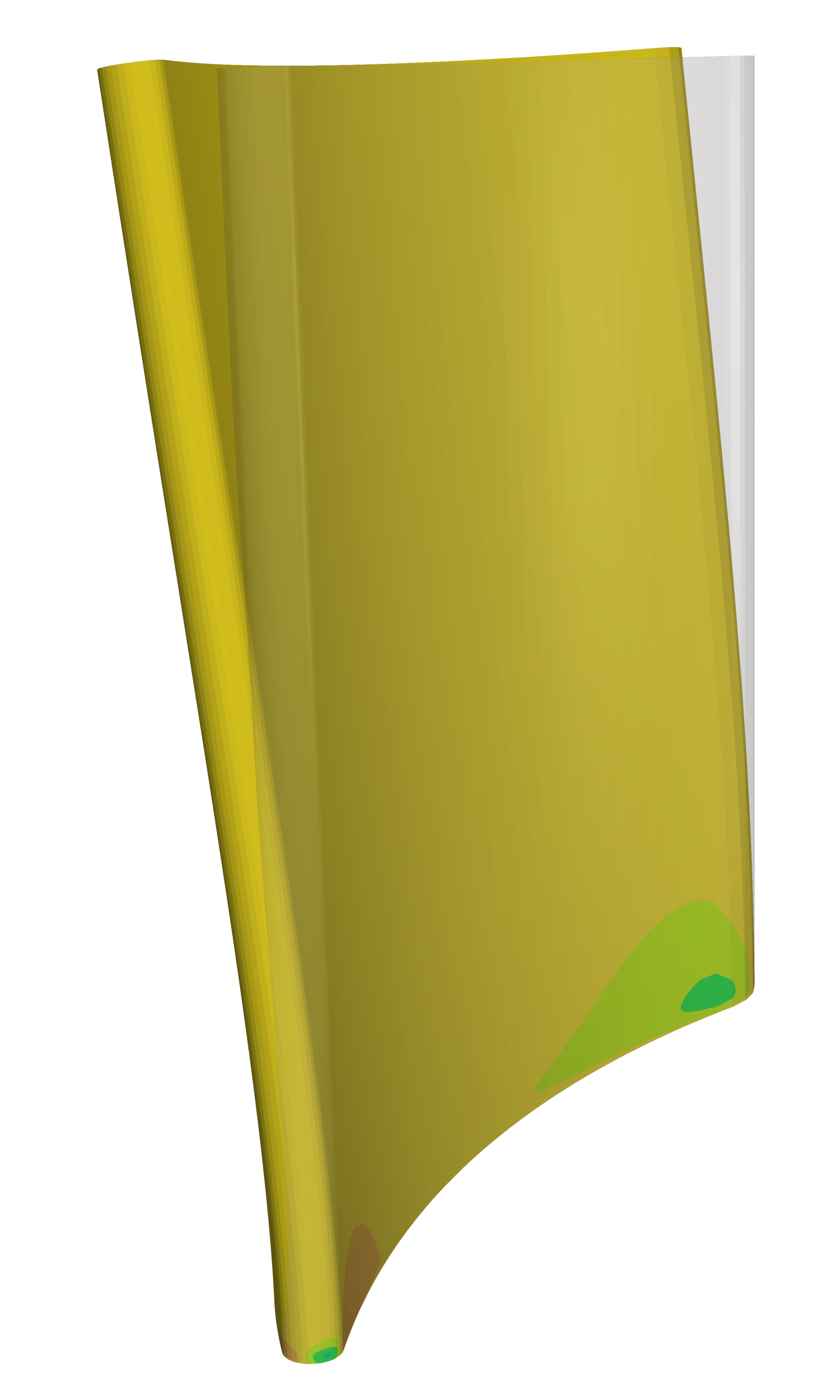}
  \end{subfigure}%
  \begin{subfigure}{.22\textwidth} 
    \includegraphics[width=\textwidth]{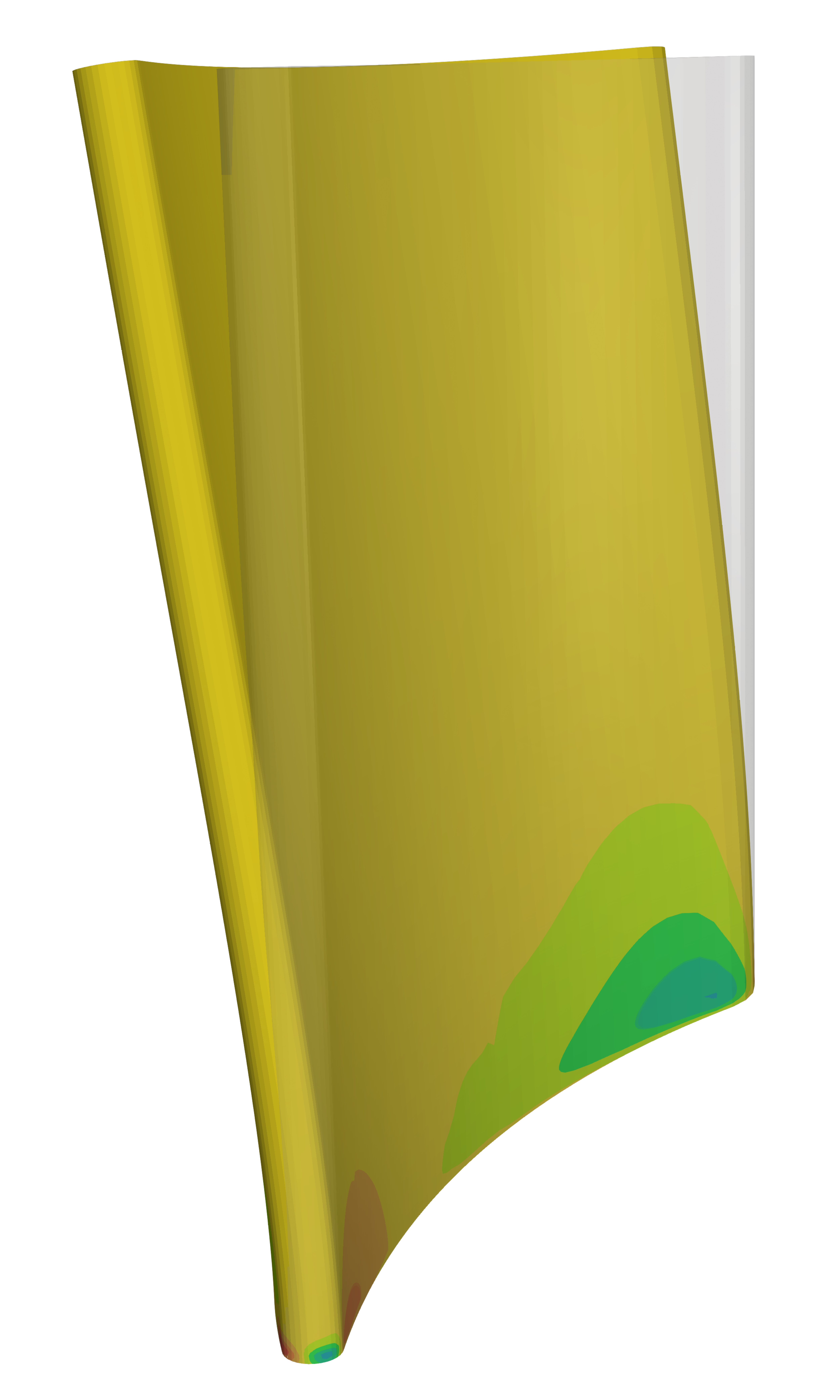}
  \end{subfigure}%
  \begin{subfigure}{.22\textwidth} 
    \includegraphics[width=\textwidth]{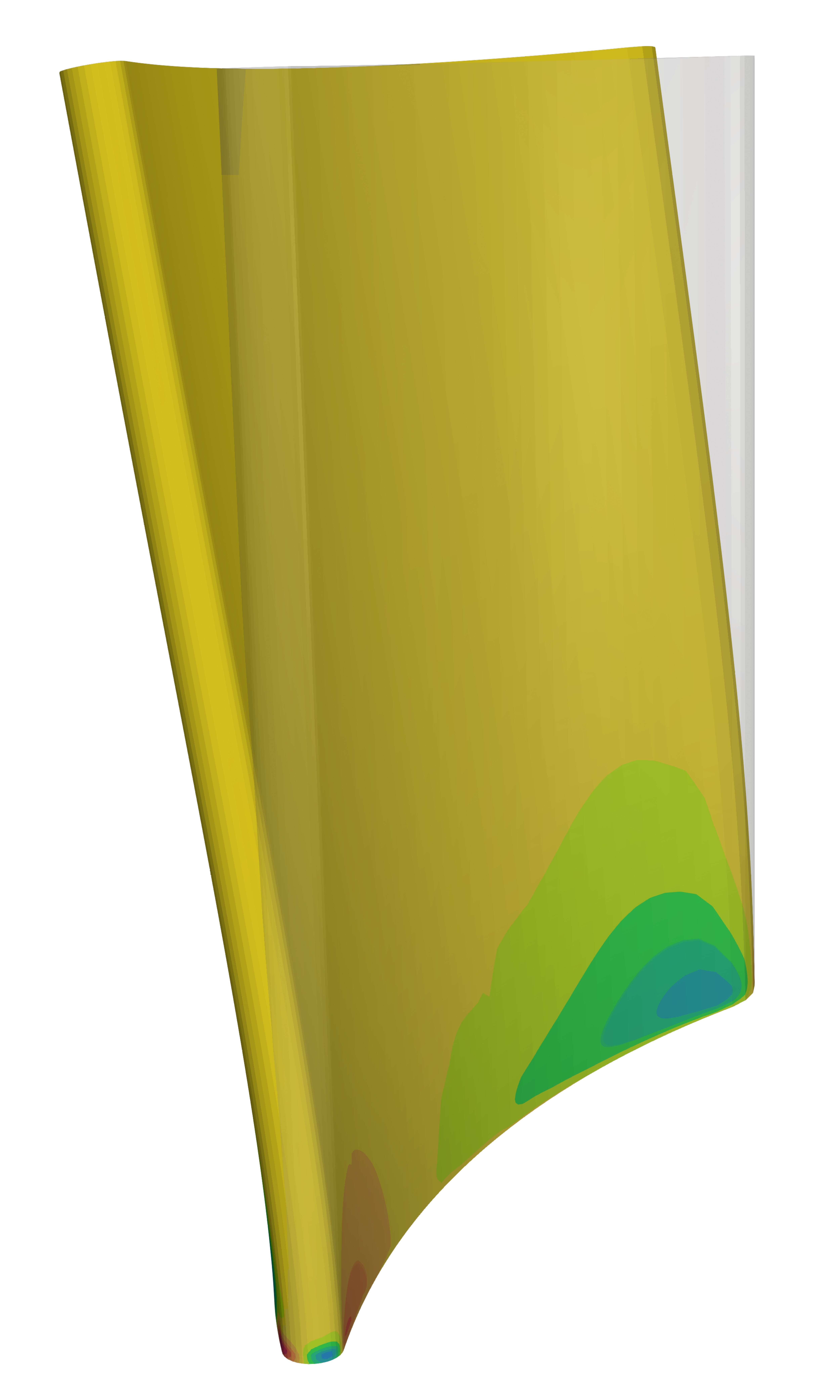}
  \end{subfigure}%
  \begin{subfigure}{.22\textwidth} 
    \includegraphics[width=\textwidth]{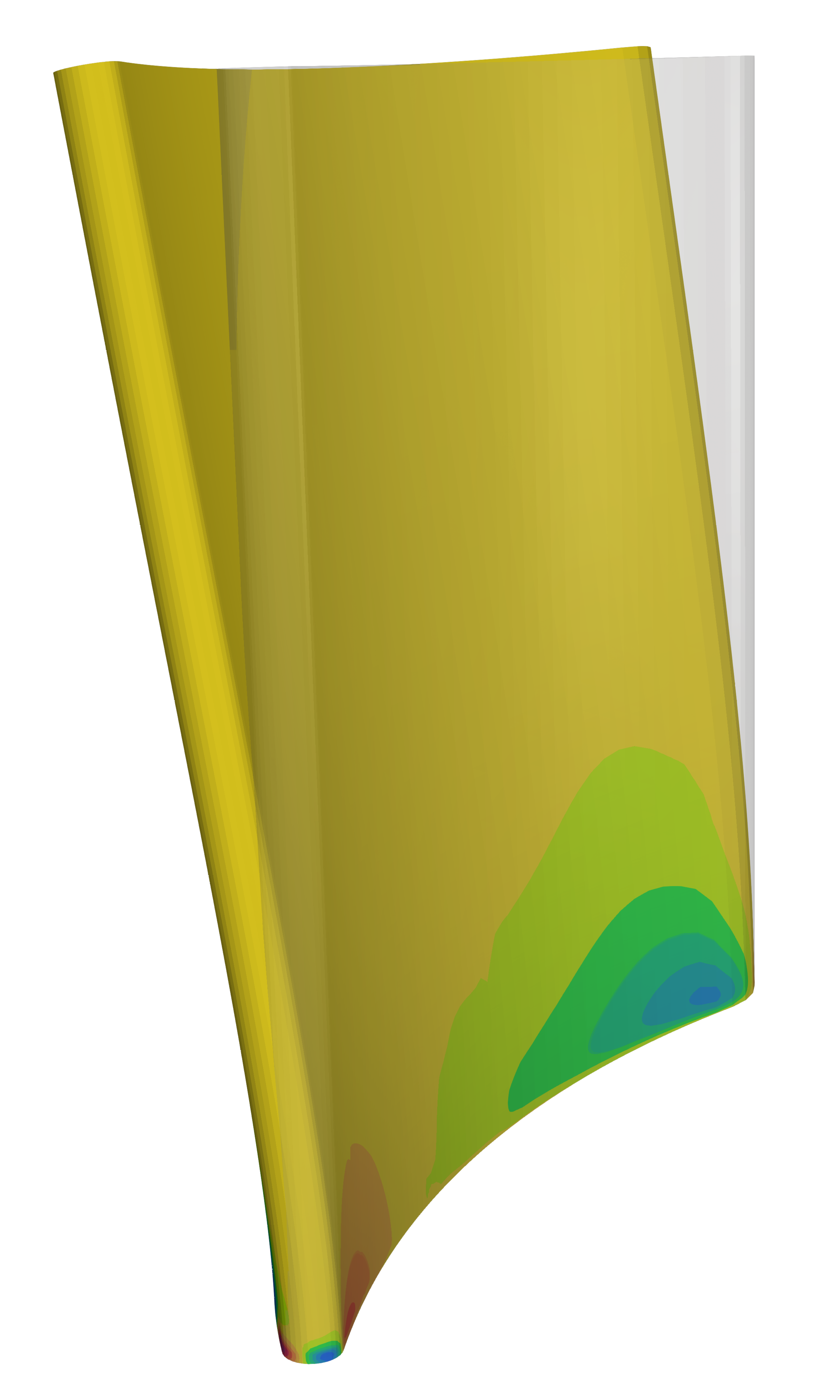}
  \end{subfigure}%
  \begin{subfigure}{.08\textwidth} 
    \centering 
    \begin{tikzpicture}
      \node[inner sep=0pt] (pic) at (0,0) {\includegraphics[height=40mm, width=5mm]
      {03_Contour/00_Color_Maps/Damage_Step_Vertical.pdf}};
      \node[inner sep=0pt] (0)   at ($(pic.south)+( 1.05, 0.15)$)  {$-0.0091$};
      \node[inner sep=0pt] (1)   at ($(pic.south)+( 1.05, 3.80)$)  {$+0.0077$};
      \node[inner sep=0pt] (d)   at ($(pic.south)+( 0.00, 4.35)$)  {$D_{xy}~\si{[-]}$};
    \end{tikzpicture} 
  \end{subfigure}
  
  \vspace{2mm}
  
  \begin{subfigure}{.22\textwidth} 
    \includegraphics[width=\textwidth]{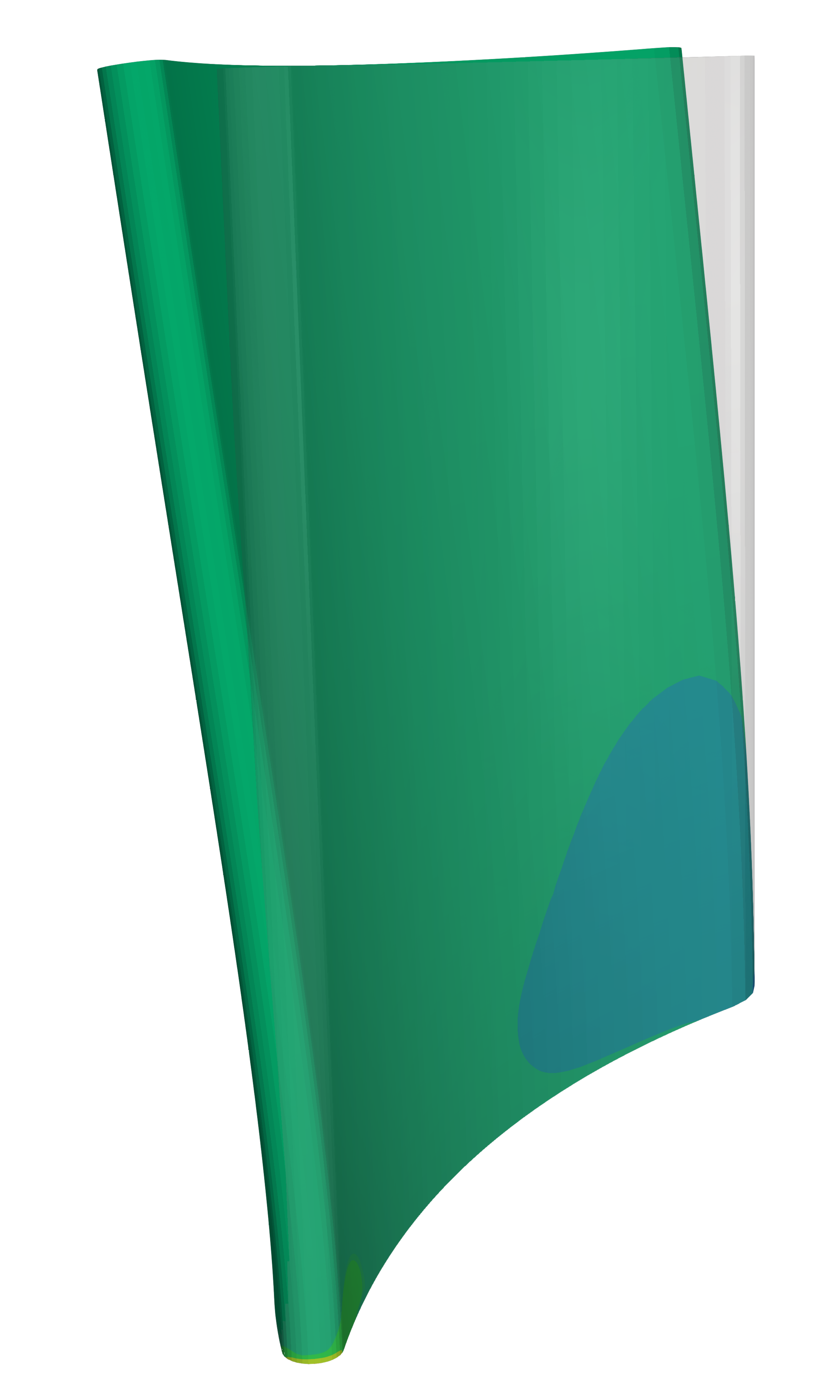}
  \end{subfigure}%
  \begin{subfigure}{.22\textwidth} 
    \includegraphics[width=\textwidth]{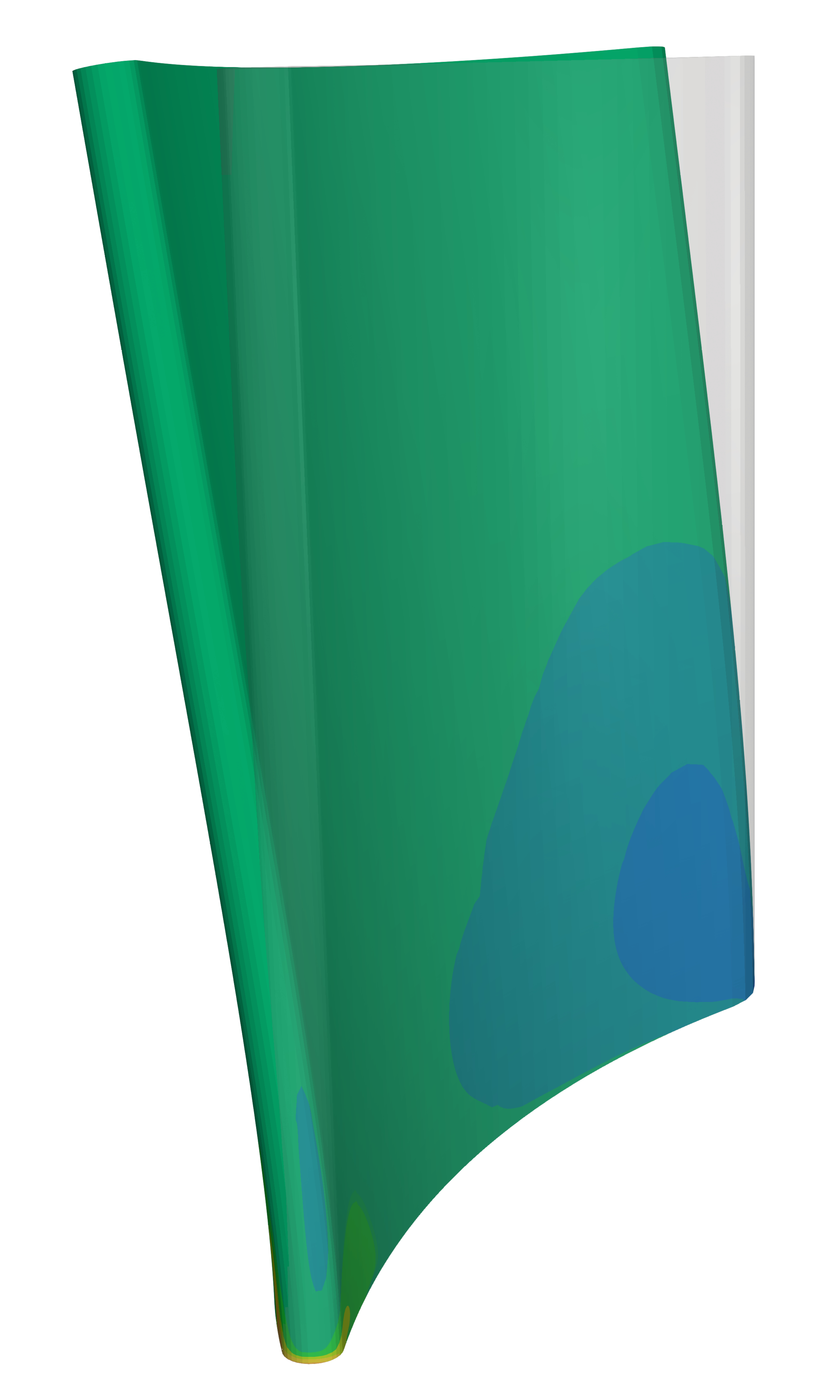}
  \end{subfigure}%
  \begin{subfigure}{.22\textwidth} 
    \includegraphics[width=\textwidth]{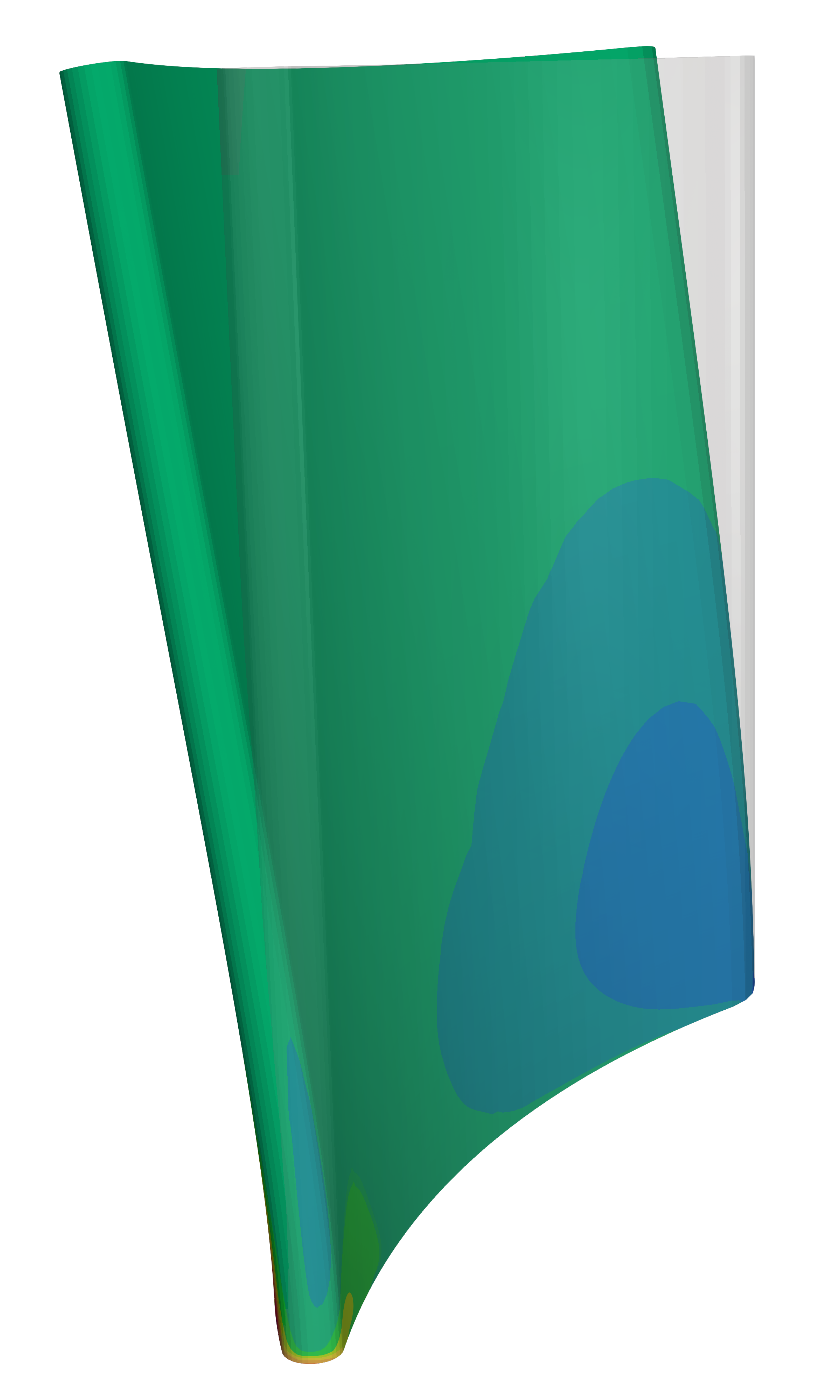}
  \end{subfigure}%
  \begin{subfigure}{.22\textwidth} 
    \includegraphics[width=\textwidth]{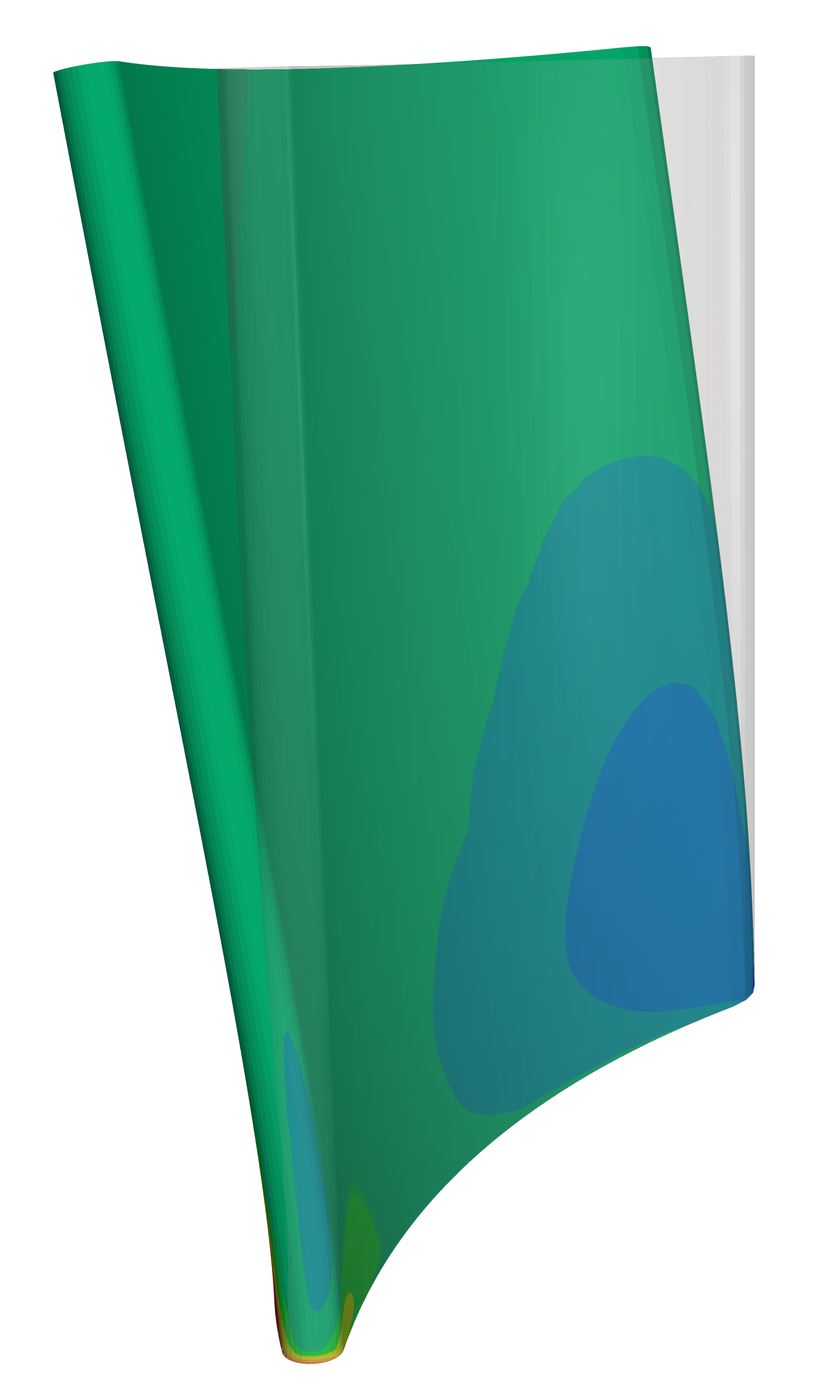}
  \end{subfigure}%
  \begin{subfigure}{.08\textwidth} 
    \centering 
    \begin{tikzpicture}
      \node[inner sep=0pt] (pic) at (0,0) {\includegraphics[height=40mm, width=5mm]
      {03_Contour/00_Color_Maps/Damage_Step_Vertical.pdf}};
      \node[inner sep=0pt] (0)   at ($(pic.south)+( 1.05, 0.15)$)  {$-0.0189$};
      \node[inner sep=0pt] (1)   at ($(pic.south)+( 1.05, 3.80)$)  {$+0.0521$};
      \node[inner sep=0pt] (d)   at ($(pic.south)+( 0.00, 4.35)$)  {$D_{xz}~\si{[-]}$};
    \end{tikzpicture} 
  \end{subfigure}
  
  \vspace{2mm}
  
  \begin{subfigure}{.22\textwidth} 
    \includegraphics[width=\textwidth]{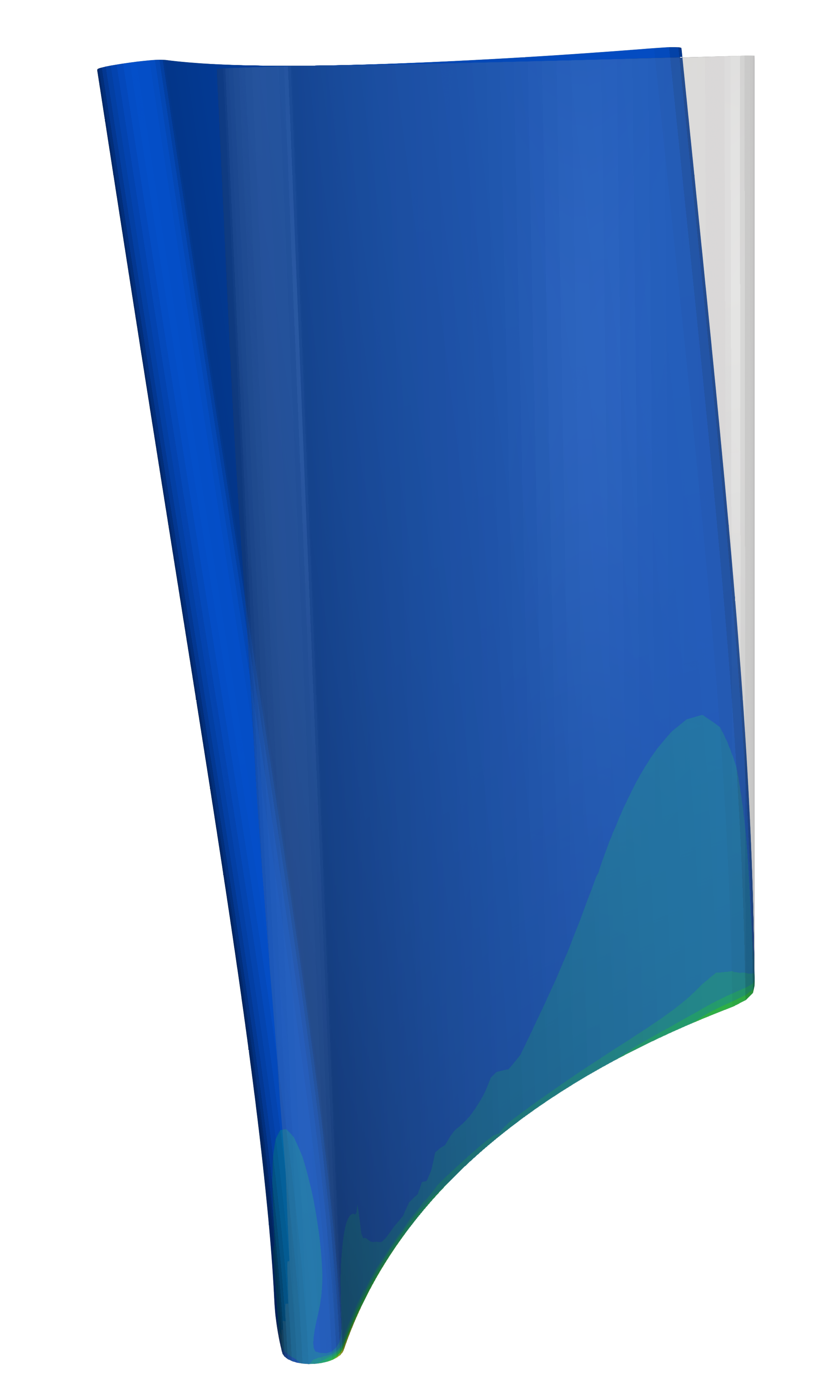}
    \centering
    \caption{damage initiation}
    \label{fig:p4_Dshear_evolution_damage_initiation}
  \end{subfigure}%
  \begin{subfigure}{.22\textwidth} 
    \includegraphics[width=\textwidth]{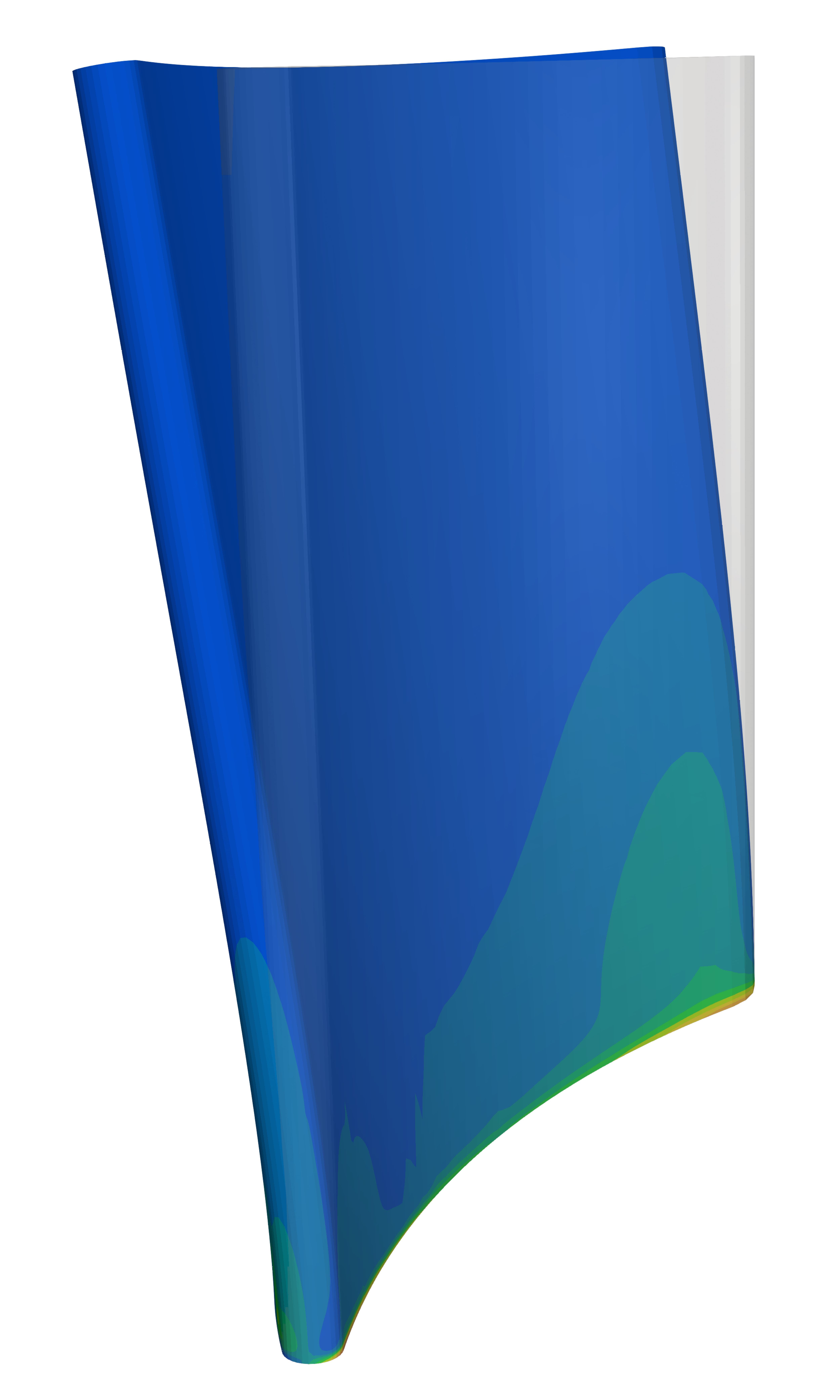}
    \centering
    \caption{intermediate~1}
    \label{fig:p4_Dshear_evolution_intermediate_1}
  \end{subfigure}%
  \begin{subfigure}{.22\textwidth} 
    \includegraphics[width=\textwidth]{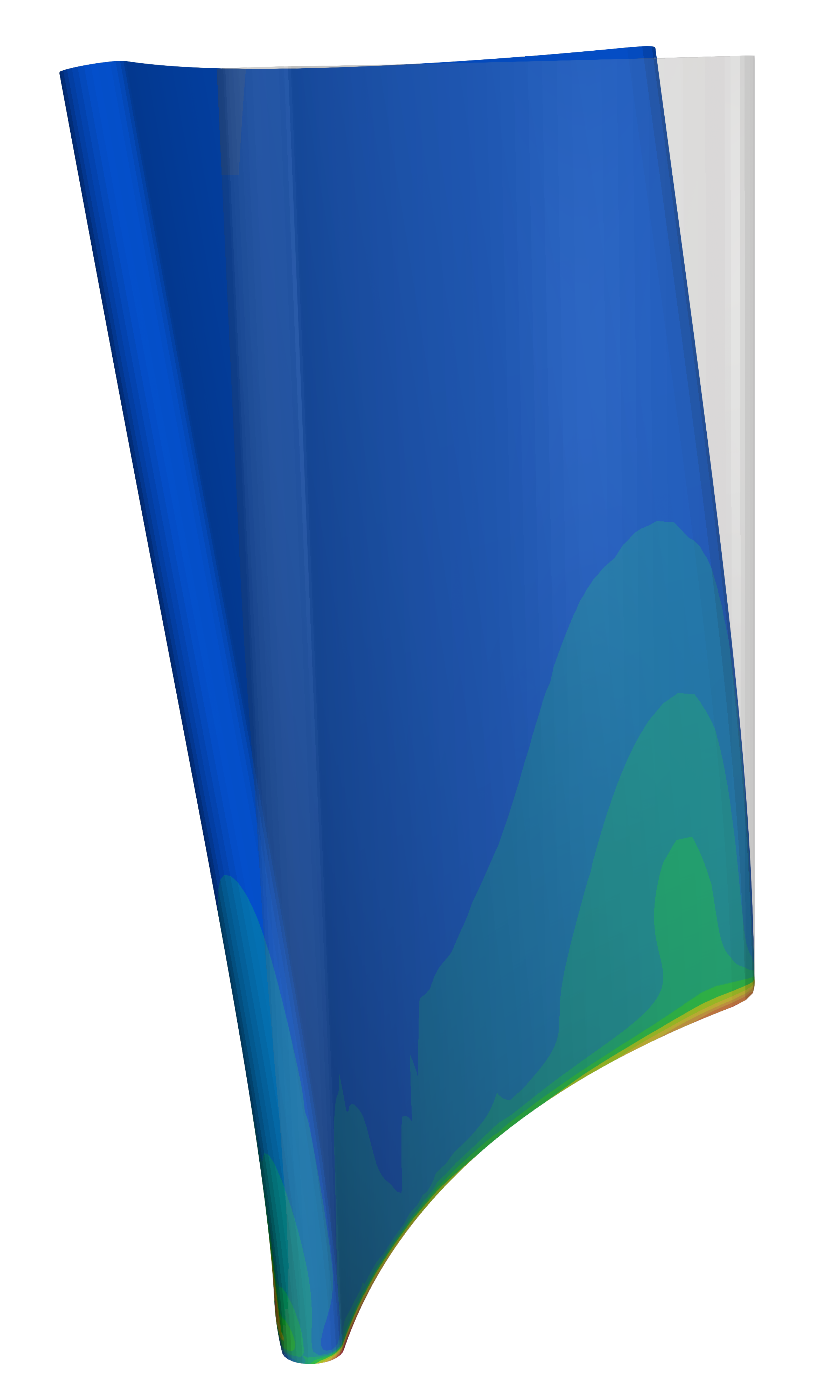}
    \centering
    \caption{intermediate~2}
    \label{fig:p4_Dshear_evolution_intermediate_2}
  \end{subfigure}%
  \begin{subfigure}{.22\textwidth} 
    \includegraphics[width=\textwidth]{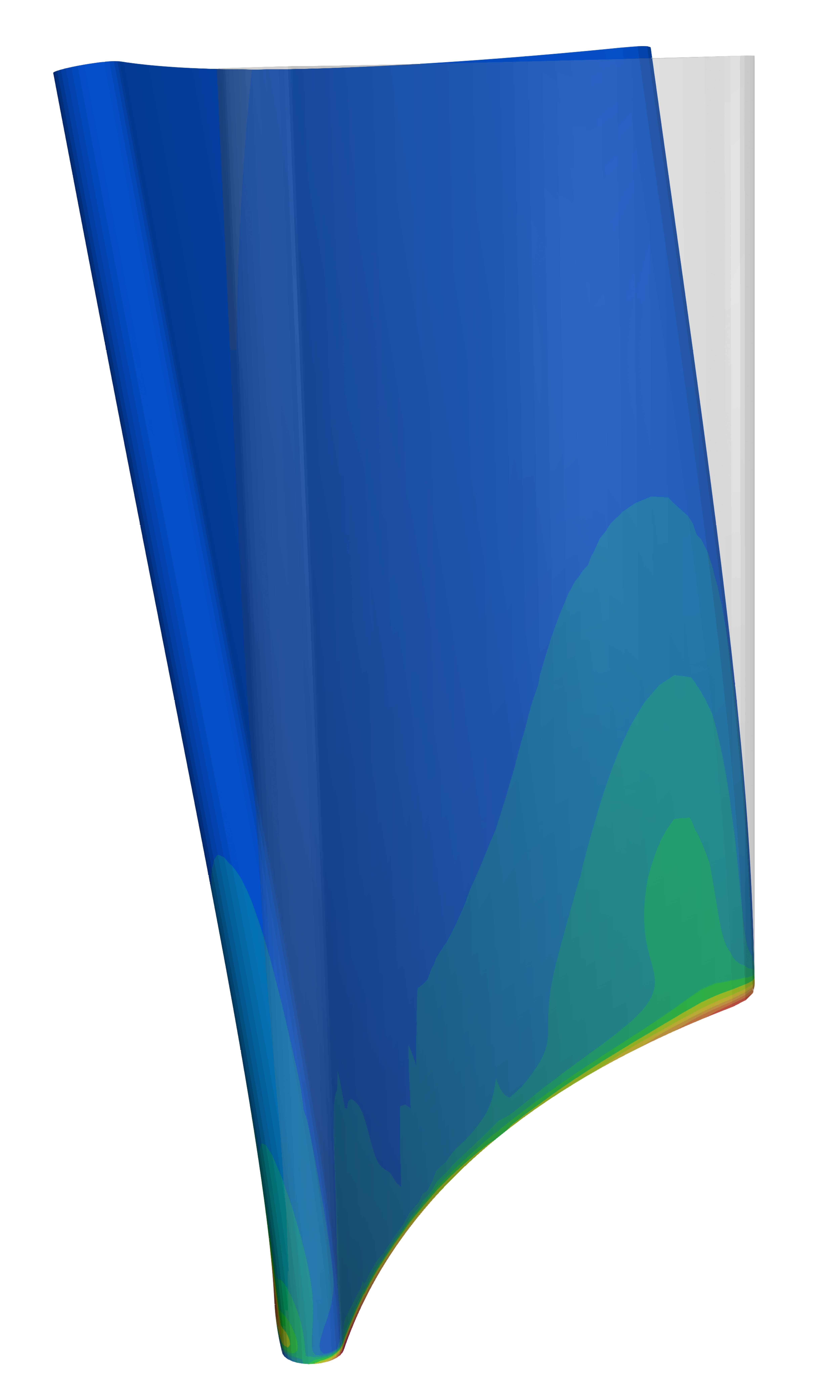}
    \centering
    \caption{crack opening}
    \label{fig:p4_Dshear_evolution_crack_opening}
  \end{subfigure}%
  \begin{subfigure}{.08\textwidth} 
    \centering 
    \begin{tikzpicture}
      \node[inner sep=0pt] (pic) at (0,0) {\includegraphics[height=40mm, width=5mm]
      {03_Contour/00_Color_Maps/Damage_Step_Vertical.pdf}};
      \node[inner sep=0pt] (0)   at ($(pic.south)+( 1.05, 0.15)$)  {$-0.0024$};
      \node[inner sep=0pt] (1)   at ($(pic.south)+( 1.05, 3.80)$)  {$+0.0347$};
      \node[inner sep=0pt] (d)   at ($(pic.south)+( 0.00, 4.35)$)  {$D_{yz}~\si{[-]}$};
    \end{tikzpicture} 
    \vphantom{dp}
  \end{subfigure}

  \caption{Contour plots for the evolution of the shear components of the damage tensor for the rotor blade specimen (side view). The contours are plotted on the deformed configuration and the opaque solid shapes indicate the reference configuration.}
  \label{fig:p4_Dshear_evolution}
\end{figure}

The damage evolution process is presented in Figs.~\ref{fig:p4_Dnormal_evolution} and \ref{fig:p4_Dshear_evolution} from damage initiation (Figs.~\ref{fig:p4_Dnormal_evolution_damage_initiation} and \ref{fig:p4_Dshear_evolution_damage_initiation}) to crack opening (Figs.~\ref{fig:p4_Dnormal_evolution_crack_opening} and \ref{fig:p4_Dshear_evolution_crack_opening}) on the deformed configuration. For all normal components, damage initiates at the clamped edge (Fig.~\ref{fig:p4_Dnormal_evolution_damage_initiation}) and, upon further deformation, progresses with a diffuse damage zone in $z$-direction (Figs.~\ref{fig:p4_Dnormal_evolution_intermediate_1} and \ref{fig:p4_Dnormal_evolution_intermediate_2}). Finally, crack opening occurs by a significant concentration of $D_{zz}$ at the intersection of the free outer edge with radius $r_4$ and the clamped edge (Fig.~\ref{fig:p4_Dnormal_evolution_crack_opening}). The corresponding evolution of the shear components is presented in Fig.~\ref{fig:p4_Dshear_evolution} and shows that their peak values occur at the clamped edge.

\subsection*{Summary of the numerical results}

The most important findings of the numerical examples with the local formulation, model~A (full regularization), model~B (reduced principal traces regularization), and model~C (reduced volumetric-deviatoric regularization) include:
\begin{itemize}
  \item The material strength of the local model increases with the degree of damage anisotropy for this specific choice of the Neo-Hookean elastic energy formulation (Fig.~\ref{fig:p4_SEFu}).
  \item Models~A, B, and C prove excellent coarse mesh accuracy (Figs.~\ref{fig:p4_ANotchedFuA}, \ref{fig:p4_ANotchedFuB}, \ref{fig:p4_ANotchedFuC}, and \ref{fig:p4_RBFu}).
  \item Models~A and C yield high consistency in the structural force-displacements curves (Fig.~\ref{fig:p4_ANotchedFuComp}) and damage contour plots (Figs.~\ref{fig:p4_AnotchedDfinalA} and \ref{fig:p4_AnotchedDfinalC}), while Model~B yields a delayed force reduction (Fig.~\ref{fig:p4_ANotchedFuComp}) with higher energy dissipation and a thicker damage zone (Fig.~\ref{fig:p4_AnotchedDfinalB}).
  \item Utilizing the local anisotropic damage model without gradient-extension results in a false crack path prediction (Fig.~\ref{fig:p4_AnotchedDlocal}).
  \item A single component regularization of the damage tensor proves to be inadequate and yields different degrees of crack localization (Figs.~\ref{fig:p4_ANotchedFuStudyAi}, \ref{fig:p4_AnotchedDstudyAi_u09}, \ref{fig:p4_AnotchedDstudyAi_u20}).
  \item With two nonlocal degrees of freedom, model~C is an efficient and accurate formulation with extraordinary coarse mesh accuracy (Figs.~\ref{fig:p4_RBFu}, \ref{fig:p4_Dmeshconbottom}, \ref{fig:p4_Dmeshcontop}, \ref{fig:p4_Dnormal_evolution}, and \ref{fig:p4_Dshear_evolution}).
\end{itemize}

\section{Conclusion}
\label{sec:p4_conclusion}

In this work, we successfully introduced a universal framework for nonlocal anisotropic damage at finite strains. Due to the design of the degradation functions, arbitrary established hyperelastic finite strain energy formulations can be incorporated into the anisotropic damage model. Furthermore, the generic micromorphic gradient-extension provides a versatile regularization that can prevent any desired number of local quantities from localization.

Initially, the behavior of a specific local anisotropic damage model utilizing a Neo-Hookean energy was investigated in single element studies and yielded an increased material strength for an increase of the degree of damage anisotropy. Thereafter, the model was applied with different gradient-extensions for the simulation of an asymmetrically notched specimen. It confirmed the accurate regularization capabilities of the volumetric-deviatoric regularization steming from previous works for a new local anisotropic damage model. Moreover, the regularization of single components of the damage tensor was studied for the asymmetrically notched specimen and yielded localization regardless of the selection of the gradient-extended quantity. Eventually, the anisotropic damage model with the volumetric-deviatoric regularization was used for the simulation of a rotor blade specimen that was subjected to a pressure load.

Subsequent works could investigate the anisotropic damage behavior of further hyperelastic finite strain energy formulations utilizing the presented framework by straightforwardly replacing the Neo-Hookean energy in the elastic term.

\subsection*{Acknowledgements}

Funding granted by the German Research Foundation (DFG) is gratefully acknowledged by T.~van der Velden, S.~Reese and T.~Brepols for project 453715964, by S.~Reese and H.~Holthusen for project 495926269, by S.~Reese, H.~Holthusen and T.~Brepols for project 417002380 (A01) and by S.~Reese and T.~Brepols for project 453596084 (B05).

%% file: 01_Sections/appendix.tex
\clearpage

\appendix

\section{Appendix}
\label{sec:p4_appendix}

\subsection{General fulfillment of the damage growth criterion}
\label{sec:p4_app1}

According to the damage growth criterion of \cite{WulfinghoffFassinEtAl2017}, the elastic energy has to decrease with any positive semi-definite increment of the damage tensor $\mathrm{d}\bm{D}$ for arbitrary stretch tensors $\bm{\tilde{C}}$. This is mathematically expressed by
\begin{equation}
  \psi_e ( \bm{\tilde{C}},\bm{D}+\mathrm{d}\bm{D} ) \leq
  \psi_e ( \bm{\tilde{C}},\bm{D} )\quad \forall \, \bm{\tilde{C}}
  \label{eq:p4_DGC}
\end{equation}
or
\begin{equation}
  \frac{\partial}{\partial\bm{D}} \, \psi_e ( \bm{\tilde{C}},\bm{D} ) : \dot{\bm{D}} \leq 0 \quad \forall \, \bm{\tilde{C}}.
  \label{eq:p4_DGC2}
\end{equation}
Since $\dot{\bm{D}}$ is always positive semi-definite, the partial derivative $\partial \, \psi_e ( \bm{C},\bm{D} ) / \partial\bm{D}$ must always satisfy negative semi-definiteness due to the following derivation with $\partial \psi_e ( \bm{C},\bm{D} ) / \partial\bm{D} =: \bm{A}$.

Any symmetric second-order tensor $\bm{A}$ is negative semi-definite if
\begin{equation}
    \bm{x} \cdot \bm{A} \cdot \bm{x} \leq 0 \quad \forall \, \bm{x} \, \backslash \, \{ \bm{0} \}
    \label{eq:p4_Anegsd}
\end{equation}
holds. Furthermore, employing the spectral decomposition for the rate of the damage tensor yields 
\begin{equation}
\dot{\bm{D}} =  \dot{D}_1 \, \bm{n}_1^{\scriptscriptstyle \re{\dot{D}}} \otimes \bm{n}_1^{\scriptscriptstyle \re{\dot{D}}} + \dot{D}_2 \, \bm{n}_2^{\scriptscriptstyle \re{\dot{D}}} \otimes \bm{n}_2^{\scriptscriptstyle \re{\dot{D}}} + \dot{D}_3 \, \bm{n}_3^{\scriptscriptstyle \re{\dot{D}}} \otimes \bm{n}_3^{\scriptscriptstyle \re{\dot{D}}}
\label{eq:p4_Ddot_sd}
\end{equation}
where $\dot{D}_i \geq 0$ and $\bm{n}_i^{\scriptscriptstyle \re{\dot{D}}}$ $(i=1,2,3)$ denote the eigenvalues and eigenvectors of $\dot{\bm{D}}$. Inserting Eq.~\eqref{eq:p4_Ddot_sd} into Eq.~\eqref{eq:p4_DGC2}  yields
\begin{equation}
    \bm{A} : \dot{\bm{D}} =  \bm{A} : \left(  \dot{D}_1 \, \bm{n}_1^{\scriptscriptstyle \re{\dot{D}}} \otimes \bm{n}_1^{\scriptscriptstyle \re{\dot{D}}} + \dot{D}_2 \, \bm{n}_2^{\scriptscriptstyle \re{\dot{D}}} \otimes \bm{n}_2^{\scriptscriptstyle \re{\dot{D}}} + \dot{D}_3 \, \bm{n}_3^{\scriptscriptstyle \re{\dot{D}}} \otimes \bm{n}_3^{\scriptscriptstyle \re{\dot{D}}}  \right).
    \label{eq:p4_ADdot1}
\end{equation}
Using $\bm{A} : (\bm{a} \otimes \bm{b}) = \bm{a} \cdot \bm{A} \cdot \bm{b}$, it follows that
\begin{equation}
  \bm{A} : \dot{\bm{D}} =
  \dot{D}_1 \, \underbrace{\bm{n}_1^{\scriptscriptstyle \re{\dot{D}}} \cdot \bm{A} \cdot \bm{n}_1^{\scriptscriptstyle \re{\dot{D}}}}_{\leq 0} +
  \dot{D}_2 \, \underbrace{\bm{n}_2^{\scriptscriptstyle \re{\dot{D}}} \cdot \bm{A} \cdot \bm{n}_2^{\scriptscriptstyle \re{\dot{D}}}}_{\leq 0} +
  \dot{D}_3 \, \underbrace{\bm{n}_3^{\scriptscriptstyle \re{\dot{D}}} \cdot \bm{A} \cdot \bm{n}_3^{\scriptscriptstyle \re{\dot{D}}}}_{\leq 0}
  \leq 0
  \label{eq:p4_ADdot2}
\end{equation}
is generally fulfilled, if $\bm{A}$ is negative semi-definite (cf.~Eq.~\eqref{eq:p4_Anegsd}).

Thus, the elastic energy of Eq.~\eqref{eq:p4_psie} must be analyzed with respect to the negative semi-definiteness of $\partial \, \psi_e / \partial\bm{D}$. The elastic energy reads with its functional arguments
\begin{equation}
  \psie(\C,\D) = \left( \left( 1 - \kani \right) \fiso(\D) + \kani \, \fani(\C,\D) \right) \psi_\star(\C)
\end{equation}
and the partial derivative with respect to the damage tensor
\begin{equation}
  \frac{\partial}{\partial \D} \, \psie(\C,\D) = \left( \left( 1 - \kani \right) \frac{\partial \fiso(\D)}{\partial \D} + \kani \, \frac{\partial \fani(\C,\D)}{\partial \D}  \right) \psi_\star(\C).
\end{equation}
Hence, the analysis of the sensitivity of the degradation functions, Eqs.~\eqref{eq:p4_fiso} and \eqref{eq:p4_fani}, suffices and yields
\begin{equation}
  \pd{\fiso(\D)}{\D}
  =
  \underbrace{ \ed \left( 1 -\frac{\tr{\D}}{3} \right)^{\ed-1} }_{\geq 0} \left( - \, \frac{1}{3} \, \I \right)
  \label{eq:p4_dfisodD}
\end{equation}
and
\begin{equation}
  \pd{\fani(\C,\D)}{\D}
  =
  \underbrace{ \fd \left( 1 -\frac{\tr{\C^2 \D}}{\tr{\C^2}} \right)^{\fd-1} }_{\geq 0}
  \left( - \frac{\C^2}{\tr{\C^2}} \right)
  \label{eq:p4_dfanidD}
\end{equation}
that are both negative semi-definite and, thus, the elastic energy in Eq.~\eqref{eq:p4_psie} generally fulfills the damage growth criterion.

\subsection{Isochoric violation of the damage growth criterion}
\label{sec:p4_app2}

The split of the elastic strain energy into a volumetric and an isochoric part is an established procedure in nonlinear solid mechanics (see e.g.~\cite{Holzapfel2000}). In damage modeling, isotropic material degradation can be associated with volumetric deformations and anisotropic material degradation with isochoric deformations. This results in the following composition for the elastic strain energy
\begin{equation}
  \psi_e = f_\mathrm{dam}(\bm{D}) \, \psi_\mathrm{volumetric}(J) + \psi_\mathrm{isochoric}(\bar{\bm{C}}, \bm{D})
  \label{eq:p4_psievoliso}
\end{equation}
where $f_\mathrm{dam}$ denotes an isotropic degradation function analogously to Eq.~\eqref{eq:p4_fiso}, $\psi_\mathrm{volumetric}$ the volumetric energy contribution, $\psi_\mathrm{isochoric}$ the isochoric energy contribution, $J$ the determinant of the deformation gradient $\F$, and $\bar{\bm{C}}$ the isochoric right Cauchy-Green stretch tensor with $\bar{\bm{C}} = \mathrm{det}(\bm{C})^{-1/3} \bm{C}$. Considering the strong form of the damage growth criterion requires the individual summands of Eq.~\eqref{eq:p4_psievoliso} to fulfill negative semi-definiteness in their partial derivatives with respect to the damage tensor separately. For the volumetric part, this proof is straightforward analogously to Eq.~\eqref{eq:p4_dfisodD}.

However, for the isochoric part, the analysis of a given energy will yield the violation of the damage growth criterion. An isochoric energy may read
\begin{equation}
  \psi_\mathrm{isochoric}(\bar{\bm{C}}, \bm{D}) = \frac{\mu}{2} \, \mathrm{tr} \left((\bar{\bm{C}}-\bm{I})(\bm{I}-\bm{D}) \right)
  \label{eq:psiiso}
\end{equation} 
with
\begin{equation}
  \frac{\partial}{\partial\bm{D}} \, \psi_\mathrm{isochoric}(\bar{\bm{C}}, \bm{D}) = - \frac{\mu}{2} \left( \bar{\bm{C}} - \bm{I} \right)
  \label{eq:dpsiisodD}
\end{equation} 
where Eq.~\eqref{eq:dpsiisodD} is required to be negative semi-definite, i.e.~to posses only negative or zero eigenvalues, to fulfill the damage growth criterion. The eigenvalues are analyzed with the spectral form of Eq.~\eqref{eq:dpsiisodD}
\begin{equation}
  - \frac{\mu}{2} \left( \bar{\bm{C}} - \bm{I} \right)
  =
  - \frac{\mu}{2} \, \sum_{i=1}^3 \left( \bar{\lambda}_i^2 - 1 \right) \bm{N}_i \otimes \bm{N}_i
  \label{eq:dpsiisodDspectral}
\end{equation} 
where $\bm{N}_i$ denote the principal directions of $\bm{C}$ and $\bar{\bm{C}}$ and $\bar{\lambda}_i$ the modified principal stretches (cf.~\cite{Holzapfel2000}) with 
\begin{equation}
  \bar{\lambda}_i = \frac{\lambda_i}{(\lambda_1 \lambda_2 \lambda_3)^{1/3}} , \quad i=1,2,3.
\end{equation} 
Now, assuming a uniaxial stress state with $\lambda_1 > 1$ and $\lambda_2 = \lambda_3 < 1$ yields for the eigenvalue associated with $\bm{N}_2 \otimes \bm{N}_2$ a positive value, viz.
\begin{align}
  -\frac{\mu}{2} \left( \bar{\lambda}_2^2 \, - \, 1 \right)
  &=
  -\frac{\mu}{2} \left( \left( \frac{\lambda_2}{(\lambda_1 \lambda_2 \lambda_3)^{1/3}} \right)^2 - \, 1 \right) \\
  &=
  -\frac{\mu}{2} \left( \left( \frac{\lambda_2^3}{\lambda_1 \lambda_2 \lambda_3} \right)^{2/3} - \, 1 \right) \\
  &\stackrel{\lambda_2=\lambda_3}{=}
  \underbrace{-\frac{\mu}{2}}_{<0} \underbrace{\Bigg( \bigg( \underbrace{\frac{\lambda_2}{\lambda_1}}_{<1} \bigg)^{2/3} - \, 1 \Bigg)}_{<0}
  > 0
\end{align} 
and, thus, violates the damage growth criterion.

\subsection{Derivative of anisotropic degradation function}
\label{sec:p4_app3}

The anisotropic degradation function employed in this work reads with the definition in Eq.~\eqref{eq:p4_fani}
\begin{equation*}
  \fani = \left( 1 - \frac{ \bigtr{ \C^2 \D } }{ \bigtr{ \C^2 } } \right)^\fd
\end{equation*}
and its partial derivative with respect to the right Cauchy-Green stretch tensor $\C$ follows as
\begin{equation}
  \pd{\fani}{\C}
  =
  \fd \left( 1 - \frac{ \bigtr{ \C^2 \D } }{ \bigtr{ \C^2 } } \right)^{\fd-1} \left( - \left( \D\C + \C\D \right) \frac{1}{\tr{\C^2}} + \tr{\C^2\D} \frac{1}{\tr{\C^2}^2} \, 2 \, \C \right).
  \label{eq:p4_dfanidC}
\end{equation}
Evaluating Eq.~\eqref{eq:p4_dfanidC} in the undamaged state yields
\begin{equation}
  \left. \pd{\fani}{\C} \right|_{\D=\bm{0}}
  =
  \fd \left( 1 - \frac{ \bigtr{ \bm{0} } }{ \bigtr{ \C^2 } } \right)^{\fd-1} \left( - \left( \bm{0} + \bm{0} \right) \frac{1}{\tr{\C^2}} + \tr{\bm{0}} \frac{1}{\tr{\C^2}^2} \, 2 \, \C \right) = \bm{0}
  \label{eq:p4_dfanidC_D0}
\end{equation}
and in the completely damaged state
\begin{equation}
  \left. \pd{\fani}{\C} \right|_{\D=\bm{\I}}
  =
  \fd \left( 1 - \frac{ \bigtr{ \C^2 } }{ \bigtr{ \C^2 } } \right)^{\fd-1} \left( - 2 \, \C \frac{1}{\tr{\C^2}} + \tr{\C^2} \frac{1}{\tr{\C^2}^2} \, 2 \, \C \right) = \bm{0}
  \label{eq:p4_dfanidC_D1}
\end{equation}
and, therewith, $\fani$ and its derivative fulfill the requirements of \cite{ReeseBrepolsEtAl2021} for the anisotropic degradation function.